\documentclass[10pt, journal, compsoc]{IEEEtran}

\ifCLASSOPTIONcompsoc
\usepackage[nocompress]{cite}
\else
\usepackage{cite}
\fi

\ifCLASSINFOpdf
\usepackage[pdftex]{graphicx}
\graphicspath{{../pdf/}{../jpeg/}}
\else
\usepackage[dvips]{graphicx}
\graphicspath{{../eps/}}
\fi

\usepackage[numbers]{natbib}
\usepackage{amsmath,amsfonts}
\usepackage{array}
\usepackage{url}
\usepackage{textcomp}
\usepackage{verbatim}
\usepackage{CJK}
\usepackage{indentfirst}
\usepackage{amsthm}
\usepackage{graphicx}
\usepackage{subfigure}
\usepackage{tabularx}
\usepackage{algorithm}  
\usepackage{algpseudocode}
\usepackage{xcolor}
\usepackage{makecell}
\usepackage{ragged2e}
\usepackage[normalem]{ulem}
\usepackage{tabu}              
\usepackage{multirow}                 
\usepackage{multicol}               
\usepackage{float}                    
\usepackage{makecell}                
\usepackage{booktabs}

\begin{document}
%
\title{Dynamic Evolution of Complex Networks: A Reinforcement Learning Approach Applying Evolutionary Games to Community Structure}
%
%
%

\author{Bin Pi, ~\IEEEmembership{Student Member,~IEEE,}
        Liang-Jian Deng, ~\IEEEmembership{Senior Member,~IEEE,}
        Minyu Feng, ~\IEEEmembership{Senior Member,~IEEE,}
        Matja\v{z} Perc,~\IEEEmembership{Member,~IEEE,}
        and J\"{u}rgen Kurths

\thanks{B. Pi, L.-J. Deng, and M. Feng were supported in part by the National Nature Science Foundation of China (NSFC) under Grant Nos. 12271083, 62206230, and 62273077, in part by the Sichuan Province's Science and Technology Empowerment for Disaster Prevention, Mitigation, and Relief Project under Grant No. 2025YFNH0001, and in part by the Natural Science Foundation of Chongqing under Grant No. CSTB2023NSCQ-MSX0064. M.P. was supported by the Slovenian Research and Innovation Agency (Javna agencija za znanstvenoraziskovalno in inovacijsko dejavnost Republike Slovenije) under Grant Nos. P1-0403 and N1-0232.}

\thanks{Bin Pi and Liang-Jian Deng are with the School of Mathematical Sciences, University of Electronic Science and Technology of China, Chengdu 611731, China (e-mail: liangjian.deng@uestc.edu.cn).

Minyu Feng is with the College of Artificial Intelligence, Southwest University, Chongqing 400715, China (e-mail: myfeng@swu.edu.cn).

Matja\v{z} Perc is with Faculty of Natural Sciences and Mathematics, University of Maribor,
Koro{\v s}ka cesta 160, 2000 Maribor, Slovenia, with Community Healthcare Center Dr. Adolf Drolc Maribor, Ulica talcev 9, 2000 Maribor, Slovenia, with Department of Physics, Kyung Hee University, 26 Kyungheedae-ro, Dongdaemun-gu, Seoul 02447, Republic of Korea, with Complexity Science Hub, Metternichgasse 8, 1030 Vienna, Austria, and with University College, Korea University, 145 Anam-ro, Seongbuk-gu, Seoul 02841, Republic of Korea.

J\"{u}rgen Kurths is with the Department of Complexity Science, Potsdam Institute for Climate Impact Research, 14473 Potsdam, Germany, and also with the Institute of Physics, Humboldt University of Berlin, 12489 Berlin, Germany.}

\thanks{Corresponding authors: Liang-Jian Deng and Minyu Feng.}}


%
%

\markboth{IEEE TRANSACTIONS ON PATTERN ANALYSIS AND MACHINE INTELLIGENCE}%
{Shell \MakeLowercase{\textit{et al.}}: Bare Demo of IEEEtran.cls for IEEE Journals}
%



\IEEEtitleabstractindextext{

\begin{abstract}
\justifying
Complex networks serve as abstract models for understanding real-world complex systems and provide frameworks for studying structured dynamical systems. This article addresses limitations in current studies on the exploration of individual birth-death and the development of community structures within dynamic systems. To bridge this gap, we propose a networked evolution model that includes the birth and death of individuals, incorporating reinforcement learning through games among individuals. Each individual has a lifespan following an arbitrary distribution, engages in games with network neighbors, selects actions using Q-learning in reinforcement learning, and moves within a two-dimensional space. The developed theories are validated through extensive experiments. Besides, we observe the evolution of cooperative behaviors and community structures in systems both with and without the birth-death process. The fitting of real-world populations and networks demonstrates the practicality of our model. Furthermore, comprehensive analyses of the model reveal that exploitation rates and payoff parameters determine the emergence of communities, learning rates affect the speed of community formation, discount factors influence stability, and two-dimensional space dimensions dictate community size. Our model offers a novel perspective on real-world community development and provides a valuable framework for studying population dynamics behaviors.

\end{abstract}

\begin{IEEEkeywords}
Complex networks, Evolutionary games, Reinforcement learning, Community structure, Stochastic process.
\end{IEEEkeywords}}

\maketitle

\IEEEdisplaynontitleabstractindextext

%
\IEEEpeerreviewmaketitle

\IEEEraisesectionheading{\section{Introduction}\label{sec:introduction}}

\IEEEPARstart{F}{rom} the internet to mobile communication networks \cite{onnela2007structure}, \cite{ramanathan1996survey}, social relationship networks \cite{heaney2008social}, \cite{berkman1985relationship}, biological networks \cite{girvan2002community}, \cite{pavlopoulos2011using}, and transportation networks \cite{banavar1999size}, \cite{bast2016route}, etc. \cite{lu2021localdrop}, \cite{schwikowski2000network}, \cite{zhang2021self}, complex networks have become ubiquitous and have an increasing impact on human production and life. How to establish a dynamic evolution mechanism for complex networks that accurately reflects real-world conditions is a key challenge in the field of complex networks. The origin of complex networks dates back to 1,960 when Erd{\H{o}}s and R{\'e}nyi \cite{erdHos1960evolution} established the theory of random graphs, which pioneered the study of complex networks. The ``six degrees of separation'' small-world experiment conducted by Milgram \cite{milgram1967small} in 1,967 is a classic example of an empirical study of complex networks. At the end of the 20th century, Watts and Strogatz \cite{watts1998collective} proposed small-world networks, Barab{\'a}si and Albert \cite{barabasi1999emergence} developed scale-free networks. They respectively revealed the small-world and scale-free properties of complex networks, built corresponding models to elaborate the mechanisms of these properties, and successfully explained the widespread small-world properties \cite{sen2003small}, \cite{rubinov2009small} and power-law distributions \cite{clauset2009power}, \cite{adamic2000power} in reality. Since then, network science has started to emerge, and various novel complex network models have sprung up in people's vision.

An extensively developed complex network model not only provides insights into the evolutionary patterns of real network structures but also facilitates the investigation of various collective behaviors within structured populations, including synchronization \cite{lu2019quad}, \cite{wang2009synchronization}, evolutionary games \cite{pi2022evolutionary}, \cite{wang2015evolutionary}, propagation \cite{li2022limited}, \cite{hens2019spatiotemporal}, and so on \cite{lu2011link}, \cite{du2008community}, \cite{travenccolo2009border}. This capability makes it a valuable tool for studying and understanding complex phenomena in diverse fields. In 1,992, Nowak and May \cite{nowak1992evolutionary} first introduced spatial chaos into the prisoner's dilemma, and their findings revealed that the cooperation frequency in the network remained stable at around 0.374, regardless of the specific values of the game parameters. Subsequent studies \cite{challet1997emergence}, \cite{ebel2002coevolutionary} have further demonstrated that spatial structure has a positive impact on the evolution of cooperation in evolutionary games. However, the study published by Hauert and Doebeli \cite{hauert2004spatial} challenged this notion by suggesting that spatial structure actually inhibits the evolution of cooperation in the snowdrift game. Later, Santos and Pacheco \cite{santos2005scale} conducted studies on scale-free networks and demonstrated that cooperators play a dominant role in both the prisoner's dilemma game and snowdrift game, providing a plausible explanation for the existence of cooperation in real-world scenarios. With the development of complex network theory and evolutionary game theory, various mechanisms have been proposed to promote cooperative behavior. One of the most renowned contributions is the five rules introduced by Nowak \cite{nowak2006five}. Furthermore, more recent researches have uncovered that reputation \cite{han2022role}, \cite{perc2017statistical}, reward and punishment mechanisms \cite{chica2023rewarding}, \cite{zhang2023evolutionary} and game transitions \cite{su2019evolutionary}, \cite{feng2023evolutionary} also play significant roles in facilitating the persistence of cooperation. These findings have shed new light on the diverse strategies and mechanisms that contribute to the survival and evolution of cooperative behavior in various contexts.

Complex network modeling based on real-world phenomena is a classical idea and has been strongly studied. After the efforts in recent years, substantial progress has been made in the study of complex networks, numerous statistical features of complex networks have been discovered, and a growing number of complex network models that are more realistic have been proposed one after another \cite{wen2021gravity}, \cite{zhang2021topological}, \cite{li2022complex}. For example, in recent years, some researchers found that the number of individuals in the network should not only increase but also decrease. Feng et al. \cite{feng2022heritable} noticed this point and proposed a novel evolving network model that takes the growing and decreasing process into account based on the queueing system. Ref. \cite{battiston2020networks} discussed the relationship between higher-order interactions and collective behavior based on the fact that interactions can often occur in groups of three or more individuals, which cannot be expressed simply in terms of pairwise interactions.

Despite the significant attention and fruitful results achieved in complex network modeling, existing research falls short in deriving the formation and evolutionary mechanisms of community structures from the game interactions among individuals. This phenomenon is pervasive in practice, ranging from interactions among small microorganisms \cite{west2007social} to large-scale strategic games among nations \cite{axelrod1985achieving}. Moreover, community structures are a key feature of complex networks \cite{zhang2007uncovering}. Numerous empirical studies have shown that many networks exhibit heterogeneity, i.e., they consist of various types of nodes rather than random connections among a vast number of nodes of the same nature, where there are more connections of the same type of nodes and relatively few connections of different types of nodes \cite{dunne2002network}, \cite{he2021survey}. We refer here to the subgraphs formed by nodes of the same type and the edges between these nodes as communities in the network. Communities observed in real-world networks represent distinct collections of different specific objects. For example, communities in citation networks consist of related papers on the same topic, communities in the World Wide Web are groups of websites discussing similar subjects, communities in social networks represent real social groups based on interests or backgrounds, and communities in biochemical networks or electronic circuit networks can denote specific functional units. The identification and analysis of communities in these networks provide valuable insights for understanding their structures and dynamics and optimizing their functionality.

\begin{center}
\begin{figure}[htbp]
\centering
\includegraphics[width=8cm,height=6cm]{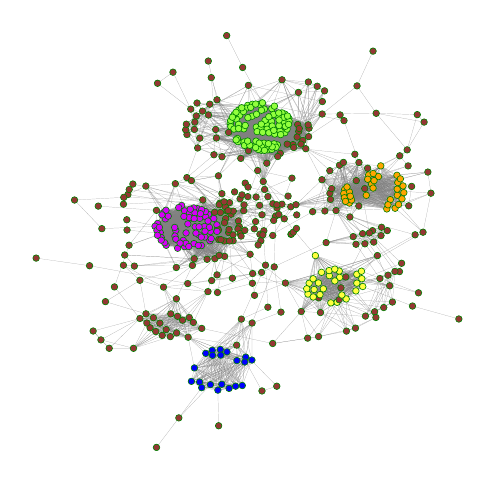}
\caption{\textbf{An example of the network structure generated according to our model.} The figure shows a network structure generated based on our model, where different colors denote different community structures, suggesting that the model proposed in this paper can provide an explanation for the formation and development of community structures in reality.}
\label{introduction figure}
\end{figure}
\end{center}

To fill this gap and address the shortcomings of previous studies, this paper primarily applies networked evolutionary game theory and Q-learning in reinforcement learning to uncover the formation and development mechanisms of community structures in real-world systems. To the best of our knowledge, this is the first study to model complex networks based on individual game interactions using reinforcement learning. Concretely, we assume that the individuals within a system are intelligent and have the potential to alter their locations in the two-dimensional space at any given moment to maximize their payoffs. They learn from past experiences to determine which mobile actions will yield the highest rewards, but occasionally make random choices due to limited intelligence or knowledge. Simultaneously, we develop a mechanism for transforming the distribution of individuals in a two-dimensional space to the network structure in a high-dimensional space. We note that individual rewards are determined through the games they are engaged in with their network neighbors. To model this process, we apply Q-learning in reinforcement learning, a model-free reinforcement learning approach, which identifies the optimal action plan based on the individual's current state. The typical structure of the network generated according to our model is shown in Fig. \ref{introduction figure}, which exhibits clear community structures, where different colors indicate different communities. Moreover, we conduct a comparative analysis of the evolution of community and cooperative behaviors both in systems with and without the birth-death process, and we apply them to real populations and networks, yielding promising results. Additionally, we perform a comprehensive analysis of the model's influence on the community and assess our model in comparison to the heuristic model and classical networks.

\vspace{-1.5\baselineskip}
\vspace{-2.5\baselineskip}

The model proposed in this paper offers a novel perspective for understanding the emergence and evolution of communities, bridges the gap left by previous studies in this aspect, advances the study of complex network modeling, and introduces a novel framework for examining the dynamics on structured populations, such as evolutionary games, propagation, and synchronization.

In summary, the major contributions of this paper can be outlined as follows:

\begin{itemize}
\item We propose an innovative model for the dynamic evolution of complex networks based on a game between individuals with reinforcement learning. Besides, we establish a transition mechanism that maps the distribution of individuals in the two-dimensional space to the network structure in a high-dimensional space.
\item We introduce a birth-death process of individuals to the system and determine the scale and distribution of the system at the steady state through both theoretical and experimental analyses. Furthermore, we compare it with systems without a birth-death process in numerical simulations.
\item The practicality of the model is validated by fitting the population of different countries and the degree distribution of real networks. The comparison between the model proposed in this paper and classical networks in terms of network structure, degree distribution, clustering coefficient, etc. is also conducted.
\item The model proposed in this paper provides a novel perspective on the emergence and development of community structure in reality, and the impacts of the parameters on community structure have been investigated from various aspects.
\end{itemize}

The remainder of this paper is structured as follows: In Section \ref{Model}, we display the details of the complex network model based on the game of individuals with reinforcement learning. In Section \ref{Simulations}, we conduct some simulations to validate our theory, observe the evolution of community structure and population cooperative behavior in systems both with and without the birth-death process, and fit the proposed model to population data and real networks. Subsequently, Section \ref{Analysis of the community structure} offers a comprehensive analysis of the formation and evolution of community structures from various perspectives. Then, we provide a comparative analysis of the proposed models in Section \ref{Model comparisons}. Finally, in Section \ref{Conclusion}, we summarize our study and provide some points for improvement.

\vspace{-0.5\baselineskip}
\section{Evolving networks based on a Game between Individuals with Reinforcement Learning}
\label{Model}
\small

In this section, we introduce a novel evolving network that considers the birth-death of individuals using the queuing theory. We assume that each individual with self-learning will move in the two-dimensional space based on their past experiences and interactions with neighbors in a game and we transform these individual dynamics into the corresponding network structure in a high-dimensional space through the proposed mechanism.

\vspace{-0.5\baselineskip}
\subsection{Birth-Death Process of Individual}
\label{Birth and Death Process of Individual}

First, we consider a system with $r\times r$ locations, where each location can be represented by a lattice in a two-dimensional space (containing $r\times r$ lattices). Each individual in the system has a lifetime that follows a general distribution. Besides, new individuals are input to the system at a specific rate $\lambda$ that follows an exponential distribution \cite{feng2022heritable} with the probability distribution:
\begin{small}
\begin{equation}
f_1(x; \lambda) = \left\{\begin{array}{l}
\lambda e^{-\lambda x}, \,\,x \geq 0 \\
0, \,\,x<0
\end{array}\right..
\end{equation}
\end{small}

The exponential distribution is employed to model individual births due to its memoryless property and mathematical simplicity, which make it an effective representation of independent random events in various natural processes. Additionally, the time intervals between events in many biological and physical phenomena align well with the assumptions of the exponential distribution \cite{bailey1991elements}, \cite{chakrabarti2020statistical}. For instance, in large natural populations, the birth times of individuals typically occur as uncorrelated random events, with intervals that often exhibit the characteristics of an exponential distribution.

\begin{center}
\begin{figure}[htbp]
\centering
\includegraphics[scale=0.26]{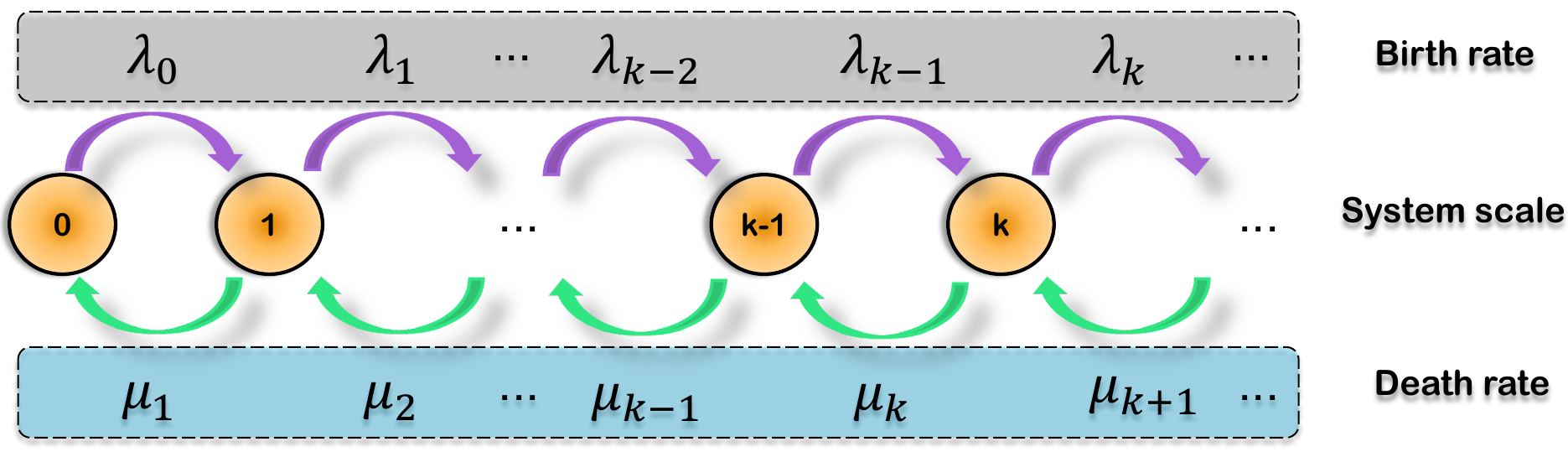}
\caption{\textbf{Phase transition of system scale.} The system scale undergoes a transition from $k - 1$ to $k$ at an exponential rate of $\lambda_{k-1}$ or from $k$ to $k - 1$ at a general rate of $\mu_k$, and the system size is constrained within the range of $[0, \infty)$.}
\label{birth and death}
\end{figure}
\end{center}

Then, the size of the system $N(t)$, i.e., the number of individuals in the system at time $t$, can be regarded as a continuous-time Markov chain with the state space $E=\{0, 1, 2, \cdots\}$. In addition, we assume that each individual in the system is a customer, and its behavior before leaving the system at the end of its lifespan is a service. Thus, this dynamic process of the input and output of individuals can be described by a $M/G/\infty$ queueing system, where $M$ represents a Poisson process for the input of individuals, $G$ indicates the lifespan of an individual obeying a general distribution, and $\infty$ denotes the number of individuals in the system belonging to $[0, \infty)$. We illustrate the impact of the birth and death of individuals on the system size in Fig. \ref{birth and death}. The first and third rows depict the birth and death rates of the system, respectively, with birth intervals following an exponential distribution and lifespans adhering to a general distribution. The second row indicates that the size of the system falls within the range of $[0, \infty)$.

\vspace{-1.5\baselineskip}

Next, we derive the scale at which the system evolves to reach a plateau. Before presenting the results, we first give some illustrations. We suppose that $S$ is the service time of an individual and that an individual staying in the system at time $t$ indicates that the arrival time $x$ of the individual obeys $t-S<x<t$. The individual arrives before time $t$ is a uniform distribution with range $[0, t]$ since the arrival of individuals is a Poisson process. Therefore, we can deduce the probability that an individual arrives before time $t$ and still remains in the system is:
\begin{small}
\begin{equation}
X(t)=\int_0^t \frac{P[S>t-x]}{t} d x=\int_0^t \frac{1-G(t-x)}{t} d x,
\end{equation}
\end{small}
where $G$ represents the distribution function of the service time.

We consider that the probability of the number of individuals at time $t$ and those that arrive before $t$ follows binomial distribution and the arrival time of each individual is i.i.d. Then, for those $m\le n$, we get
\begin{small}
\begin{equation}
\begin{aligned}
P_{n,m}(N\mid A,t)&=C_{n}^{m}\left[ X\left( t \right) \right] ^m\left[ 1-X\left( t \right) \right] ^{n-m}
\\
&=\frac{n!\left[ X(t) \right] ^m[1-X(t)]^{n-m}}{m!(n-m)!}.
\end{aligned}
\end{equation}
\end{small}
Otherwise, i.e., for those $m>n$, it is obviously that
\begin{small}
\begin{equation}
P_{n,m}(N\mid A,t)=0.
\end{equation}
\end{small}

Therefore, the number of individuals in the system at time $t$ ($N(t)$) and those that arrive before $t$ ($A(t)$) follows the conditional probability:
\begin{small}
\begin{equation}
\begin{aligned}
	P_{n,m}(N\mid A,t)&=P\{N(t)=m\mid A(t)=n\}\\
	&=\left\{ \begin{matrix}{}
	\frac{n!X(t)^m[1-X(t)]^{n-m}}{m!(n-m)!},&m\le n\\
	0, \,\,otherwise&\\
\end{matrix} \right. .\\
\end{aligned}
\end{equation}
\end{small}

Subsequently, we investigate the limiting probability distribution of the system size.

\newtheorem{thm}{Theorem}
\begin{thm}
\label{Theorem 1}
For the continuous-time Markov chain $N(t)$ with the state space $E$, assume that the expectation of death process $\{G(t), t\geq 0\}$ exists, its limiting probability $\{\pi_i, i = 1, 2, \cdots\}$ exists and follows
\begin{small}
\begin{equation}
\begin{aligned}
\pi _i&=\underset{t\rightarrow \infty}{\lim}p_i\left( t \right) =\underset{t\rightarrow \infty}{\lim}P\left\{ N\left( t \right) =i \right\} \\
&=\frac{\{\lambda E[G(t)]\}^i}{i!}e^{-\lambda E[G(t)]}.
\end{aligned}
\end{equation}
\end{small}
\end{thm}

The proof of Thm. \ref{Theorem 1} is given in Section 1 of the supplementary material for those who are interested. According to Thm. \ref{Theorem 1}, we can derive the probability distribution of the system size when it evolves to a steady state. Furthermore, we deduce certain statistical properties of the system size based on the limiting probability demonstrated in Thm. \ref{Theorem 1}.

\begin{thm}
\label{Theorem 2}
The average scale of the system is 
\begin{small}
\begin{equation}
\label{EN}
E[N(t)]=\lambda E[G(t)],
\end{equation}
\end{small}
the variance of the scale is
\begin{small}
\begin{equation}
\label{variance}
D[N(t)]=\lambda E[G(t)],
\end{equation}
\end{small}
and the average staying time of each individual is
\begin{small}
\begin{equation}
E(T)=E[G(t)].
\end{equation}
\end{small}
\end{thm}

The proof of Thm. \ref{Theorem 2} can be found in Section 2 of the supplementary material for those who are interested. Therefore, we can obtain the expected scale of the system, the dispersion of a set of scales, and the lifespan of an individual when the parameter $\lambda$ and the death process $\{G(t), t\geq 0\}$ are known.

\vspace{-0.5\baselineskip}
\subsection{Individual Movement Based on Reinforcement Learning}

We consider the case that each individual in the system has the potential to change its location at each moment to pursue a greater profit. Besides, every individual is intelligent, i.e., can get information from past experiences to determine which mobile actions will yield the highest benefits. In other words, the individual's choice of mobile behavior is based on learning experience, rather than obeying some simple rules. However, the intelligence of each individual is limited, i.e., at each time step, the individual can choose to move to a location that can bring the maximum payoff with the probability of $\delta$ (this behavior is called exploitation) and also has a random probability $\epsilon$ (this behavior is called exploration) to make a random choice of location to move to, where $\epsilon + \delta = 1$. Therefore, we employ Q-learning with an $\epsilon$-greedy strategy to model the aforementioned process in this paper. This is a model-free (does not require a model of the environment) reinforcement learning that will find the best action according to the individual's current state. Concretely, an individual performing an action in a particular state provides it with a reward, and the goal of the individual is to maximize the cumulative reward at all steps. Each individual has a utility table called Q-table learned from its action experience through the Bellman equation, which leads the individual to take the best action in each state.

Hereby, we provide a detailed description of the process of individual movement. In the system, the state of each individual is the location it lives in, which can be represented by dividing the two-dimensional space into equal-sized grids, where a grid is regarded as a location. Each individual will choose to move during a one-time step based on its current state and past experience. Each individual has six optional movement actions: move left, move right, move up, move down, stay, and random move. The first four actions mean that the individual will move to a location adjacent to the location it is currently in, ``stay'' indicates that the individual will not change its location, and ``random move'' suggests that the individual will move to a location randomly selected from all locations. When an individual performs a specific action, it will receive a certain amount of reward, which is significant for the individual's future action choice.

As we described before, each individual has a dynamic utility table to guide it in choosing the best action for each state, and the updating rule of the utility table at each time step can be expressed as
\begin{small}
\begin{equation}
\label{q-learning}
Q_{S_t}^{A_t}(t+1)=Q_{S_t}^{A_t}(t)+\eta[ R_{S_t}^{A_t}(t+1)+\gamma \max_{a\in A} Q_{S_{t+1}}^{a}(t)-Q_{S_t}^{A_t}(t)],
\end{equation}
\end{small}
where $Q_{S_t}^{A_t}(t)$ represents the utility gained by the individual in state $S_t$ when executing action $A_t$ at time $t$. $R_{S_t}^{A_t}(t+1)$ denotes the immediate payoff obtained by the individual in state $S_t$ at time $t+1$ when performing action $A_t$ at time $t$, and the calculation of the individual's immediate payoff will be explained in detail in the next subsection. In addition, the $\max_{a\in A} Q_{S_{t+1}}^{a}(t)$ means the maximum utility received by the individual in the future after moving to the state $S_{t+1}$. Moreover, $\eta$ and $\gamma$ are the learning rate and discount factor, respectively. A greater $\gamma$ will cause the individual to focus more on past experiences, otherwise, the individual will focus more on immediate payoff. We note that the Q-table possessed by each individual is not static, and its internal values are evolving according to Eq. \ref{q-learning}.

\vspace{-0.5\baselineskip}
\subsection{Payoff Calculation and Strategy Evolution of Individual}

In this subsection, we introduce the calculation of the immediate payoff, which is based on the game between individuals and their neighbors. Before then, we illustrate the transition mechanism from the previously described movement of individuals between locations in the two-dimensional space to complex networks in high-dimensional space, which is primarily designed based on theories of memory decay \cite{brown1958some} and strength of ties \cite{newman2003structure}. Specifically, there are two main situations:

i) One is individuals $i$ and $j$ are in distinct locations at time $t$. In this case, the rules for edge formation and edge weight updating are as follows:

(a) If two individuals have no edge at the last time $t-1$, then there will also be no edge between them at the current time. It reflects the principle in social networks that the formation of ties typically requires a history of interaction. In practice, connections between individuals are generally established based on previous interactions. Therefore, if no prior interactions or connections exist, it is unlikely that a new connection will form between them.

(b) If there is an edge with weight $w_{i,j}^{t-1}$ between individuals $i$ and $j$ at the previous time $t-1$, then the weight of the edge will be updated to $w_{i,j}^{t}=w_{i,j}^{t-1}/\beta$, where $\beta$ is a decay factor. The decay mechanism aligns with the real-life principle of relationship maintenance, where relationships typically weaken over time, particularly when interactions between individuals diminish. If the updated weight exceeds the generation threshold of the edge $\sigma$, the edge will be retained; otherwise, it will be removed.

Therefore, the edge weight updating rule in this case can be expressed as
\begin{small}
\begin{equation}
w_{i,j}^{t+1}=\begin{cases}
	0, \,\,\text{\small no\,\,edge\,\,between\,\,$i$\,\,and\,\,$j$\,\,at\,\,$t$}\\
	w_{i,j}^{t}/\beta, \,\,\text{\small otherwise}\\
\end{cases}.
\end{equation}
\end{small}

ii) The other scenario is individuals $i$ and $j$ are in the same state at time $t$. In this case, the rules for edge formation and the updating of edge weights are as follows:

(a) If there is no edge between the two individuals at the last time $t-1$, then a new edge will be established between them with an initial weight $\tau$. This mechanism models the creation of new interactions and the formation of relationships. For example, in social networks, individuals might form new connections when they meet through common activities or shared interests.

(b) If individuals $i$ and $j$ have an edge with weight $w_{i,j}^{t-1}$ at the previous time $t-1$, then the edge will be preserved and its weight will be updated to $w_{i,j}^{t-1}/\beta + \tau$. Here, $w_{i,j}^{t-1}/\beta$ denotes the natural decay of the edge weight, simulating the natural weakening of the relationship over time. The addition of $\tau$ reflects the strengthening of the relationship due to new interactions or joint activities when $i$ and $j$ are in the same state. The updating mechanism aligns with the theory of the strength of ties, which posits that frequent interactions enhance the strength of connections between individuals.

Therefore, the edge weight updating rule in this case can be denoted as
\begin{small}
\begin{equation}
w_{i,j}^{t+1}=\begin{cases}
	\tau, \,\,\text{\small no\,\,edge\,\,between\,\,$i$\,\,and\,\,$j$\,\,at\,\,$t$}\\ 
	w_{i,j}^{t}/\beta +\tau, \,\,\text{\small otherwise}\\
\end{cases}.
\end{equation}
\end{small}

It is important to highlight that the decay factor $\beta$ governs the gradual attenuation of edge weights over time. A larger value of $\beta$ results in faster decay of the weights and a shorter memory effect of the network. This decay mechanism mirrors the natural weakening of relationships in real-world contexts or the diminishing impact of reduced interactions over time on the strength of relationships. Conversely, the weight $\tau$ represents the additional contribution from interactions when a relationship is either established or maintained between individuals. It reflects either the initial strength of a newly formed edge or the reinforcement of an existing relationship. In summary, the configuration of edge weights updating among individuals aligns with actual social behavior patterns, ensuring that the evolution of the network is consistent with real-world dynamics.

At each discrete time step, after all individuals have synchronously taken actions to reach new states, the networked structure is updated based on the newly adjusted weights among individuals. In the new network, each node represents an individual, and the edges between nodes indicate the interactions between pairs of individuals. Subsequently, each individual synchronously engages in the snowdrift game with its neighbors in the updated network to obtain payoffs. The interactions between different strategy pairs result in different payoffs, which can be represented by the following payoff matrix:
\begin{small}
\begin{equation}
\label{payoff matrix}
M=\left( \begin{matrix}
	1&		1-r\\
	1+r&		0\\
\end{matrix} \right),
\end{equation}
\end{small}
where $r$ belongs to $[0,1]$, which is an adjustable payoff parameter for the snowdrift game. We utilize the strategy notation of two-dimensional unit vectors as $s_i = [1,0]^T$ for cooperate and $s_i = [0,1]^T$ for the defect. Accordingly, the payoff $U_i$ obtained by individual $i$ can be expressed as follows:
\begin{small}
\begin{equation}
U_i=\sum_{j\in \Omega}{s_{i}^{T}}Ms_j,
\end{equation}
\end{small}
where $\Omega$ means the set of neighbors for individual $i$ in the updated network structure. We emphasize that the cumulative payoff for all individuals in the same state corresponds to the immediate payoff obtained when an individual takes a specific action. Next, each individual performs the strategy evolution according to the Fermi function \cite{hu2021unfixed}, \cite{chiba2024social}. Specifically, individuals synchronously and randomly select one of their neighbors for payoff comparison and adopt the neighbor's strategy in the next discrete time step with a probability given by:
\begin{small}
\begin{equation}
\label{strategy updating}
P\left( s_i\gets s_j \right) =\frac{1}{1+\exp \left[ \left( U_i-U_j \right) /\kappa \right]},
\end{equation}
\end{small}
where $\kappa$ denotes the noise intensity, which is utilized to characterize the uncertainty associated with the process of strategy evolution. A larger value of $\kappa$ implies a higher tendency for individuals to adopt strategies in a more stochastic manner.

\begin{center}
\begin{figure}[htbp]
\centering
\includegraphics[scale=0.3]{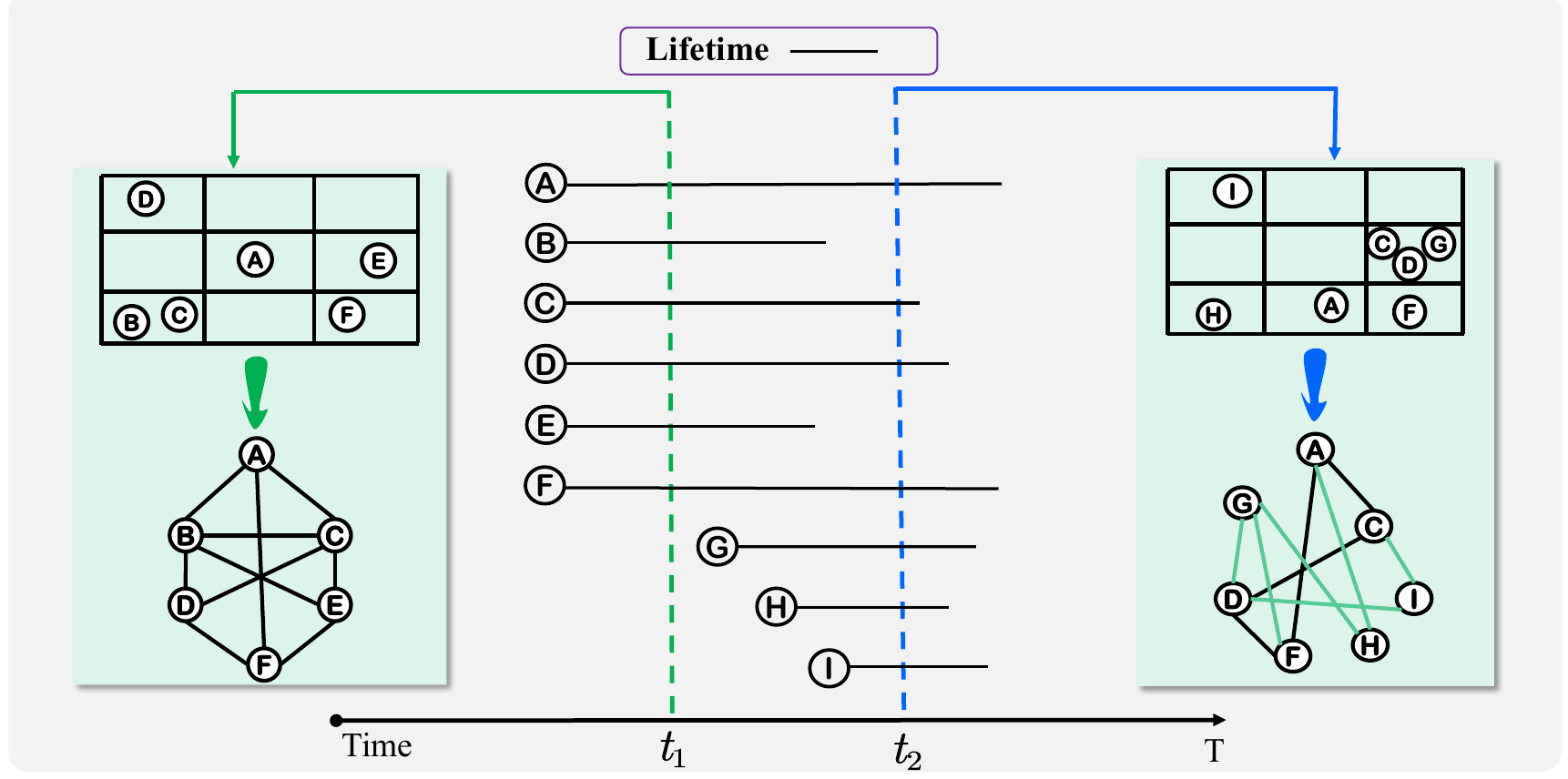}
\caption{\textbf{An illustration of the model.} The figure shows the evolution example of the complex network model. Black duration indicates the lifetime of the individual, which follows a general distribution. The individual will be removed from the system once its lifetime ends. We select $t_1$ and $t_2$ to observe the snapshots of the two-dimensional space and the network, which have a one-to-one correspondence between them. When $t=t_2$, individuals B and E in $t_1$ have died since their lifetimes are over, and new individuals G, H, and I are born in the system. The connections between newly born individuals and individuals already in the network are shown in green.}
\label{example}
\end{figure}
\end{center}
\vspace{-0.5\baselineskip}

\vspace{-1.2\baselineskip}
To enhance the clarity of our model, we give an example in Fig. \ref{example}. The black line indicates the lifetime of each individual. We choose evolution times $t_1$ and $t_2$ for observation. At $t = t_1$, there are six individuals in the system, and the corresponding networked structure can be obtained based on the locations and weights between individuals in the two-dimensional space. At $t = t_2$, individuals B and E in $t_1$ have passed away, because their lifespans ended. New individuals G, H, and I join the system and create a new network, since the addition of new individuals and individual movements, where interactions between new individuals and individuals already in the network are highlighted in green.

\vspace{-0.5\baselineskip}
\section{Simulations and results}
\label{Simulations}

In this section, we conduct extensive simulation experiments to validate the previous theory. To begin with, we introduce the methods of simulation experiments. Subsequently, we study the evolution and statistical distribution of individual numbers under different death processes to validate the theoretical analysis. Next, we compare the evolution of cooperative behaviors, the emergence and development of community structures, and the effects of payoff parameters and exploitation rates on community structures in systems both with and without birth-death process (SWBD and SWOBD)\footnote{The code of the SWBD and SWOBD is available at https://github.com/BinPi123/Dynamic-Evolution-of-Complex-Networks}. It is worth noting that for the sake of clarity, for systems with the birth-death process, we only choose to show results where the death process obeys a power-law distribution (SWBD with power-law distribution), and the interested reader can refer to Section 3 of the supplementary material to access all the results. Then, we apply our model to fit populations from four different countries and degree distributions of six distinct real networks.

\vspace{-0.5\baselineskip}
\subsection{Methods}

Herein, we explain some methods for our subsequent simulations. At the initial moment, each individual is uniformly distributed among different locations, the utility tables are initialized to 0, and individuals select cooperation or defection as their initial strategy with the same probability. We divide the two-dimensional space into $10 \times 10$ grids of equal size, with each location corresponding to one grid. The initial network is a complete graph and then evolves dynamically as individuals move around. For the generation of systems with the birth-death process, we generate a Poisson process based on the fact that it has the property of an exponential time interval, which is generated by the function $expovariate$ of the package $random$. Once an individual is born, it will be connected to all individuals in the same state and is assigned a lifespan that obeys a specific probability distribution. In this study, we explore four distinct death processes that follow different distributions, all of which are widely used in practice. The first one is the power-law distribution, often observed in urban population dynamics, where a small number of large cities have significantly higher populations compared to the majority of smaller cities, which have relatively modest populations \cite{gabaix2004evolution}. The second distribution is the uniform distribution, frequently employed in simple random sampling, where each individual has an equal chance of being selected for death without distinction \cite{olken1995random}. The third is the exponential distribution, which is typically used to model the lifetimes of devices. The non-memory property of this distribution makes it appropriate for scenarios where the failure rate of a device remains constant over time \cite{bisquert2009electron}. Lastly, we consider the lognormal distribution, which is commonly observed in certain biological characteristics, such as the sizes of human organs, cell dimensions, and organism lifespans \cite{limpert2001log}.

Next, we describe the generation for the four distributions. The power-law distribution is generated using the inverse transform technique. Concretely, the probability distribution of the power-law distribution \cite{newman2005power} is defined as follows:
\begin{small}
\begin{equation}
\label{powerlaw}
f_1(x; \alpha) =\frac{\alpha -1}{x_{min}^{1-\alpha}}x^{-\alpha}, \,\,\alpha>2,
\end{equation}
\end{small}
where $x_{min}$ and $\alpha$ indicate the lower bound and exponent of the distribution, respectively. Accordingly, its distribution function can be derived as follows:
\begin{small}
\begin{equation}
F\left( x \right) =P\left\{ X\le x \right\} =\int_{x_{min}}^x{f_1\left( x; \alpha \right)}dx=1-\frac{x^{1-\alpha}}{x_{\min}^{1-\alpha}}.
\end{equation}
\end{small}
Denote $u\sim U(0,1)$ as a random variable following a uniform distribution in the range $(0, 1)$ and $u=F(x)$, then we have
\begin{small}
\begin{equation}
\label{inverse}
x=F^{-1}(u)=x_{\min}\left( 1-u \right) ^{\frac{1}{1-\alpha}}.
\end{equation}
\end{small}
According to the inverse transform technique, the variable $x$ in Eq. \ref{inverse} follows a power-law distribution with a lower bound $x_{min}$ and exponent $\alpha$. It can be further simplified to $x=x_{\min}u^{\frac{1}{1-\alpha}}$, given that $1-u$ also follows a uniform distribution within the range $(0, 1)$. Consequently, a power-law distribution can be generated using $x_{\min}u^{\frac{1}{1-\alpha}}$, where $u\sim U(0,1)$. For the generation of uniform and exponential distributions, we utilize the $uniform$ and $expovariate$ functions from the $random$ package, respectively. Besides, the lognormal distribution is generated using the $random.lognoraml$ function from the $numpy$ package.

Once the lifespan of an individual is over, it is removed from the network, along with its connections to its neighbors. In the subsequent simulations, we conduct experiments on both systems with and without the birth-death process to demonstrate the effect of this process on the results. All final results are carried out on $Python$ and are averaged over ten independent simulations to ensure high accuracy.

\vspace{-1.5\baselineskip}
\begin{center}
\begin{figure*}[htbp]
\centering
\subfigure[Evolutionary curves]{
\includegraphics[scale=0.25]{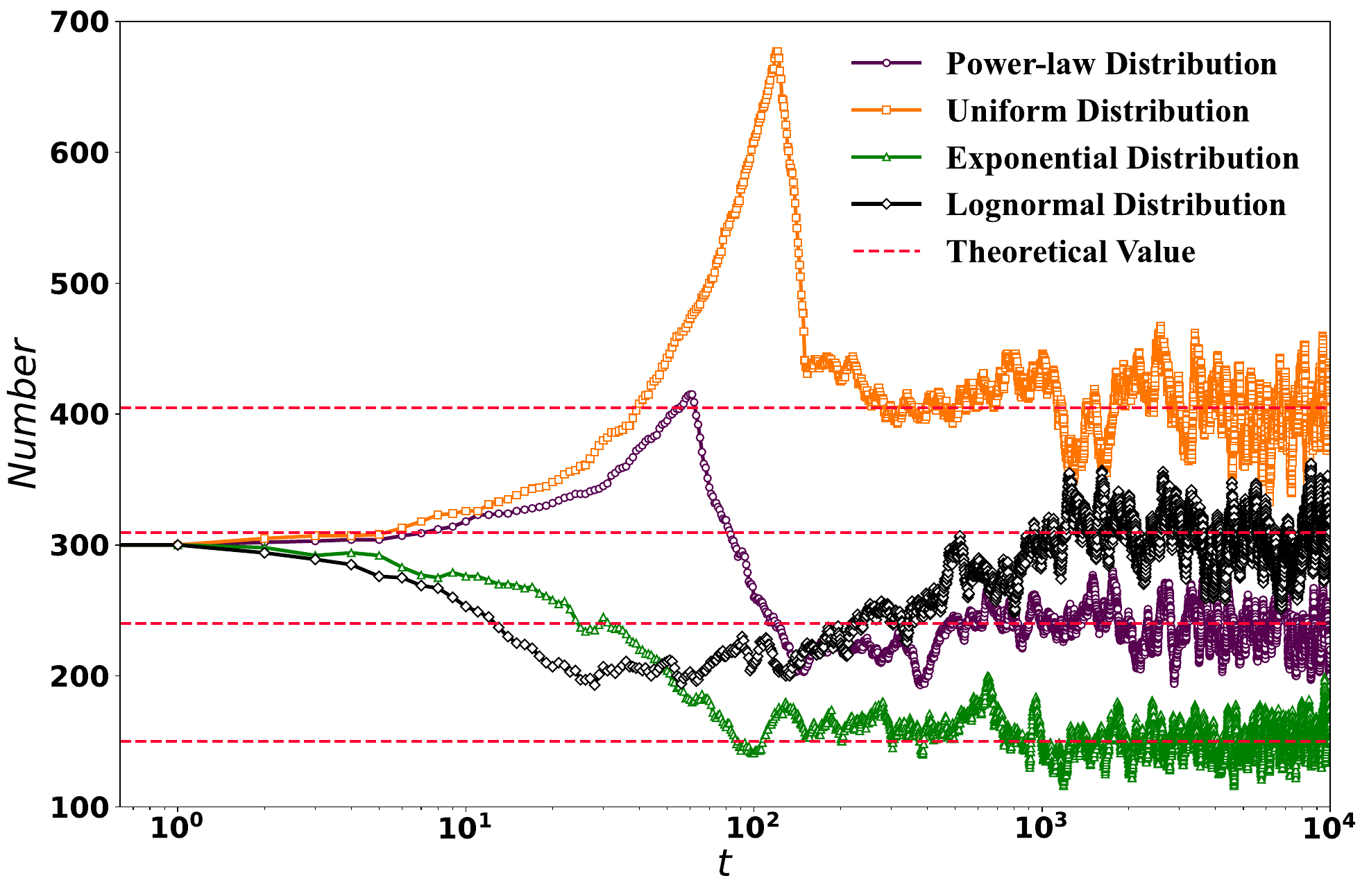}
\label{t_number}}
\hspace{0.5cm}
\subfigure[Statistical distributions]{
\includegraphics[scale=0.31]{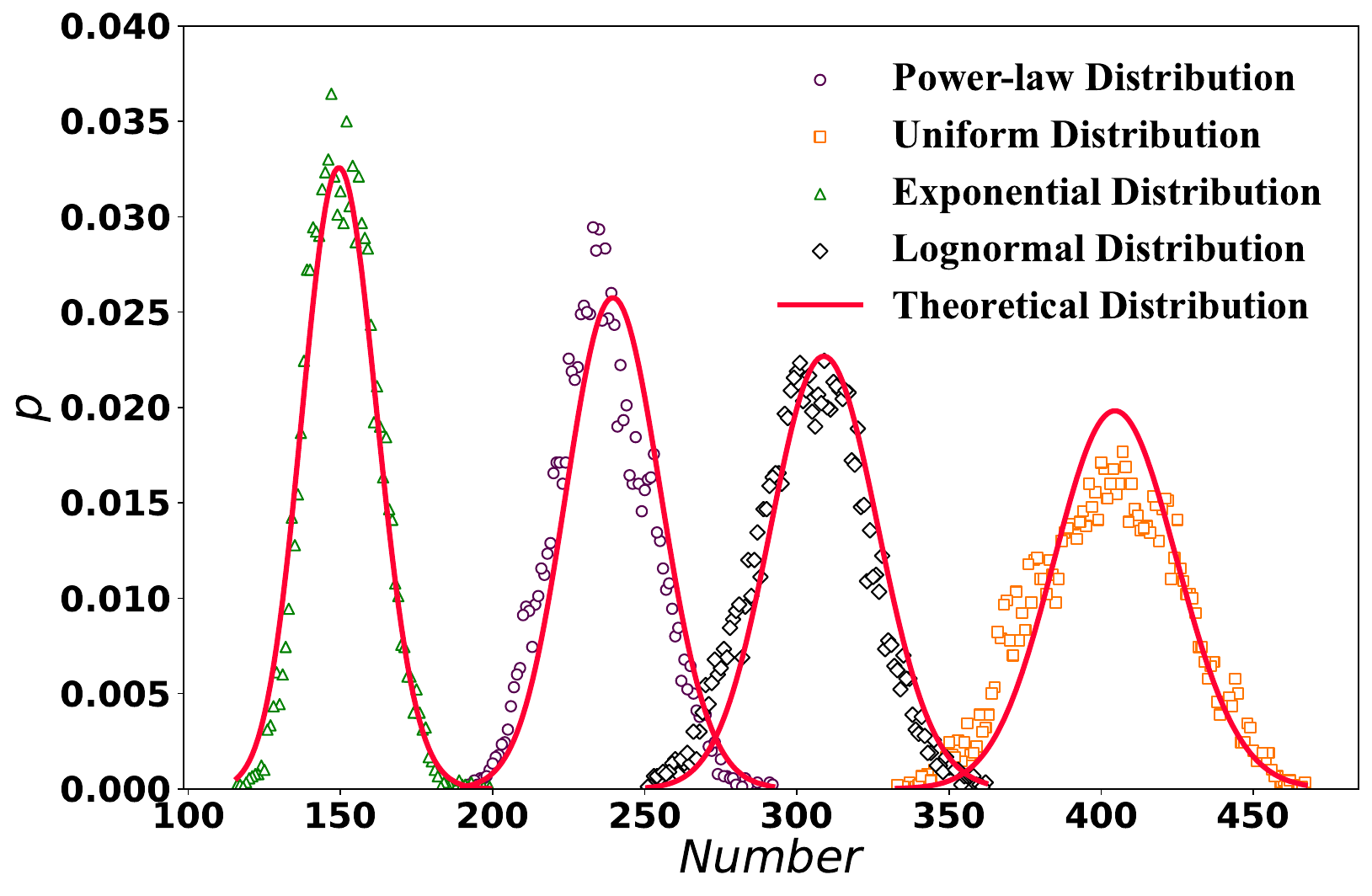}
\label{t_number_statistics}}
\caption{\textbf{Evolutionary curves and statistical distributions of the number of individuals.} In this figure, we show both the (a) evolutionary curves and (b) statistical distribution of the individual number in the system with the birth-death process under various death processes, including power-law, uniform, exponential, and lognormal distributions. In addition, we utilize different markers and colors for comparison among different death processes. The red line indicates the theoretical value for each specific death process.}
\label{t}
\end{figure*}
\end{center}

In brief, Algor. \ref{Monte Carlo Simulation} illustrates the Monte Carlo process for the complex network modeling based on the game between individuals with Q-learning in reinforcement learning. It is important to note that, at each discrete time step, all individuals first move synchronously and then play the snowdrift game with their neighbors in the updated network structure. Following this, the Q-table and strategy of each individual are updated synchronously.

\begin{algorithm}[htp]
\caption{\textbf{Monte Carlo Simulation}}
\label{Monte Carlo Simulation}
\begin{algorithmic}[1]
\State \textbf{Initialize} the current time $t_{now}=0$, the update time $t_{update}=0$, the weights among individuals $weight$, the network structure $\mathcal{G}$, each individual's state $S$, action $a$, utility table $Q(S, a)$, strategy $s$, and payoff $U$. 
\State \textbf{Initialize} the death time of each individual and the birth time of the next individual.
    \While {$t_{now}< MC_{step}$}
		\While {$t_{update}< t_{now}$}
            \For {each individual $x$}
                \State Choose action $a$ from $Q$-table based on \indent \indent \hspace{0.1cm} $\epsilon$-greedy policy.
                \State Take action $a$ and observe the new state $S'$.
            \EndFor
            \State Update the weights among individuals $weight$ \indent \hspace{0.25cm} and obtain a new network structure $\mathcal{G}$.
            \State Calculate each individual's payoff $U$ based on the \indent \hspace{0.25cm} game interactions.
            \For {each individual $x$}
                \State Update $Q(S, a)$: $Q(S, a) \leftarrow Q(S, a) + \eta[R + \indent \indent \hspace{0.2cm} \gamma \max_{a'} Q(S', a') - Q(S, a)]$.
                \State Randomly select a neighbor $y$ and generate a \indent \hspace{0.75cm} random number $p$ between 0 and 1.
                \If {$p < P(s_{x} \leftarrow s_{y})$}
                    \State Update strategy: $s_{x} \leftarrow s_{y}$.
                \EndIf
            \EndFor
        \State Increase update time: $t_{update}\leftarrow t_{update}+1$.
        \EndWhile
    \State Determine the time $t_{death}$ of the first individual to \indent \hspace{-0.25cm} die and the time $t_{birth}$ when the next individual $j$ is \indent \hspace{-0.2cm} born into the system.
    \If{$t_{birth} < t_{death}$}
        \State Update current time: $t_{now} \leftarrow t_{birth}$.
        \State Determine the death time of individual $j$ and the \indent \hspace{0.25cm} birth time of the subsequent new individual.
        \State Update the weights among individuals $weight$ \indent \hspace{0.25cm} and the network structure $\mathcal{G}$.
    \Else
        \State Update current time: $t_{now} \leftarrow t_{death}$.
        \State Update the weights among individuals $weight$ \indent \hspace{0.25cm} and the network structure $\mathcal{G}$.
    \EndIf
\EndWhile
\end{algorithmic}
\end{algorithm}

\vspace{-1\baselineskip}
\subsection{Evolution and Statistics of the Number of Individuals}
\vspace{-0.5\baselineskip}

\begin{center}
\begin{table*}[htbp]
\renewcommand{\arraystretch}{1.5}
\caption{The results of the stationary number of individuals under different death processes}
\begin{center}
\begin{tabular}{c|c|c|c|c}
\Xhline{1.5pt}
Results & \makecell{power-law distribution} & \makecell{uniform distribution} & \makecell{exponential distribution} & \makecell{lognormal distribution}\\
\hline
Theoretical result & 240 & 405 & 150 & 309.339 \\
Experimental result & 236.233 & 402.169 & 150.856 & 304.971 \\
Relative error & 1.569\% & 0.699\% & 0.571\% & 1.412\% \\
Variance & 242.141 & 545.242 & 127.547 & 336.829 \\
Skewness & 0.141 & -0.062 & 0.208 & -0.027 \\
Kurtosis & -0.241 & -0.481 & -0.099 & -0.152 \\
\Xhline{1.5pt}
\end{tabular}
\label{t_number_table}
\end{center}
\end{table*}
\end{center}

To validate the theory presented in subsection \ref{Birth and Death Process of Individual}, we begin by plotting the evolutionary curves and statistical distributions of the number of individuals in the system under different death processes over time, and the results are shown in Fig. \ref{t}. According to the theoretical results, we know that the final size of the system is determined solely by the expected value of the death process distribution and the input rate $\lambda$. Therefore, our analysis primarily focuses on the impact of various death process distributions on the number of individuals, including power-law, uniform, exponential, and lognormal distributions. For the power-law distribution, the parameters are set to $\lambda = 2, \alpha = 3$, and $x_{min} = 60$. For the uniform distribution, we set the birth rate to $\lambda = 3$, with the starting and ending values set to $a = 120$ and $b = 150$, respectively. In the case of the exponential distribution, the birth rate is $\lambda = 3$ and the death rate is $\kappa = 0.02$. Lastly, for the lognormal distribution, the birth rate, mean, and standard deviation are set as $\lambda = 5, \upsilon = 3$, and $\phi = 1.5$, respectively.

\begin{center}
\begin{table*}[htbp]
\renewcommand{\arraystretch}{1.5}
\caption{The comparison between the theoretical and experimental distributions under different death processes}
\begin{center}
\begin{tabular}{c|c|c|c|c}
\Xhline{1.5pt}
Results & \makecell{power-law distribution} & \makecell{uniform distribution} & \makecell{exponential distribution} & \makecell{lognormal distribution}\\
\hline
Kullback-Leibler divergence & 0.039 & 0.045 & 0.022 & 0.038 \\
Jensen-Shannon divergence & 0.009 & 0.012 & 0.005 & 0.009 \\
Pearson correlation coefficient & 0.942 & 0.957 & 0.989 & 0.965 \\
Cosine similarity & 0.975 & 0.977 & 0.994 & 0.985 \\
\Xhline{1.5pt}
\end{tabular}
\label{t_number_statistics_table}
\end{center}
\end{table*}
\end{center}
\vspace{-0.5\baselineskip}

We set the temporal coordinates of Fig. \ref{t_number} to logarithmic coordinates to better observe the rising and smooth phases of individual population evolution. As time progresses, the evolutionary curves of the number of individuals become stable around $t=1000$ in all four cases and subsequently fluctuate steadily around the theoretical value, as indicated by the red dashed line. However, the magnitude of these fluctuations varies among the different curves. To show the similarity between theoretical and experimental results more intuitively, we present the theoretical and experimental results for different death processes in Tab. \ref{t_number_table}, where the theoretical results are calculated according to Eq. \ref{EN}, and we have $E[N(t)] = \frac{\lambda(\alpha - 1)}{\alpha - 2} x_{min}$ for power-law distribution, $E[N(t)] = \frac{\lambda}{2} (a + b)$ for uniform distribution, $E[N(t)] = \frac{\lambda}{\kappa}$ for the exponential distribution, and $E[N(t)] = \lambda e^{\upsilon + \phi ^2 / 2}$ for lognormal distribution. The experimental results are averaged using the last 9,000 steps of a total of 10,000 evolutionary steps, and the relative error is obtained by $e=\left| x^*-x \right|/x$, where $x^*$ is the experimental result and $x$ is the theoretical one. Besides, we calculate the skewness to characterize the direction and degree of skewness of the data distribution, and the kurtosis to describe the steepness of the data distribution. It can be seen that the maximum relative error in Tab. \ref{t_number_table} is just 1.569\%, which illustrates that the experimental results are well matched with the theoretical ones. Moreover, the variance becomes larger as the size of the system increases, which is consistent with Eq. \ref{variance} and the phenomenon shown in Fig. \ref{t_number}, where the curve exhibits greater fluctuations as the scale of the system grows. By comparing the skewness and kurtosis of four data distributions, we get that the data distributions obtained with uniform and lognormal distributions are left-skewed, while the other two are right-skewed, and all the data distributions are thin-tailed.

Then, we capture the number of individuals in the last 9,000 steps of the total of 10,000 evolutionary steps, compute the frequency of each specific number separately, and plot the probability distribution of the four death processes in Fig. \ref{t_number_statistics} by treating the frequencies as probabilities according to the law of large numbers. The solid red line is the theoretical distribution derived from Thm. \ref{Theorem 1}. It is evident that the theoretical distributions fit well with the experimental ones. To illustrate this more intuitively, we respectively utilize the Kullback-Leibler (KL) divergence (Eq. \ref{KL}), the Jensen-Shannon (JS) divergence (Eq. \ref{JS}), the Pearson correlation coefficient (Eq. \ref{Pearson}), and the cosine similarity (Eq. \ref{Cosine similarity}) to capture the distance between the theoretical and experimental distributions, and the results are demonstrated in Tab. \ref{t_number_statistics_table}.

\begin{small}
\begin{equation}
\label{KL}
KL(A \vert\vert B)=\sum A_i\log\frac{A_i}{B_i},
\end{equation}
\end{small}
where $A$ and $B$ respectively denote the theoretical and experimental distributions.

\begin{small}
\begin{equation}
\label{JS}
JS(A \vert\vert B)=\frac{1}{2}KL(A \vert\vert M)+\frac{1}{2}KL(A \vert\vert M),
\end{equation}
\end{small}
where $M=\frac{1}{2}(A+B)$.

\begin{small}
\begin{equation}
\label{Pearson}
r=\frac{E[(A-E(A))(B-E(B))]}{S(A)S(B)},
\end{equation}
\end{small}
where $E(X)$ and $S(X)$ represent the expectation and standard deviation of probability distribution $X$.

\begin{small}
\begin{equation}
\label{Cosine similarity}
cos=\frac{\sum A_i B_i}{\sqrt{\sum A_i^2}\sqrt{\sum B_i^2}}.
\end{equation}
\end{small}

It can be observed that the KL divergence (all less than 0.045) and JS divergence (all less than 0.012) are very small, while the Pearson correlation coefficient and cosine similarity are very close to 1 in all four death processes, which shows that the theoretical and experimental distributions are in very good agreement with each other, which also proves the validity of our theoretical analysis in subsection \ref{Birth and Death Process of Individual}.

\subsection{Evolution of Network Cooperation Behaviors}

\begin{center}
\begin{figure}[htbp]
\centering
\subfigure[\scalebox{0.88}{SWBD with power-law distribution}]{
\includegraphics[scale=0.28]{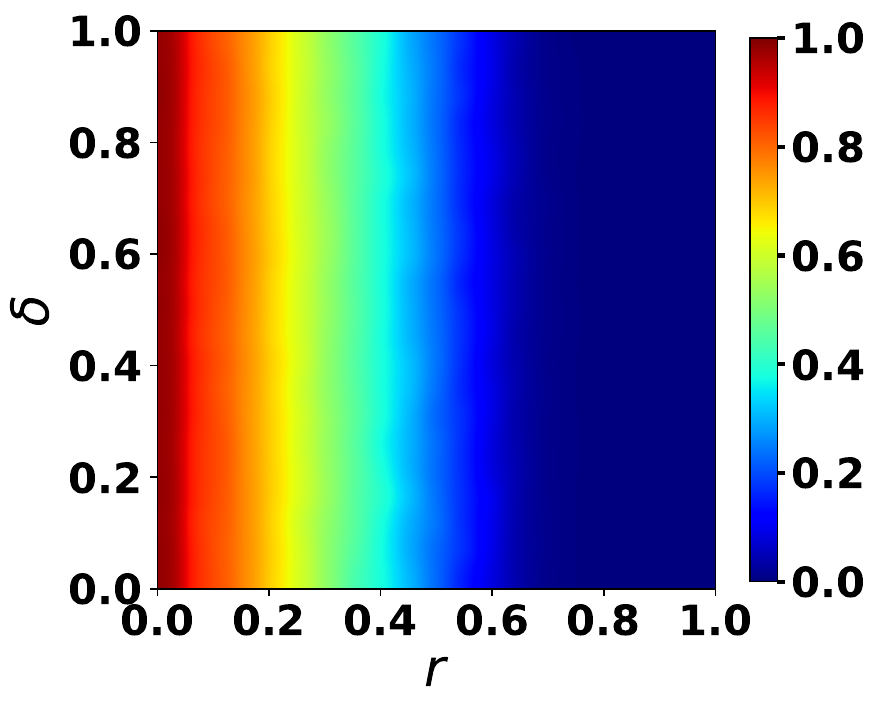}
\label{SWBD_delta_r_fc}}
\subfigure[\scalebox{0.88}{SWOBD}]{
\includegraphics[scale=0.28]{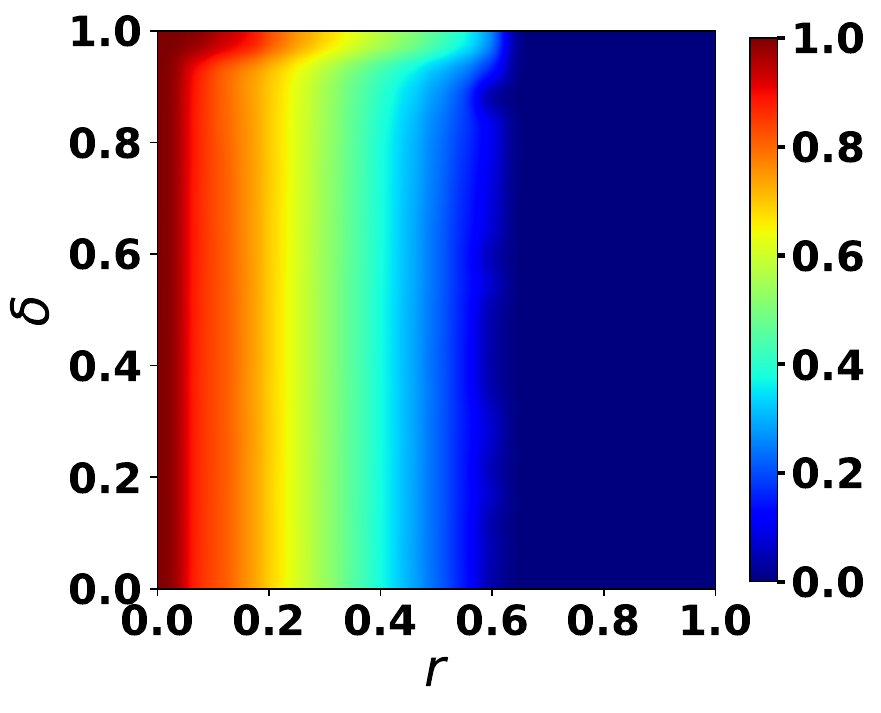}
\label{SWOBD_delta_r_fc}}
\caption{\textbf{Heat maps of cooperation fraction about payoff parameter $r$ and exploitation rate $\delta$.} By setting the $y$-axis as the exploitation rate $\delta$ with a range [0, 1] and the $x$-axis as the payoff parameter $r$ with a range [0, 1], we demonstrate the heat maps of cooperation fraction on the system with (in panel (a)) and without (in panel (b)) birth-death process with respect to payoff parameter $r$ and exploitation rate $\delta$. The learning rate, discount factor, and weight fading factor of the edge in both subfigures are all set to 0.7, 0.3, and 2.}
\label{heatmaps_fc}
\end{figure}
\end{center}

In order to explore the effect of exploitation rate $\delta$ and payoff parameter $r$ on the cooperative behaviors of the system, we present the heat maps of the proportion of cooperators on SWBD with power-law distribution and SWOBD about $\delta$ and $r$ in Figs. \ref{SWBD_delta_r_fc} and \ref{SWOBD_delta_r_fc}, respectively. Both simulations share identical parameter settings, differing only in the presence or absence of the birth-death process. Concretely, we respectively set the learning rate, discount factor, and weight fading factor of the edge of the two simulations to 0.7, 0.3, and 2 to ensure the comparability of the results. In addition, the total evolutionary time for both simulations is set to 5,000 long enough to allow the evolution of the cooperation ratio to reach stability.

\vspace{-1.5\baselineskip}

\begin{center}
\begin{figure*}[htbp]
\centering
\subfigure[Individual distribution of SWBD with power-law distribution]{
\includegraphics[scale=0.19]{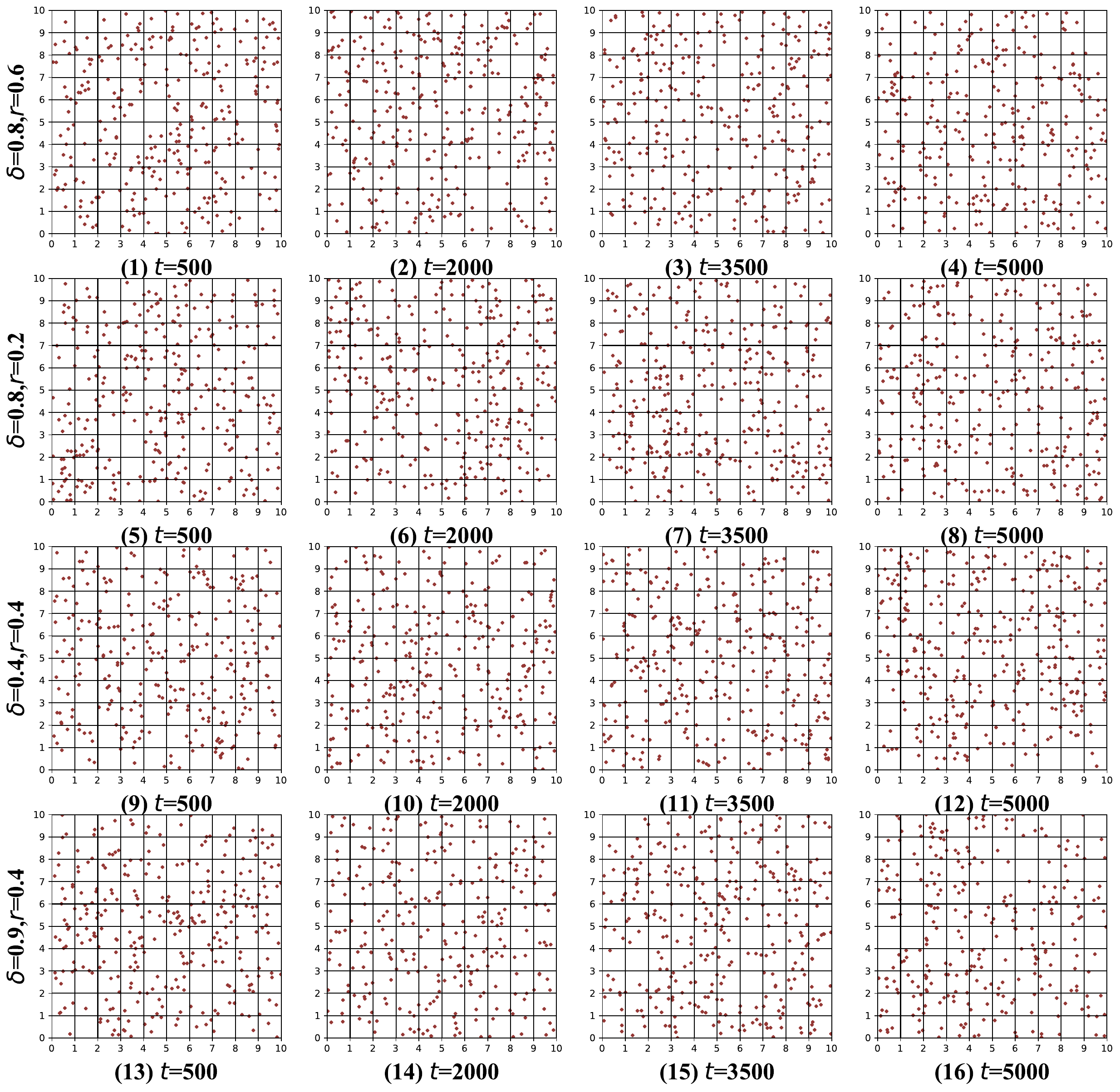}
\label{SWBD_snapshot}}
\subfigure[Individual distribution of SWOBD]{
\includegraphics[scale=0.19]{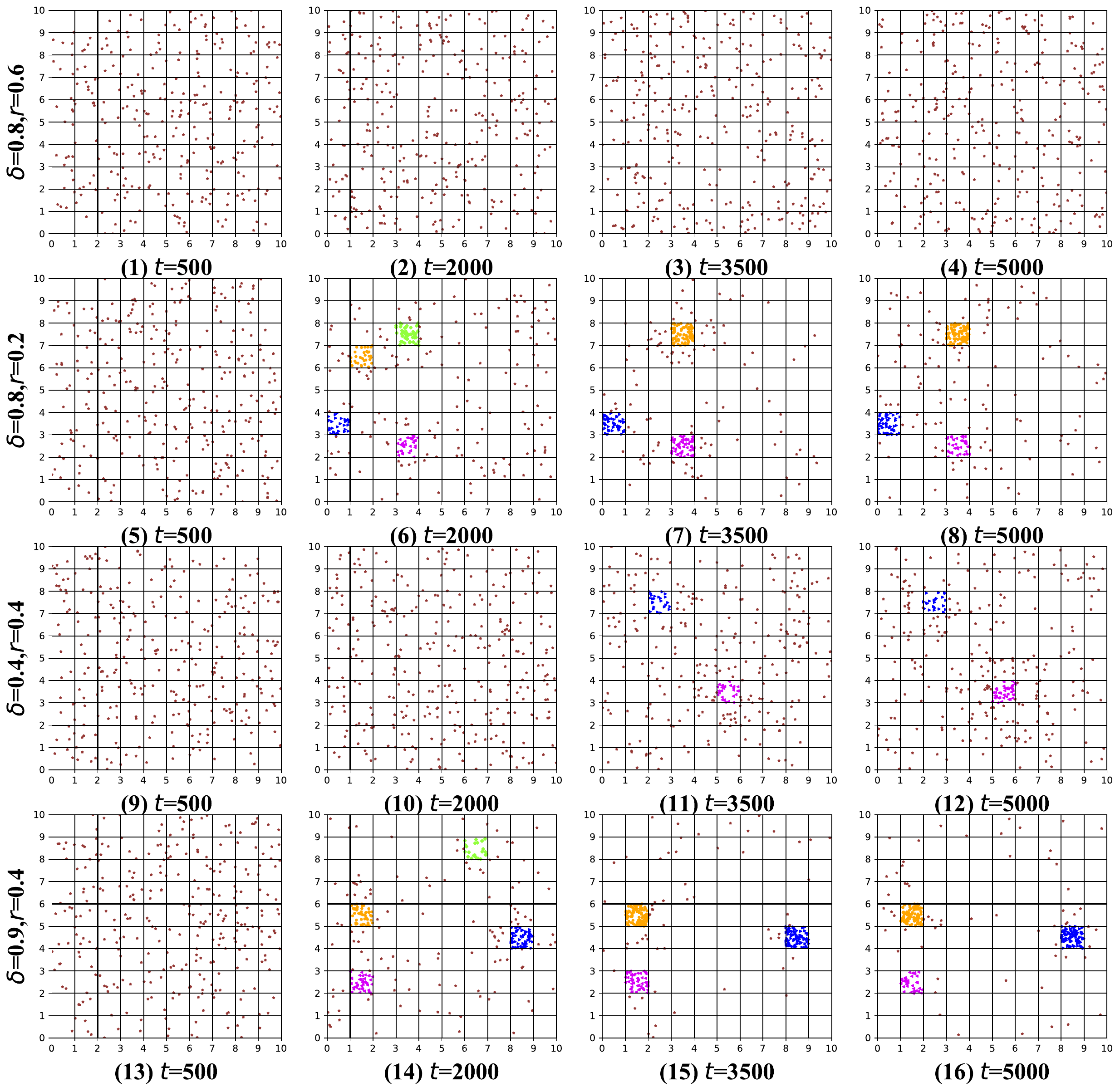}
\label{SWOBD_snapshot}}
\caption{\textbf{Evolutionary snapshots in the two-dimensional space under different parameter pairs ($\delta$, $r$).} By setting the parameter pairs ($\delta$, $r$) as (0.8, 0.6), (0.8, 0.2), (0.4, 0.4), and (0.9, 0.4) from top to bottom and fixing the time step at 500, 2000, 3500, and 5000 from left to right, we present the evolutionary snapshots of individual movement in the two-dimensional space on the system with and without birth-death process in subplots (a) and (b), respectively. All other parameters in the two subplots remain consistent for comparison.}
\label{snapshots}
\end{figure*}
\end{center}
\vspace{-2\baselineskip}

From Fig. \ref{SWBD_delta_r_fc}, we can see that the number of cooperators varies relatively uniformly about the payoff parameter, whereas the exploitation rate has almost no effect on the cooperative behavior for a particular payoff parameter $r$. However, in Fig. \ref{SWOBD_delta_r_fc}, a high exploitation rate can promote the emergence of cooperation at a specific $r$, which is a significant difference between the cooperative behavior of the system with the birth-death process and without. Furthermore, both Figs. \ref{SWBD_delta_r_fc} and \ref{SWOBD_delta_r_fc} exhibit that a smaller payoff parameter $r$ will be more favorable to the emergence and maintenance of cooperation. This is because according to the payoff matrix of the snowdrift game indicated in Eq. \ref{payoff matrix}, a higher $r$ results in a larger payoff to the defector and a smaller payoff to the cooperator, which in turn encourages more individuals to prefer the defective strategy. Additionally, we show the effect of $\delta$ and $r$ on cooperative behaviors for the death process obeying uniform, exponential, and lognormal distributions, and the interested reader can refer to subsection 3.1 of the supplementary material for more details.

\vspace{-1\baselineskip}
\subsection{Emergence and Evolution of Communities}

In this subsection, we visually analyze the evolution of individual movement in a two-dimensional space to gain insights into the emergence of system communities under different parameter pairs ($\delta$, $r$) of (0.8, 0.6), (0.8, 0.2), (0.4, 0.4), and (0.9, 0.4) in Figs. \ref{SWBD_snapshot} and \ref{SWOBD_snapshot}. Simultaneously, we transform the distribution of individuals in two dimensions to the corresponding network structure based on the weights between them in Figs. 2(a) and 2(b) of the supplementary material. We find that the formation and evolution of the communities in the network structure correspond well to the two-dimensional space. The scale of the system without birth-death process comprises 300 individuals, while the parameters associated with the scale of the system with birth-death process are set to $\lambda = 3$, $x_{min} = 80$, and $\alpha = 5$, i.e., the size of the system as it evolves to a steady state is 320, which can be obtained from Eq. \ref{EN} and is very close to the scale of the system without birth-death process.

As displayed in Fig. \ref{SWBD_snapshot}, the individuals are consistently evenly distributed on the plane regardless of the values of the exploitation rates and payoff parameters. In addition, the individuals of the corresponding network in Fig. 2(a) of the supplementary material are also distributed randomly with no distinguishing features. However, in Fig. \ref{SWOBD_snapshot}, when the exploitation rate is high and the payoff parameter is low, we observe that individuals do not maintain an even distribution across locations. Instead, they begin to cluster, forming multiple communities as time progresses and we adopt different colors to mark these different communities that are formed. In Fig. 2(b) of the supplementary material, the communities corresponding to ($\delta$, $r$) = (0.8, 0.2) and (0.9, 0.4) are clearly identifiable, but exhibit an instability at $t = 2,000$, which are continuing to rapidly evolve and become stable again at $t = 3,500$. Furthermore, by comparing the first and second rows (where $\delta$ is the same but $r$ differs) of Fig. \ref{SWOBD_snapshot}, we conclude that a small payoff parameter can promote the emergence and maintenance of communities. Although communities are formed in both the third and fourth rows (where $\delta$ differs but $r$ is the same) of Fig. \ref{SWOBD_snapshot}, the number of individuals in each community structure in ($\delta$, $r$) = (0.9, 0.4) is clearly larger and appears earlier than that in ($\delta$, $r$) = (0.4, 0.4). It suggests that a larger exploitation rate is more conducive to the evolution and maintenance of communities. These conclusions are further corroborated by the corresponding network structures shown in Fig. 2(b) of the supplementary material, but the communities are not observed in the system with the birth-death process. In this system, individuals die and new individuals join the system at each moment but individuals cannot form a community in a short period of time, therefore the joining of new individuals is equivalent to entering a location at random. Besides, we investigate the emergence and evolution of communities under the death process following uniform, exponential, and lognormal distributions, and the interested reader can refer to subsection 3.2 of the supplementary material for further details. Furthermore, we examine the emergence and evolution of communities when the stag hunt game is adopted as the underlying game model. Our findings reveal that the SWOBD within the stag hunt game remains capable of fostering the formation of communities. This suggests that the game model employed in this study is not limited to the snowdrift game but can be readily extended to other game scenarios by modifying the payoff matrix. Readers interested in further details are referred to subsection 3.2 of the supplementary material.

\vspace{-1\baselineskip}
\subsection{Effect of Payoff Parameter and Exploitation Ratio on Community Structures}
\label{Effect of Payoff Parameters and Exploitation Ratio on Community Structure}

\begin{center}
\begin{figure}[htbp]
\centering
\subfigure[\scalebox{0.86}{SWBD with power-law distribution}]{
\includegraphics[scale=0.28]{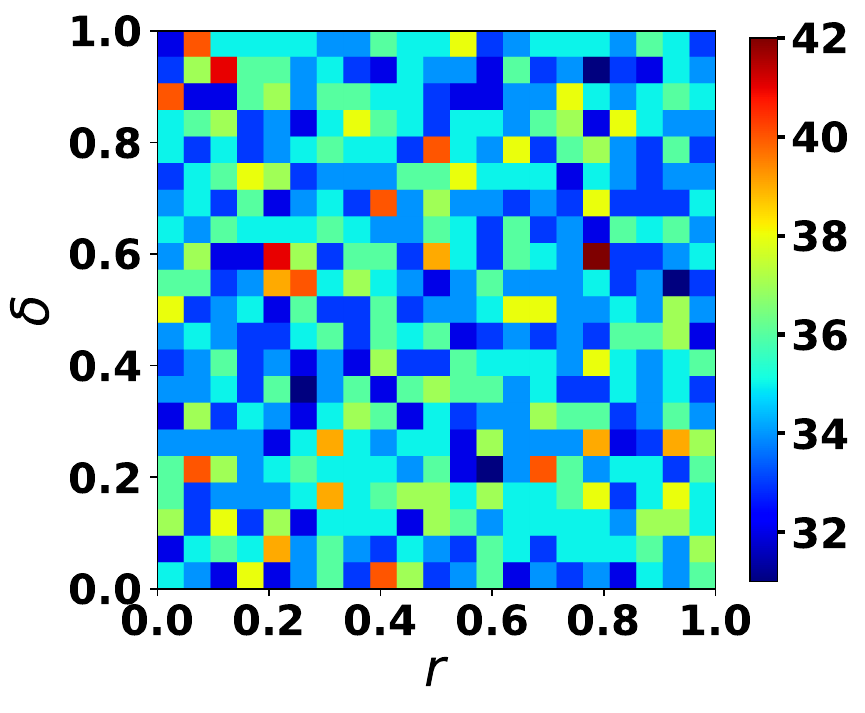}
\label{SWBD_delta_r_cluster}}
\subfigure[\scalebox{0.86}{SWOBD}]{
\includegraphics[scale=0.28]{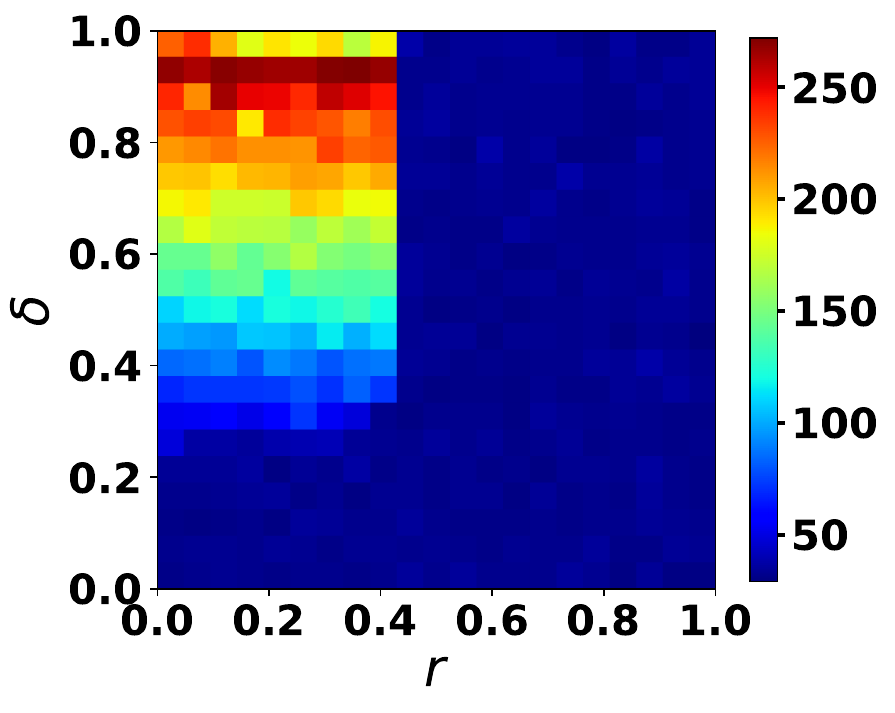}
\label{SWOBD_delta_r_cluster}}
\caption{\textbf{Heat maps of the sum of the first four highest numbers of individuals with respect to payoff parameter $r$ and exploitation rate $\delta$.} By setting the $x$-axis as the payoff parameter $r$ with a range [0, 1] and the $y$-axis as the exploitation rate $\delta$ with a range [0, 1], we show the impact of $r$ and $\delta$ on the sum of the first four highest numbers of individuals on SWBD with power-law distribution (in panel (a)) and SWOBD (in panel (b)).}
\label{heatmaps_cluster}
\end{figure}
\end{center}

Subsequently, to investigate the impact of the payoff parameter and the exploitation rate on community structures, we use the sum of the number of individuals in the four locations with the highest populations throughout the evolutionary process as a measure of community structure (denoted as $N_{c}$). In this measure, a larger value indicates a more significant community structure and vice versa. The results of SWBD with power-law distribution and SWOBD are displayed in Figs. \ref{SWBD_delta_r_cluster} and \ref{SWOBD_delta_r_cluster}.

\vspace{-1\baselineskip}
\vspace{-3.5\baselineskip}

From Fig. \ref{SWBD_delta_r_cluster}, we can see that the value of $N_{c}$ for the system with birth-death process falls within the range of 30 to 42, which is considerably small compared to the scale of the system at the steady state. It suggests that, in this scenario, individuals cannot self-organize themselves to form a community structure. Additionally, the exploitation rates and payoff parameters appear to have minimal impact on the community structure. For the system without the birth-death process, the result shown in Fig. \ref{SWOBD_delta_r_cluster} illustrates that when the payoff parameter $r<0.45$ and the exploitation rate $\delta<0.95$, the value of $N_c$ increases as the exploitation rate grows. In some cases, the maximum value can exceed 250, indicating that more than 83\% of the individuals in the system are self-organizing into communities. However, it is worth noting that the value of $N_c$ with $\delta = 1$ is smaller than that with $\delta = 0.95$, demonstrating that the community structure will be better served by an appropriate random selection of individuals as they move around the location. Meanwhile, when $r>0.45$, the value of $N_c$ remains around 50 regardless of the value of the exploitation rate adopted, i.e., the system does not form communities under these conditions. Moreover, we explore the effect of $r$ and $\delta$ on community structures under the death process obeying uniform, exponential, and lognormal distributions, and the interested reader can refer to subsection 3.3 of the supplementary material for more details.

\vspace{-1\baselineskip}
\subsection{Fit of the Proposed Model on Real Data}

To demonstrate the practical importance of the model proposed in this paper, in this subsection, we apply the scale of the system with the birth-death process under various death distributions to model the populations of four different countries. Furthermore, the degree distributions of networks generated by SWBD and SWOBD with reinforcement learning are employed to fit the degree distributions of six different real networks.

\vspace{-1\baselineskip}
\subsubsection{Fit of the Birth-Death Process on Real Populations}

\begin{center}
\begin{figure}[htbp]
\centering
\subfigure[\scalebox{0.89}{Greenland with power-law}]{
\includegraphics[width = 4.2cm, height = 3.5cm]{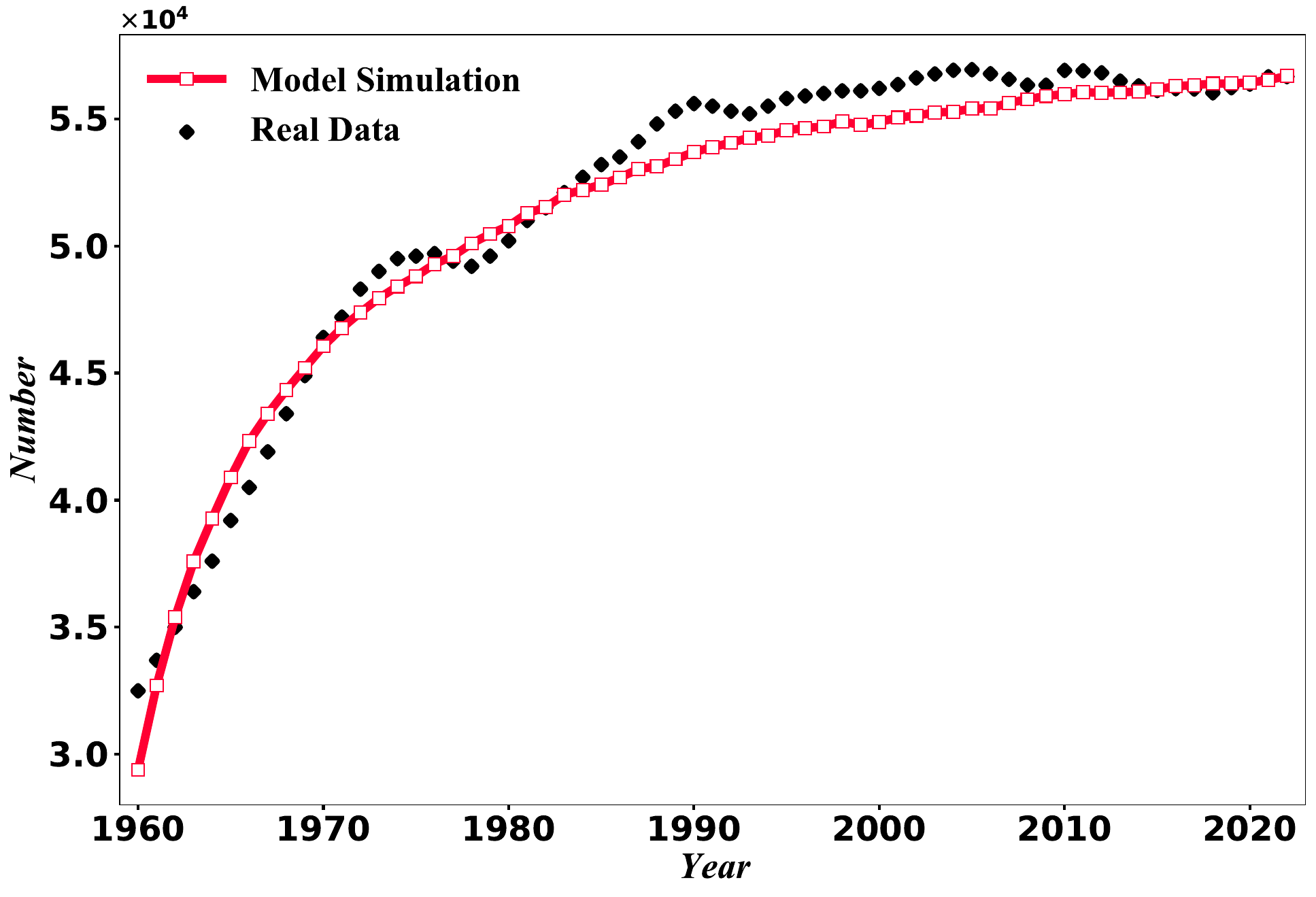}
\label{fit_Greenland}}
\subfigure[\scalebox{0.89}{Guam with uniform}]{
\includegraphics[width = 4.2cm, height = 3.5cm]{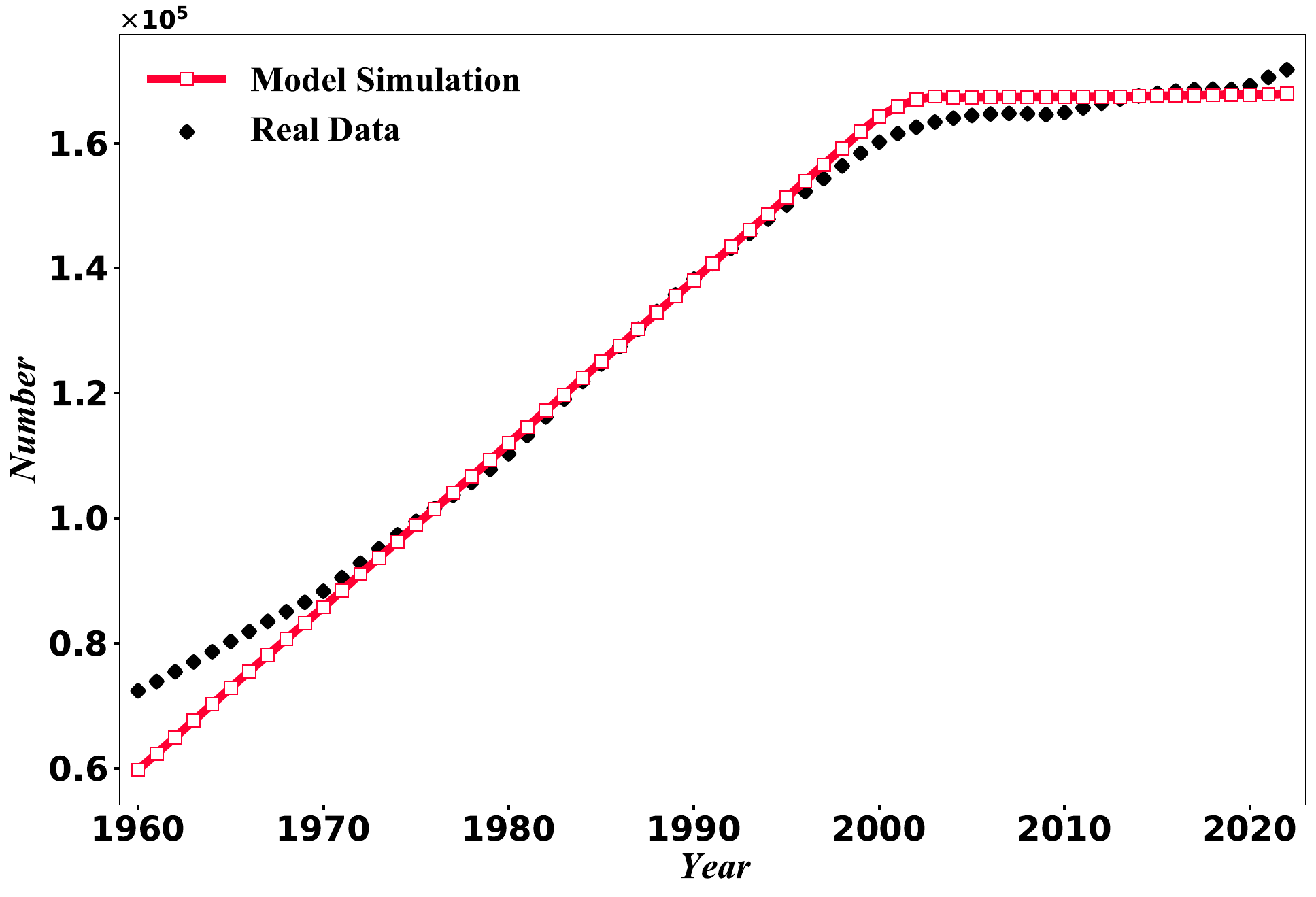}
\label{fit_Guam}}
\subfigure[\scalebox{0.89}{Slovak Republic with exponential}]{
\includegraphics[width = 4cm, height = 3.5cm]{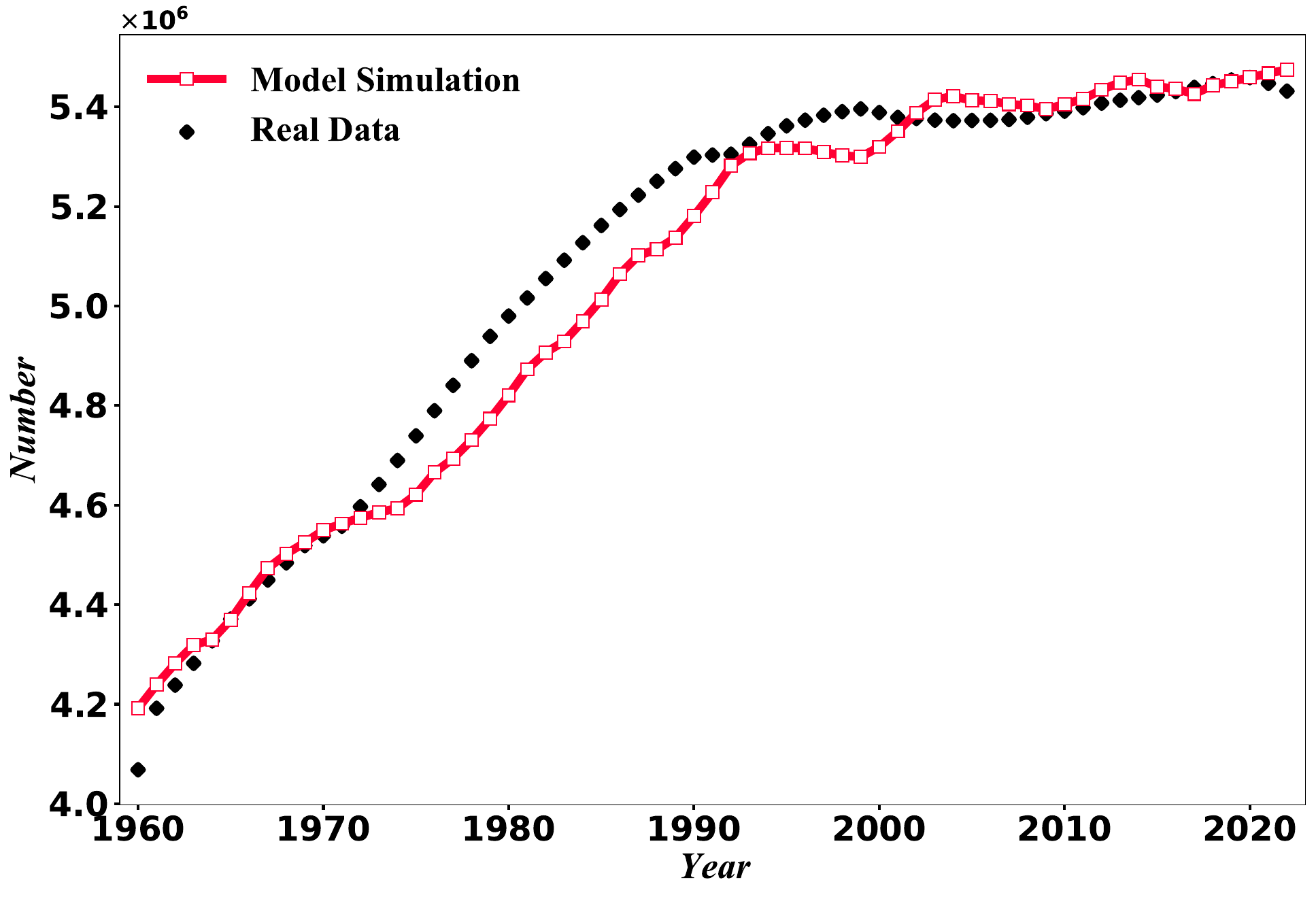}
\label{fit_Slovak}}
\subfigure[\scalebox{0.89}{Cuba with lognormal}]{
\includegraphics[width = 4cm, height = 3.5cm]{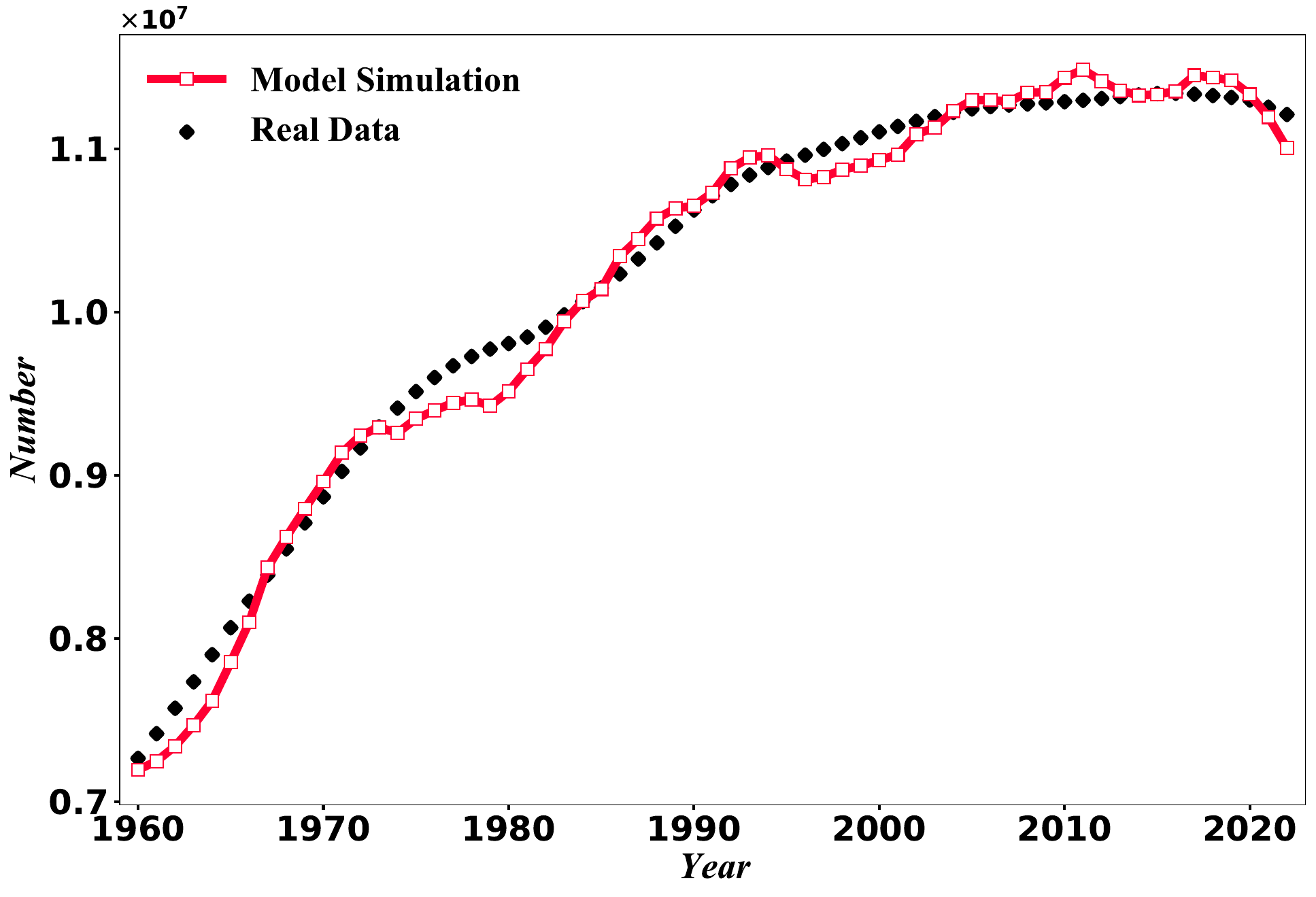}
\label{fit_Cuba}}
\caption{\textbf{Fit of the birth-death process under different death distributions on the populations of four different countries.} In this figure, we simulate the populations of four countries using different death process distributions. Specifically, we use the power-law distribution for Greenland (subplot (a)), and the uniform distribution for Guam (subplot (b)). Subplots (c) and (d) illustrate the simulations for the Slovak Republic and Cuba, where the death processes follow exponential and lognormal distributions. The actual population data is represented by black diamonds, while the simulated curves are depicted with red squares.}
\label{fit_population}
\end{figure}
\end{center}

We select the populations of Greenland\footnote{https://data.worldbank.org/indicator/SP.POP.TOTL?locations=GL}, Guam\footnote{https://data.worldbank.org/indicator/SP.POP.TOTL?locations=GU}, the Slovak Republic\footnote{https://data.worldbank.org/indicator/SP.POP.TOTL?locations=SK}, and Cuba\footnote{https://data.worldbank.org/indicator/SP.POP.TOTL?locations=CU} from 1960 to 2022, spanning a total of 62 years, and model them using the scale of the system with birth-death process under various death distributions. Specifically, we utilize a power-law distribution to model the population of Greenland, setting the input rate at $\lambda=600$, with the power-law distribution parameters $x_{min}=50$ and $\alpha=3$. For Guam's population, we employ a uniform distribution, with the input rate set to 525, and the starting and ending values set to 310 and 330, respectively. The population of the Slovak Republic is modeled using an exponential distribution, with an input rate of 89,600 and a death rate of 1/73. For the population of Cuba, we use a lognormal distribution with an input rate of 291,200, a mean of 2.5, and a standard deviation of 1.8. The fitting results for these four regions are presented and compared with the real data in Fig. \ref{fit_population}. All figures show that the populations of the four countries gradually increase and then tend to stabilize, and the simulation results closely match the real data.

\vspace{-1\baselineskip}
\vspace{-1\baselineskip}

To provide a more intuitive demonstration of the fit, we calculate the expected populations for the four countries alongside the simulation results. We then determine the relative error using the equation: $e = \mid x - x^* \mid / x$, where $x$ represents the actual population and $x^*$ denotes the simulated population. Furthermore, we compute the correlation coefficient between the simulation results and the actual populations. The results are summarized in Tab. \ref{Fitting population}, with the relative errors for Greenland, Guam, the Slovak Republic, and Cuba being 0.96\%, 0.48\%, 0.71\%, and 0.38\%, respectively, which are all below 1\%. Besides, the correlation coefficients for all four fits exceed 98.39\%, further confirming the accuracy of the fit, as illustrated in Fig. \ref{fit_population}.

\begin{center}
\begin{table*}[htbp]
\renewcommand{\arraystretch}{2}
\setlength\tabcolsep{9pt}
\caption{The comparison of models under different death processes in fitting real populations}
\begin{center}
\begin{tabular}{cccccccc}
\toprule[1.5pt]
Results       & Simulation population & Real population & Relative error & Correlation coefficient \\ \hline
Greenland    & 51,216.46  & 51,713.24 & 0.96\% & 98.96\%      \\
Guam    & 131,359.27  & 131,996.69 & 0.48\%  & 99.72\%     \\
Slovak Republic    & 5,039,330.92  & 5,075,572.91 & 0.71\%  & 98.39\%      \\
Cuba    & 10,170,945.35  & 10,209,542.11 & 0.38\%  & 99.49\%     \\
\bottomrule[1.5pt]
\end{tabular}
\label{Fitting population}
\end{center}
\end{table*}
\end{center}

\begin{center}
\begin{table*}[htbp]
\renewcommand{\arraystretch}{2}
\setlength\tabcolsep{5.2pt}
\caption{Statistical results of real networks and similarity between simulated degree distributions and real degree distributions}
\begin{center}
\begin{tabular}{cccccccc}
\toprule[1.5pt]
Results  & Nodes & Edges & Assortativity & Clustering coefficient & Density & JS divergence   & Cosine similarity    \\ \hline
HS-HT    & 2,570  & 13,691 & 0.2943        & 0.1695  & 0.0042  & 0.0268 & 0.9877 \\
TWITTER-COPEN & 761   & 1,029  & -0.099        & 0.0759  & 0.0036  & 0.0884 & 0.9732  \\
AVES-WEAVER-SOCIAL & 445   & 1,423  & 0.2295        & 0.6924  & 0.0144  & 0.0997 & 0.7563  \\
DWT-607    & 607  & 2,262 & 0.1861        & 0.4792  & 0.0123  & 0.0398 & 0.9823 \\
MAMMALIA-DOLPHIN    & 62  & 159 & -0.0436        & 0.2589  & 0.0841  & 0.0095 & 0.9852 \\
NETSCIENCE & 379   & 914  & -0.0817        & 0.7412  & 0.0128  & 0.0502 & 0.9645  \\
\bottomrule[1.5pt]
\end{tabular}
\label{Fitting degree}
\end{center}
\end{table*}
\end{center}
\vspace{-1.5\baselineskip}

\vspace{-1\baselineskip}
\subsubsection{Fit of the Degree Distribution using Reinforcement Learning on Real Networks}

Next, we analyze the degree distributions of networks generated by both SWOBD and SWBD with reinforcement learning and utilize them to fit with the degree distributions of six distinct real-world networks \cite{nr-aaai152}: HS-HT, TWITTER-COPEN, AVES-WEAVER-SOCIAL, DWT-607, MAMMALIA-DOLPHIN, and NETSCIENCE. The HS-HT network is a biological network, where nodes represent genes and edges signify connections between genes. The TWITTER-COPEN network is a retweet network consisting of 761 users as nodes and 1,029 retweets/mentions as edges. AVES-WEAVER-SOCIAL and MAMMALIA-DOLPHIN are animal social networks, where nodes denote animals and edges represent interactions between them. DWT-607 is a miscellaneous network based on a symmetric connection table from DTNSRDC, Washington. Finally, NETSCIENCE is a collaboration network containing 379 researchers as nodes and 914 co-authorships as edges. Some statistical information about these six real networks, such as the number of nodes, number of edges, assortativity, clustering coefficients, etc., are given in Tab. \ref{Fitting degree}. The statistics reveal that all networks have relatively low densities and the NETSCIENCE network has the largest clustering coefficient among the six networks, indicating that individuals in the NETSCIENCE network tend to have the closest interactions.

We utilize the network generated by the SWOBD model to fit HS-HT and TWITTER-COPEN networks, with the fitting results shown in Figs. 7(a) and 7(b) of the supplementary material, respectively. Additionally, we apply the SWBD model to fit the real networks based on different death processes, with the results displayed in Figs. 7(c)-(f) of the supplementary material, where these death processes separately follow power-law, uniform, exponential, and lognormal distributions. As observed in Fig. 7 of the supplementary material, the degree distributions of HS-HT and TWITTER-COPEN networks, along with the network generated without the birth-death process, exhibit characteristics consistent with the power-law distribution. In contrast, the degree distributions of the AVES-WEAVER-SOCIAL network and the network generated with a birth-death process that follows power-law distribution approximate the normal distribution. Notably, the network generated by SWBD can also obey a power-law distribution, as illustrated in Fig. 7(f) of the supplementary material. For the fitting of DWT-607 and MAMMALIA-DOLPHIN networks, we model the death process in the SWBD model with uniform and exponential distributions, respectively, and the degree distributions of the resulting networks display irregular patterns, as depicted in Figs. 7(d) and 7(e) of the supplementary material.

To more intuitively assess the fit of the SWOBD and SWBD models, we calculate cosine similarity (Eq. \ref{Cosine similarity}) and Jensen-Shannon divergence (Eq. \ref{JS}) to quantify the similarity between the real and simulated data, as shown in Tab. \ref{Fitting degree}. The results indicate that the Jensen-Shannon divergence between the simulated data and real data is relatively small, not exceeding 0.0997. Conversely, the cosine similarity is notably high, with values no less than 0.7563. These findings suggest that the difference between the degree distribution generated by our models and the real data is minimal.

Therefore, based on the fitting of our model to real-world data, we can conclude that our model demonstrates strong consistency with actual population dynamics and real network structures. This consistency indicates that the model proposed in this paper has significant practical value.

\vspace{-1\baselineskip}
\section{Analysis of the community structures}
\label{Analysis of the community structure}
\vspace{-0.5\baselineskip}

\begin{center}
\begin{figure*}[htbp]
\centering
\includegraphics[scale=0.68]{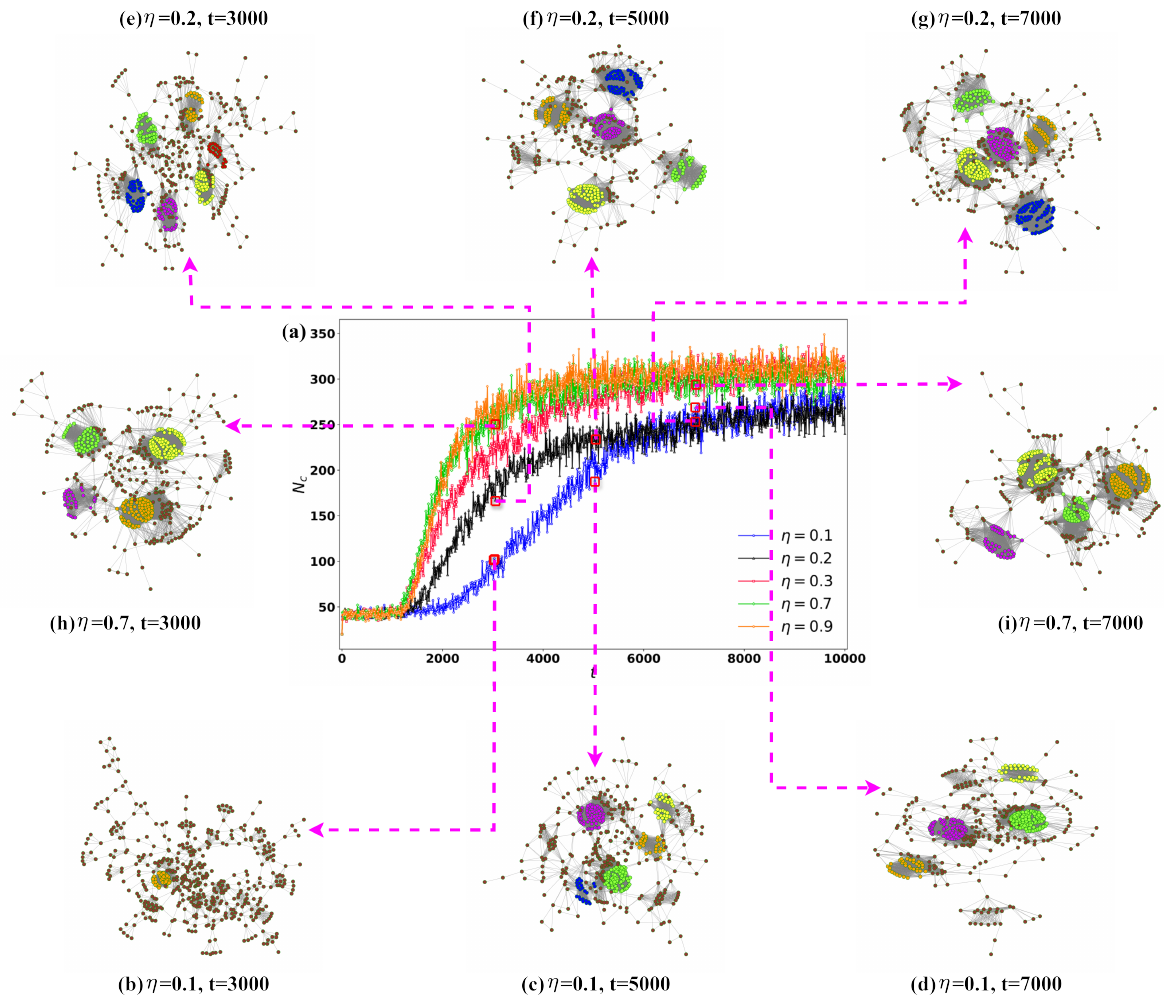}
\caption{\textbf{Impact of learning rate on the formation and evolution of community structure.} (a) The sum of the number of individuals in the four locations with the highest populations (denoted as $N_c$) as a function of time $t$ under different learning rates with $\eta = 0.1, 0.2, 0.3, 0.7$, and 0.9. (b)-(i) The corresponding networks at different given learning rates and times, utilizing distinct colors to demonstrate the community structures.}
\label{lr}
\end{figure*}
\end{center}

\begin{center}
\begin{figure*}[htbp]
\centering
\includegraphics[scale=0.48]{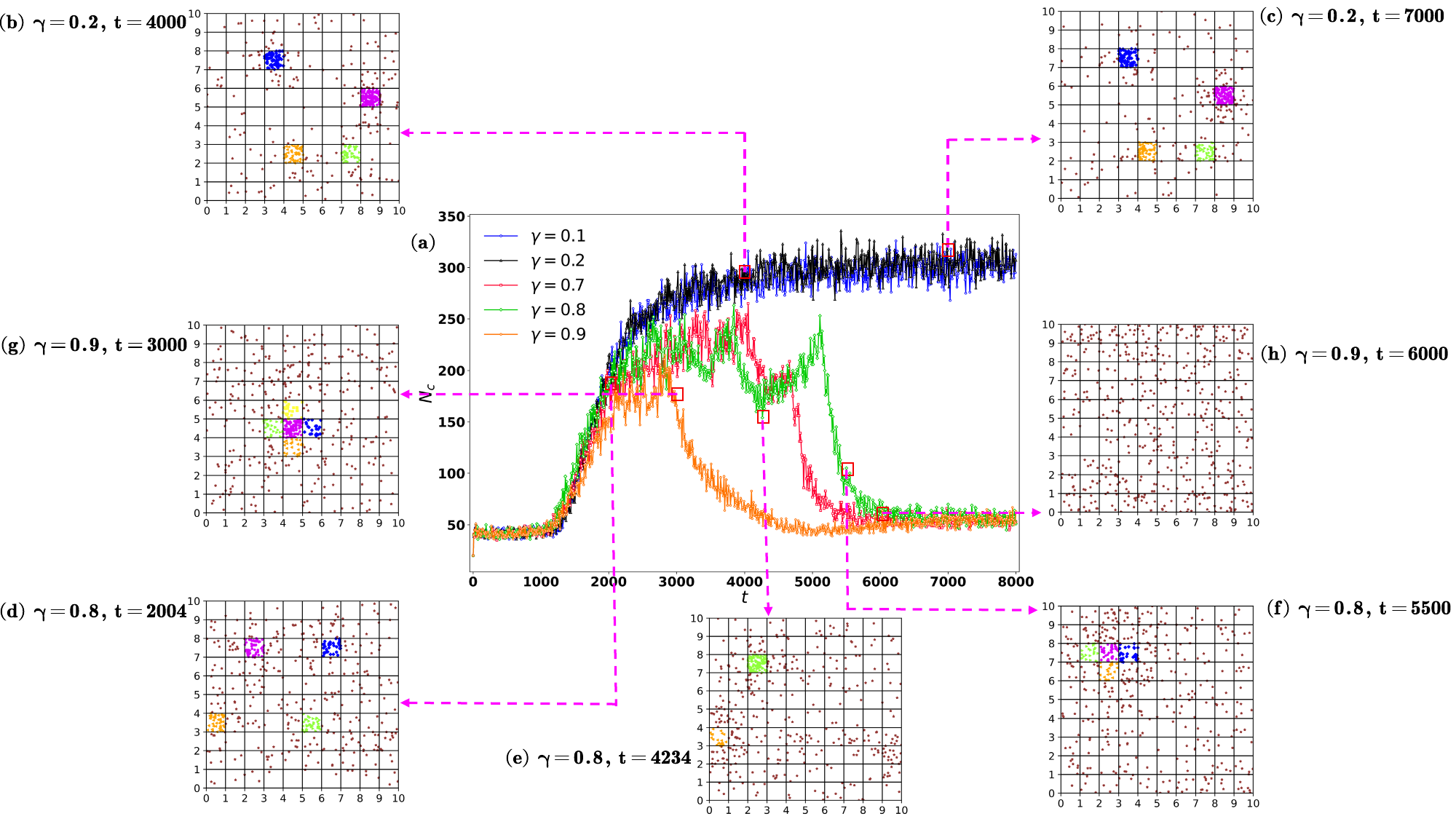}
\caption{\textbf{Effect of discount factor on the formation and development of community structure.} Subplot (a) demonstrates the evolution of $N_c$ as time evolves under different discount factors with $\gamma = 0.1, 0.2, 0.7, 0.8$, and $0.9$. The snapshots of individual distributions in the two-dimensional are depicted in (b)-(c) for $\gamma = 0.2$ at $t = 4,000$ and $t = 7,000$, in (d)-(f) for $\gamma = 0.8$ at $t = 2,004$, $t = 4,234$, and $t = 5,500$, and in (g)-(h) for $\gamma = 0.9$ at $t = 3,000$ and $t = 6,000$.}
\label{gamma}
\end{figure*}
\end{center}
\vspace{-2.8\baselineskip}

As we have shown before, community structures predominantly emerge in systems, when there is no process of birth and death. In this section, our focus shifts to analyzing the effect of several parameters, including the learning rate, the discount factor in Q-learning, and the dimensions in the two-dimensional space (number of rows and columns) on community formation and development in SWOBD. In alignment with the approach detailed in subsection \ref{Effect of Payoff Parameters and Exploitation Ratio on Community Structure}, we employ the sum of the individuals in the four locations with the highest populations as a metric for community structure (denoted as $N_{c}$). A higher value in this measure signifies a more substantial community structure, and vice versa.

\vspace{-1\baselineskip}
\subsection{Impact of Learning Rate on Community Structures}

In this subsection, we study the influence of the learning rate $\eta$ in Eq. \ref{q-learning} on the community structure. The dimension of the two-dimensional space is set to $10\times 10$, with an initial population of 5 individuals at each location, i.e., the total scale of the system is 500. The exploitation rate, discount factor, and payoff parameter are respectively set to 0.7, 0.2, and 0.1. In addition, we set the evolutionary time to 10,000, ensuring sufficient time for the evolution to stabilize. The evolutionary curve of $N_c$ over time under different learning rates is depicted in Fig. \ref{lr}(a). Figs. \ref{lr}(b)-(d), (e)-(g), and (h)-(i) illustrate the corresponding network structure at selected instants of time with $\eta=$ 0.1, 0.2, and 0.7.

\begin{center}
\begin{figure*}[htbp]
\centering
\includegraphics[scale=0.7]{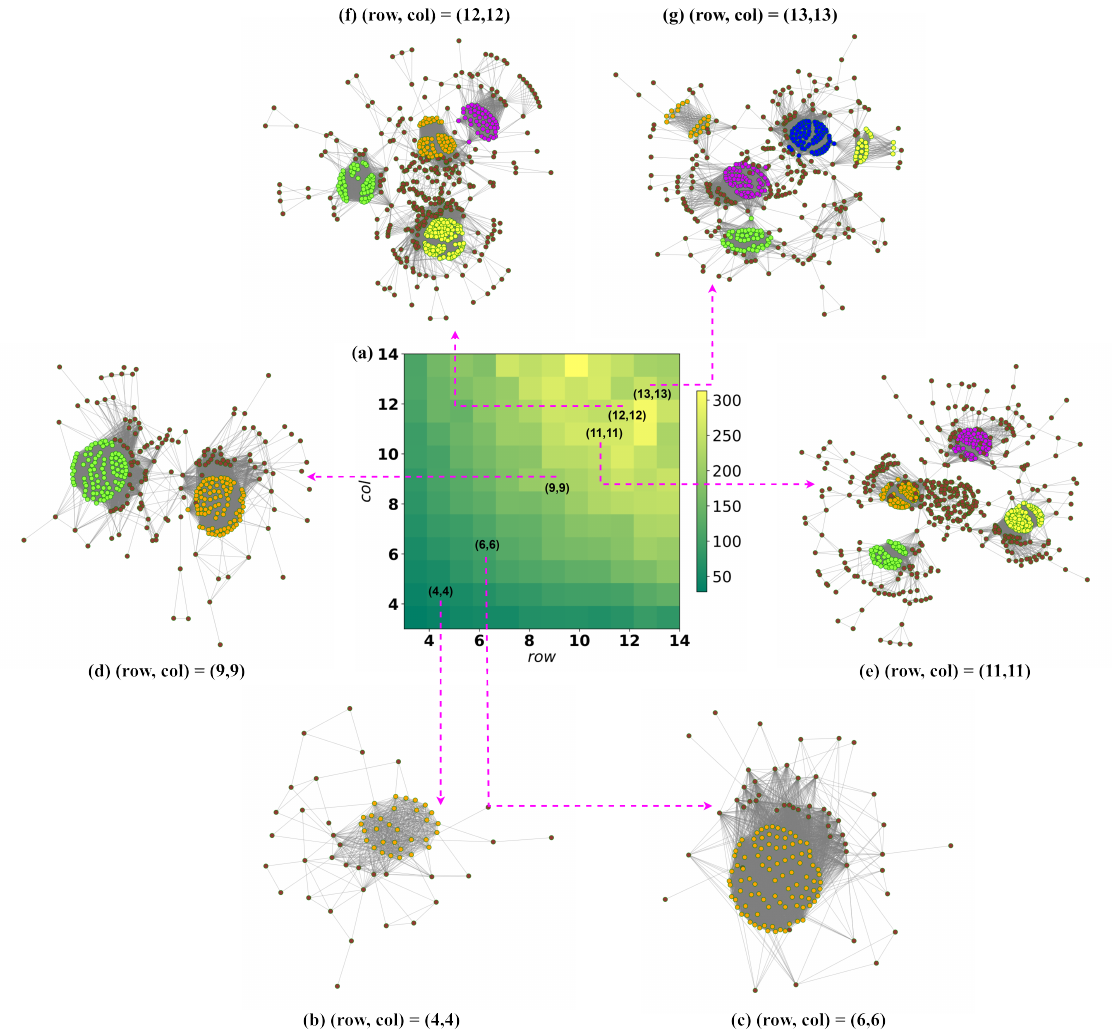}
\caption{\textbf{Influence of the number of rows and columns in the two-dimensional space on the emergence and development of community structure.} Panel (a) presents the heat map of the sum of the first four highest numbers of individuals in relation to the number of rows and columns. Panels (b)-(g) show the network structure under different values [(4, 4), (6, 6), (9, 9), (11, 11), (12, 12), (13, 13)] of the number of rows and columns $(row, col)$, respectively.}
\label{row_col}
\end{figure*}
\end{center}
\vspace{-2\baselineskip}

From Fig. \ref{lr}(a), we observe that the sum of the number of individuals in the four locations with the highest populations grows as time progresses and eventually stabilizes around a certain value. The difference between the various curves lies in the speed at which they evolve to a steady state. Specifically, the curve with $\eta=0.1$ reaches a steady state around $t=6,000$, $\eta=0.2$ achieves stability around $t=5,000$, $\eta=0.3$ reaches a steady state around $t=4,000$, while the curves with $\eta=0.7$ and 0.9 achieve stability around $t=3,000$. This trend is visually demonstrated by examining the corresponding network structures. For instance, the network structure associated with $\eta=0.1$ in Figs. \ref{lr}(b)-(d) reveals 1, 5, and 4 communities at $t=3,000$, $t=5,000$, and $t=7,000$ respectively, and the corresponding values of $N_c$ at these times are notably distinct, which suggests that its evolution has not yet reached a steady state at $t=5,000$. On the contrary, the network structure with $\eta=0.7$ in Figs. \ref{lr}(h)-(i) exhibits 4 communities at both $t=3,000$ and 7,000, with similar values for $N_c$, which indicates that it has reached a stable state already at around $t=3,000$. Hence, $\eta = 0.2$ in Figs. \ref{lr}(e)-(g) can be obtained by a similar analysis. It is noteworthy that although the rate of evolution varies among the curves, the value of $N_c$ upon reaching a stable state is similar. Therefore, we derive that the learning rate influences the speed of community structure formation and evolution, with a large learning rate facilitating a quicker emergence and stabilization of community structures. This is attributed to a larger learning rate enabling individuals to learn more rapidly from their actions, resulting in a faster convergence of their Q-tables.

\vspace{-1\baselineskip}
\subsection{Effect of Discount Factor on Community Structures}

Subsequently, we examine the impact of the discount factor $\gamma$ in Eq. \ref{q-learning} on the community structure. We present the evolutionary curves of the value of $N_c$ about time $t$ in Fig. \ref{gamma}(a), along with some individual distributions in the two-dimensional space in Figs. \ref{gamma}(b)-(h). The exploitation rate, learning rate, and payoff parameter are separately set to 0.7, 0.9, and 0.1. The scale of the two-dimensional space is $10\times 10$ and there are 5 individuals at each location at the beginning, leading to a total system scale of 500.

As depicted in Fig. \ref{gamma}(a), when the discount factor $\gamma$ is relatively small ($\gamma$ = 0.1, 0.2), $N_c$ gradually increases and then plateaus with the passage of time. Conversely, when $\gamma$ takes larger values ($\gamma$ = 0.7, 0.8, 0.9), $N_c$ exhibits an initial increase, followed by significant fluctuations over time, ultimately reducing to approximately 50. This indicates a lack of sustained community formation, a point corroborated by Fig. \ref{gamma}(h). Furthermore, the snapshots demonstrated in Figs. \ref{gamma}(b)-(c) illustrate that there is already a substantial community formation at $t = 4,000$, which remains stable as time progresses. However, for larger values of $\gamma$, although communities emerge during the evolution process, the structures are unstable and undergo significant changes, such as in Figs. \ref{gamma}(d)-(f) with $\gamma = 0.8$, where the number of communities at $t = 2,004$, $t = 4,234$, and $t = 5,500$ are 4, 2, and 4, respectively, as well as the positions of the emerging communities continuously change, and the community eventually disappears around $t = 6,000$, after which the value of $N_c$ stabilizes. This observation underscores the influence of the discount factor on the stability of community structures. A larger $\gamma$, indicative of individuals placing greater emphasis on future rewards, proves detrimental to the stability of the community.

\vspace{-1\baselineskip}
\subsection{Influence of Dimensions in the Two-Dimensional Space on Community Structures}

We note that in the preceding numerical simulations, the dimensions of the two-dimensional space were held constantly at $10\times 10$. In this subsection, we introduce variability by treating the number of rows and columns as parameters to study their influence on the community. The number of individuals at each location at $t = 0$ is 4, and the initial configuration implies that the system scale changes as the number of rows and columns varies. With the learning rate, discount factor, exploitation rate, and payoff parameter set to 0.9, 0.2, 0.7, and 0.1, respectively, we show the heat map of $N_c$ concerning the number of rows and columns in Fig. \ref{row_col}(a). Figs. \ref{row_col}(b)-(g) demonstrate the network structure corresponding to $(row, col)$ = (4, 4), (6, 6), (9, 9), (11, 11), (12, 12), and (13, 13). Different colors in the network represent different community structures formed.

According to the results shown in Fig. \ref{row_col}(a), it can be seen that $N_c$ increases steadily with the growth of the number of rows and columns, which illustrates the consistent formation of communities in the system, with the size of the community progressively expanding. Further insights can be gleaned from the networks presented in Figs. \ref{row_col}(b)-(g). Under $(row, col)$ = (4, 4) and (6, 6), only one community is observed, whereas the communities increase to 2, 4, 4, and 5 under $(row, col)$ = (9, 9), (11, 11), (12, 12), and (13, 13). This observation suggests that the number of communities grows proportionally with the number of rows and columns. When the number of rows and columns is small, the limited space constrains individual movement, resulting in a small number of communities formed but with a large number of individuals compared to the total population. Conversely, with a larger number of rows and columns, individuals have more space to choose and move, leading to the formation of more communities, albeit with fewer individuals in each community.

\vspace{-1\baselineskip}
\section{Model comparisons}
\label{Model comparisons}

In this section, we compare and analyze our proposed model from two main aspects. On the one hand, from the reinforcement learning point of view, we verify how the use of heuristic models affects the results in the absence of reinforcement learning mechanisms. On the other hand, from the perspective of the classical network model, the model proposed in this paper is compared and analyzed with the classical network in terms of network structure, clustering coefficients, and degree distribution.

\vspace{-1.5\baselineskip}
\subsection{Comparison with Heuristic Model}

Hereby, we focus on the impact of reinforcement learning on the outcomes and propose an alternative heuristic model. Specifically, each individual begins with random exploration, which can be considered as setting the initial exploration rate to $\epsilon = 1$. The exploration rate then decays by a factor of 0.99 at each time step, and each individual explores its neighboring states with a probability $\epsilon$. If an individual finds a neighboring state that offers greater benefits than its current state, it will transition to the new state; otherwise, it will retain its original state. It is important to note that state and payoff calculations remain the same as in SWOBD, except that a heuristic model is used instead of reinforcement learning.

\vspace{-1.5\baselineskip}
\begin{center}
\begin{figure}[htbp]
\centering
\includegraphics[scale=0.28]{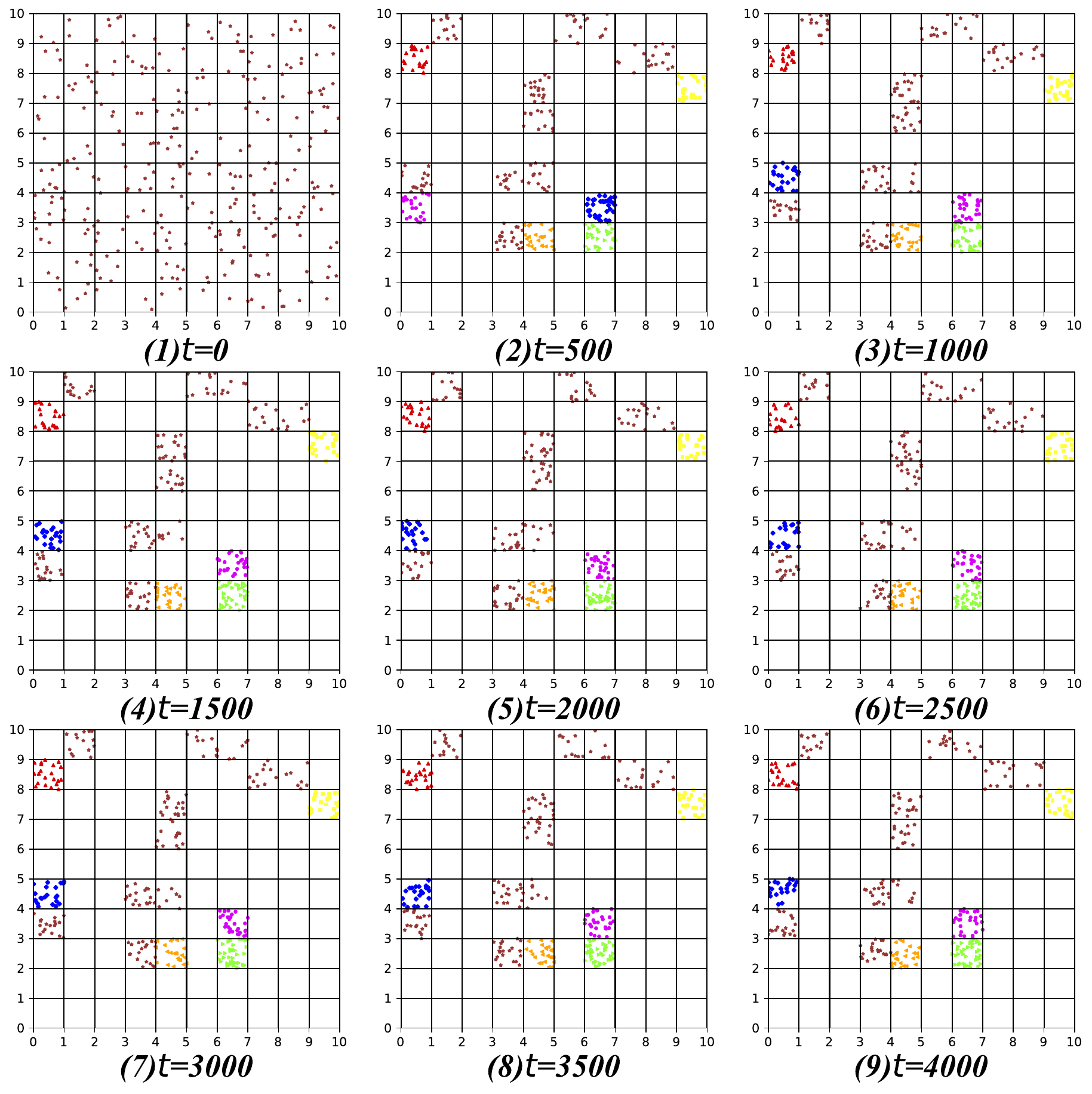}
\caption{\textbf{Evolutionary snapshots of the heuristic model.} This figure illustrates the snapshots in the two-dimensional space generated based on the heuristic model at different time steps, where the time step is taken from 0 to 4,000 at an interval of 500.}
\label{heuristic_snapshot}
\end{figure}
\end{center}
\vspace{-1.5\baselineskip}

In Fig. \ref{heuristic_snapshot}, we demonstrate evolutionary snapshots of the heuristic model in a two-dimensional space. These snapshots reveal that a community structure begins to emerge around $t = 500$, but remains nearly constant thereafter. Additionally, Fig. 8 of the supplementary material displays the performance metrics of the heuristic model, including degree distributions and state transition ratios. The state transition ratio means the proportion of individuals that change their states at each time step. The degree distribution, shown in Fig. 8(a) of the supplementary material, indicates that the network generated by the heuristic model does not follow typical normal or power-law distributions, but rather exhibits an irregular distribution. Moreover, Fig. 8(b) of the supplementary material shows that the state transition ratio of the heuristic model rapidly decreases to 0, suggesting that individuals maintain their states unchanged after stabilization. This observation contradicts the real-world scenario, where community structures are typically dynamically stabilizing and do not remain constant indefinitely. In contrast, for SWOBD with a reinforcement learning mechanism, the final state transition ratio stabilizes around 0.25, indicating that the community structure evolves dynamically and maintains a dynamic equilibrium, which aligns with real-world observations.

The results above demonstrate that SWOBD with reinforcement learning is more effective in modeling the evolution of complex networks and community structures, and it cannot be replaced by a simple heuristic model. This effectiveness arises from several key differences. In the heuristic model, individuals lack memory and learning capabilities, focusing solely on maximizing immediate benefits. Consequently, individuals in the same state take identical actions. In contrast, the reinforcement learning model provides each individual with a Q-table and the ability to learn, allowing them to consider long-term cumulative payoffs, which contributes to the fact that even individuals in the same state may act differently based on their different experiences. Thus, it can be concluded that reinforcement learning is crucial to the effectiveness of the proposed SWOBD model.

\vspace{-1.5\baselineskip}
\subsection{Comparison with Classic Networks}

\begin{center}
\begin{figure*}[htbp]
\centering
\includegraphics[width=16cm,height=13cm]{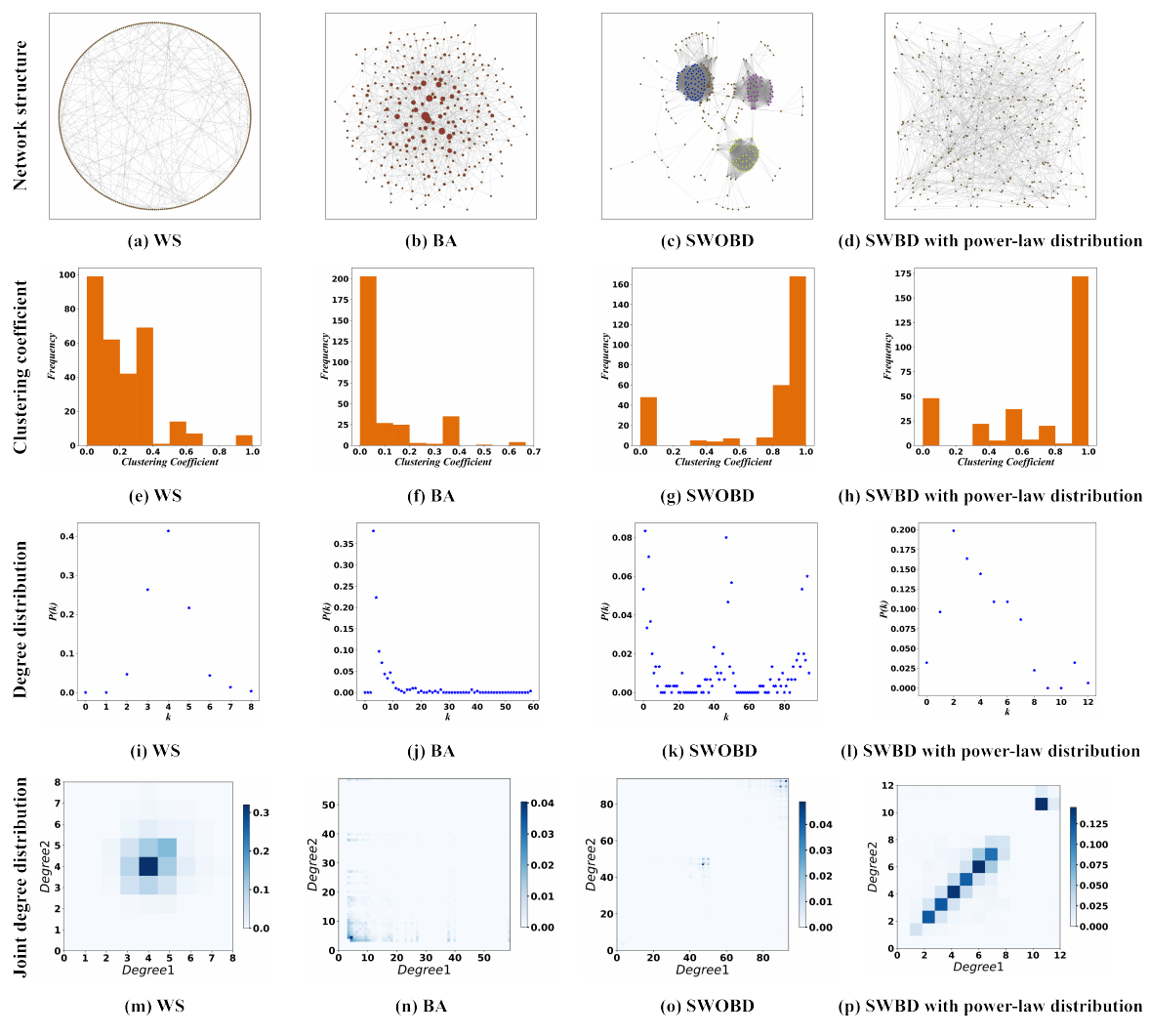}
\caption{\textbf{Comparison of the networks proposed in this paper with classical networks.} In this figure, we respectively compare our SWOBD network (third column) and the network of SWBD with power-law distribution (fourth column) with the WS small-world network (first column) and the BA scale-free network (second column) in terms of network structure in subplots (a)-(d), clustering coefficient in subplots (e)-(h), degree distribution in subplots (i)-(l), and joint degree distribution in subplots (m)-(p).}
\label{classic_networks}
\end{figure*}
\end{center}

In this subsection, we perform a comparative analysis between the model proposed in this paper (both SWBD with power-law distribution and SWOBD) with classic networks, including Watts-Strogatz small-world (WS) and Barab{\'a}si-Albert scale-free (BA) networks. The comparison spans aspects such as network structure, degree distribution, clustering coefficient, etc., and the results are demonstrated in Fig. \ref{classic_networks}. All networks have 300 nodes (SWBD with power-law distribution also reaches 300 nodes when it evolves to a stable state). In the WS network, each node initially connects to the 2 nearest nodes on its left and right sides, with a subsequent reconnection probability set to 0.3 for each edge, and no repeated edges or self-loops are permitted. The new nodes in the BA network are linked to the 3 existing nodes through the degree-preferential connection mechanism.

\vspace{-1\baselineskip}
\vspace{-1\baselineskip}

In complex networks, the clustering coefficient of node $i$ with degree $k_i$ is defined as the ratio of the number of connected edges between neighboring nodes of node $i$ to the maximum possible number of connected edges between these neighboring nodes \cite{zhou2005maximal}, which can be calculated using the following formula: $C_i=\frac{2E_i}{k_i\left( k_i-1 \right)}$, where $E_i$ represents the actual number of edges between the $k_i$ neighbors of node $i$. If node $i$ has only one neighbor or no neighbors, we denote $C_i=0$. The clustering coefficient of the network is the average of the clustering coefficient of all the nodes in the network, which reflects the degree of node aggregation in the network, with a higher clustering coefficient indicating a greater node aggregation, and vice versa. By plotting the distribution of node clustering coefficients in the networks (as shown in the second row of Figs. \ref{classic_networks}(e)-(h)) and calculating the clustering coefficients of the networks, we get that most nodes in the WS and BA networks exhibit low clustering coefficients, with overall network clustering coefficients of 0.201 and 0.079, respectively, whereas the majority of nodes in the networks of SWOBD and SWBD with power-law distribution display high clustering coefficients, with overall network clustering coefficients of 0.751 and 0.714, respectively. This phenomenon suggests that the degree of aggregation in BA and WS networks is low, while in SWOBD and SWBD with power-law distribution, it is high.

Figs. \ref{classic_networks}(i)-(l) present the degree distributions of the four networks. It is evident that the degree distributions of WS and BA networks follow normal and power-law distributions, respectively, while the degree distributions of SWBD with power-law distribution and SWOBD networks exhibit irregular patterns due to parameter settings of the model, and we emphasize that the degree distributions of SWOBD and SWBD with power-law distribution can also conform power-law and normal distributions, as demonstrated in Fig. 7 of the supplementary material. Degree correlation is another metric used to assess the connection between nodes with similar or dissimilar degrees. A network is considered degree-degree positively correlated or assortative if nodes with large degrees tend to connect with other nodes with large degrees, and vice versa for degree-degree negatively correlated or disassortative. In this paper, we employ the joint degree distribution to characterize the degree correlation. The joint probability $P(j,k)$ is defined as the probability that the two endpoints of a randomly selected edge in the network have degrees $j$ and $k$, respectively \cite{newman2002assortative}. It represents the proportion of the number of edges that exist between nodes with degrees $j$ and $k$ in the network to the total number of edges and can be expressed as: $P(j, k)=\frac{m(j, k) \mu(j, k)}{2 M}$, where $M$ is the number of edges in the network, $m(j, k)$ indicates the number of connected edges between nodes with degree $j$ and nodes with degree $k$. If $j=k$, then $\mu(j, k)=2$, otherwise, $\mu(j, k)=1$.

By observing the joint degree distribution in the fourth row of Fig. \ref{classic_networks} and calculating the assortativity coefficient of the network, we get that most edges in the BA network (assortativity coefficient of -0.082) are connected between small-degree nodes and small-degree nodes. Besides, due to the presence of hub nodes, many small-degree nodes are connected to large-degree nodes, indicating that the BA network is disassortative. However, the WS network (assortativity coefficient of -0.035) shows most connected edges from moderate-degree nodes to other moderate-degree nodes, suggesting a neutral network. Concerning SWOBD (assortativity coefficient of 0.807) and SWBD with power-law distribution (assortativity coefficient of 0.877) in Figs. \ref{classic_networks}(o) and (p), the values in and around the diagonal in the heat maps of the joint degree distribution are relatively large, suggesting that both networks are assortative and the degree correlation is high. In addition, the network structure, clustering coefficient, degree distribution, and joint degree distribution of networks generated by the death process following uniform, exponential, and lognormal distributions are studied, and the interested reader can refer to subsection 3.6 of the supplementary material for more details.

\vspace{-1\baselineskip}
\section{Conclusions and outlooks}
\label{Conclusion}

Complex network modeling has always played a significant role in network science, which not only reveals the formation mechanisms in real networks but is also crucial in studying structural dynamical systems. In this paper, we propose a novel complex network model based on the game between individuals with Q-learning in reinforcement learning and take the lifetime of individuals into account, which follows an arbitrary distribution. Besides, we derive the probability distribution of the system size when it evolves to a steady state and the expected scale of the system. In the numerical simulations, we initially study the stable distribution of the number of individuals under various death processes governed by different distributions, find that the numerical simulation results coincide with the theoretical derivation, and verify the correctness of the theoretical derivation by typical statistical indices, such as relative error and Kullback-Leibler divergence. Subsequently, we observe the evolution of network cooperative behaviors, reveal the emergence and evolution of community structure, and investigate the effects of the payoff parameter and exploitation rate on community structure on both SWBD and SWOBD. We discover that the exploitation rate and the presence of the birth-death process have no significant effect on cooperators, while decreasing the payoff parameter can effectively promote the emergence of cooperators. Besides, we gain that a smaller payoff parameter and a larger exploitation rate will be more conducive to the evolution of community structure on SWOBD and SWBD with lognormal distribution, while on SWBD with power-law, uniform, and exponential distributions, it is difficult for a community structure to emerge regardless of the values of the payoff parameter and the exploitation rate. Furthermore, we apply our model to real situations, namely, fit i) the population evolution of four distinct countries and ii) the degree distributions of six different real networks. Whether in fitting population sizes using the birth-death process or modeling degree distributions in complex networks generated through reinforcement learning, the model proposed in this paper performs well in both cases, demonstrating its excellent practical values. Next, we perform an in-depth analysis of the community structure shaped by SWOBD and reach the conclusion that the learning rate determines the speed of community formation, the discount factor governs the stability of the community, and the dimension of the two-dimensional space impacts the size of the community. Then, we demonstrate the critical role of reinforcement learning in the evolution of complex networks by comparing it with a heuristic model as an alternative approach. Furthermore, through the comparative assessment of our model with classic BA and WS networks concerning network structure, degree distribution, clustering coefficients, etc., we find that our model is capable of generating networks that conform to both power-law and normal distributions, and the networks exhibit high clustering coefficients and assortativity characteristics than BA and WS networks.

The model presented in this paper provides a fresh perspective on the formation and development of community structures in the real world. It also furnishes a valuable framework for studying the dynamic behavior of populations, encompassing phenomena, such as the spread of diseases, the evolution of cooperative behavior, and the synchronization of individuals. Additionally, it is worth emphasizing that the snowdrift game utilized in this paper is not restrictive. The proposed framework is broadly applicable and can be extended to other games by appropriately modifying the payoff matrix. However, there are still some areas for improvement. In particular, beyond the Fermi function utilized in this paper to update individual strategies, other update rules, such as best-take-over and the Moran process, are also worth considering for strategy updates. Besides, expanding the movement of individuals in the two-dimensional plane to the three-dimensional space is also an interesting point. In addition to the evolution of cooperative behavior explored in this paper, the network model developed herein can also serve as a foundational structure for investigating other forms of prosocial behavior on complex networks, including fairness \cite{cimpeanu2023social}, trust \cite{kumar2020evolution}, the safe development of artificial intelligence \cite{cimpeanu2022artificial}, technological coordination \cite{ogbo2022shake}, and open data management \cite{benko2025evolutionary}. In the near future, these issues will be the focus of our studies.


%





\ifCLASSOPTIONcaptionsoff
  \newpage
\fi

\normalem
\bibliographystyle{ieeetr}

\ifCLASSOPTIONcaptionsoff
  \newpage
\fi

\vspace{-4.5\baselineskip}
\begin{IEEEbiography}[{\includegraphics[width=1in,height=1.25in,clip,keepaspectratio]{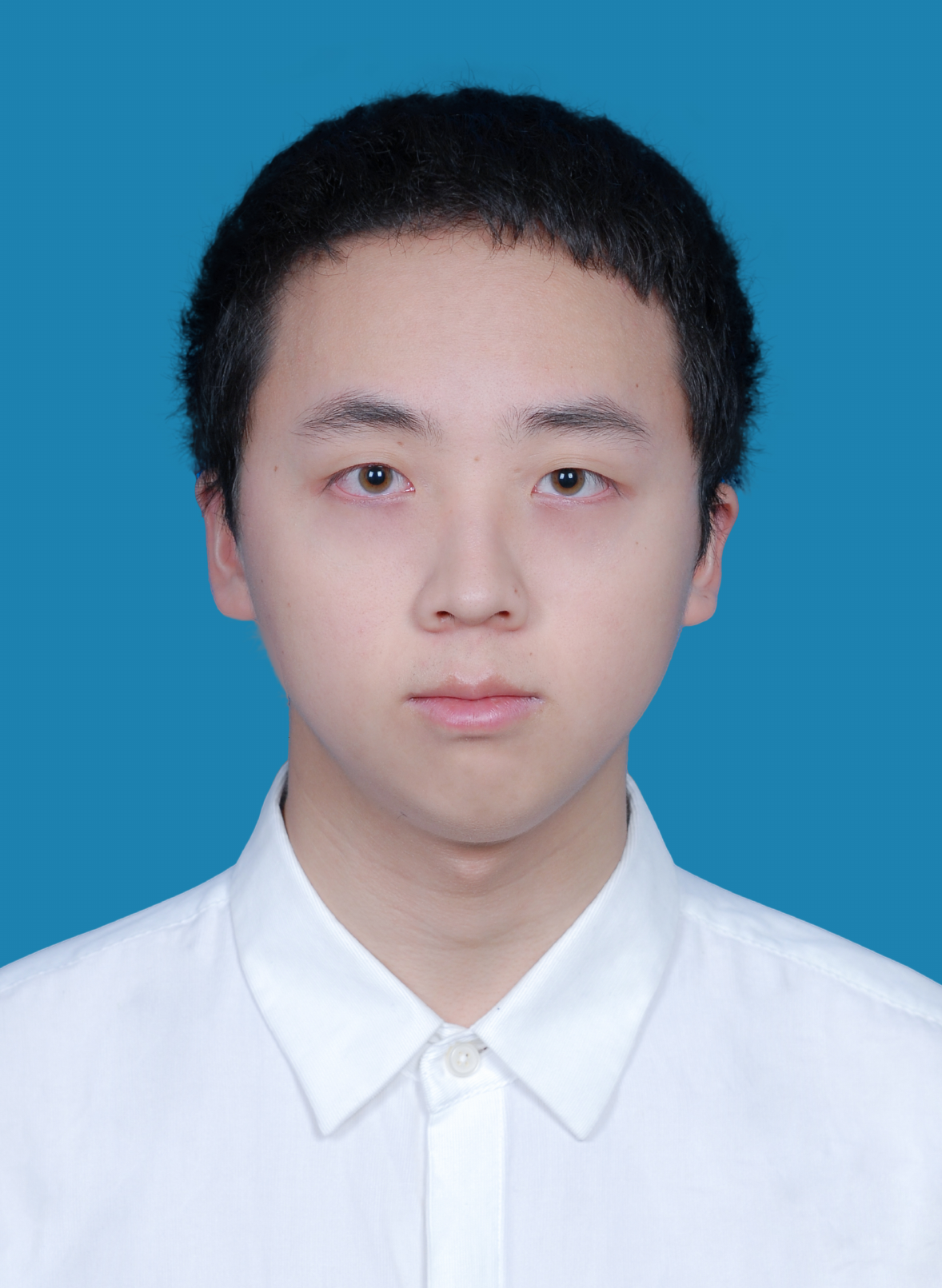}}]{Bin Pi} (Student Member, IEEE) received the B.E. degree in data science and big data technology from the College of Artificial Intelligence, Southwest University, Chongqing, China, in 2023. He is currently pursuing the M.S. degree in mathematics with the School of Mathematical Sciences, University of Electronic Science and Technology of China, Chengdu, China. His research interests include complex networks, evolutionary games, stochastic processes, and reinforcement learning.
\end{IEEEbiography}

\vspace{-3\baselineskip}
\begin{IEEEbiography}[{\includegraphics[width=1in,height=1.25in,clip,keepaspectratio]{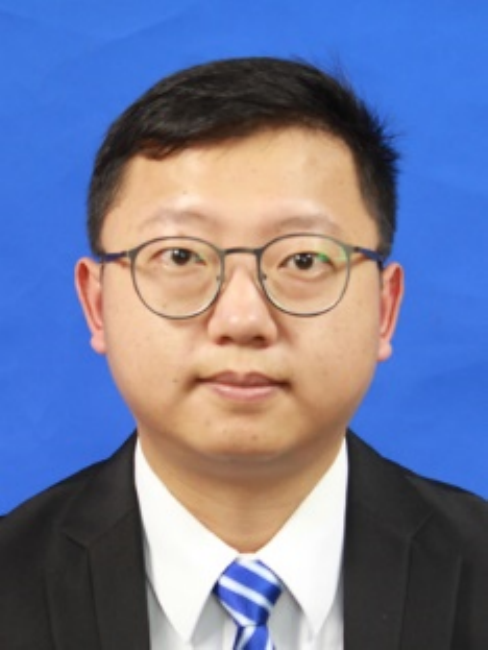}}]{Liang-Jian Deng} (Senior Member, IEEE) received the B.S. and Ph.D. degrees in applied mathematics from the School of Mathematical Sciences, University of Electronic Science and Technology of China (UESTC), Chengdu, China, in 2010 and 2016, respectively. From 2013 to 2014, he was a joint-training Ph.D. Student with Case Western Reserve University, Cleveland, OH, USA. In 2017, he was a Post-Doctoral Researcher with Hong Kong Baptist University (HKBU), Hong Kong. In addition, he has stayed at the Isaac Newton Institute for Mathematical Sciences, University of Cambridge, Cambridge, U.K., and HKBU, for short visits. He is currently a Professor with the School of Mathematical Sciences, UESTC. His research interests include use of optimization modeling, deep learning and numerical PDEs, to address several tasks in image processing and computer vision, e.g., resolution enhancement and restoration. Please visit his homepage for more info.: https://liangjiandeng.github.io/.
\end{IEEEbiography}

\vspace{-3\baselineskip}
\begin{IEEEbiography}[{\includegraphics[width=1in,height=1.25in,clip,keepaspectratio]{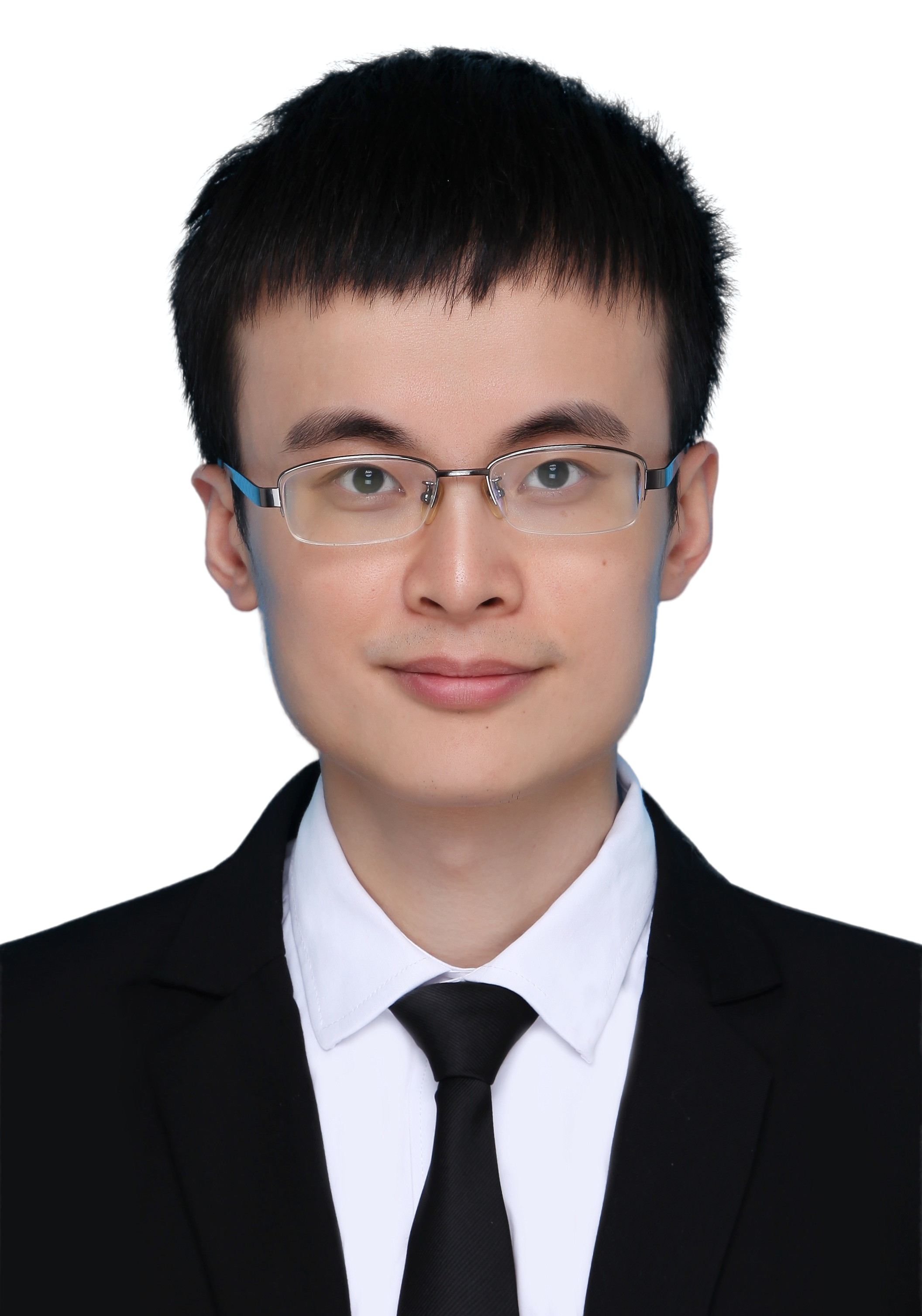}}]{Minyu Feng} (Senior Member, IEEE) received his Ph.D. degree in Computer Science from a joint program between University of Electronic Science and Technology of China, Chengdu, China, and Humboldt University of Berlin, Berlin, Germany in 2018. Since 2019, he has been an associate professor at the College of Artificial Intelligence, Southwest University, Chongqing, China. Dr. Feng is a Senior Member of IEEE, China Computer Federation (CCF), and Chinese Association of Automation (CAA). Currently, he serves as a Subject Editor for Applied Mathematical Modelling, and an Editorial Board Member for Humanities \& Social Sciences Communications, Scientific Reports, and International Journal of Mathematics for Industry. Besides, he is a Reviewer for Mathematical Reviews of the American Mathematical Society. Dr. Feng's research interests include Complex Systems, Evolutionary Game Theory, Computational Social Science, and Mathematical Epidemiology.
\end{IEEEbiography}

\vspace{-3\baselineskip}
\begin{IEEEbiography}[{\includegraphics[width=1in,height=1.25in,clip,keepaspectratio]{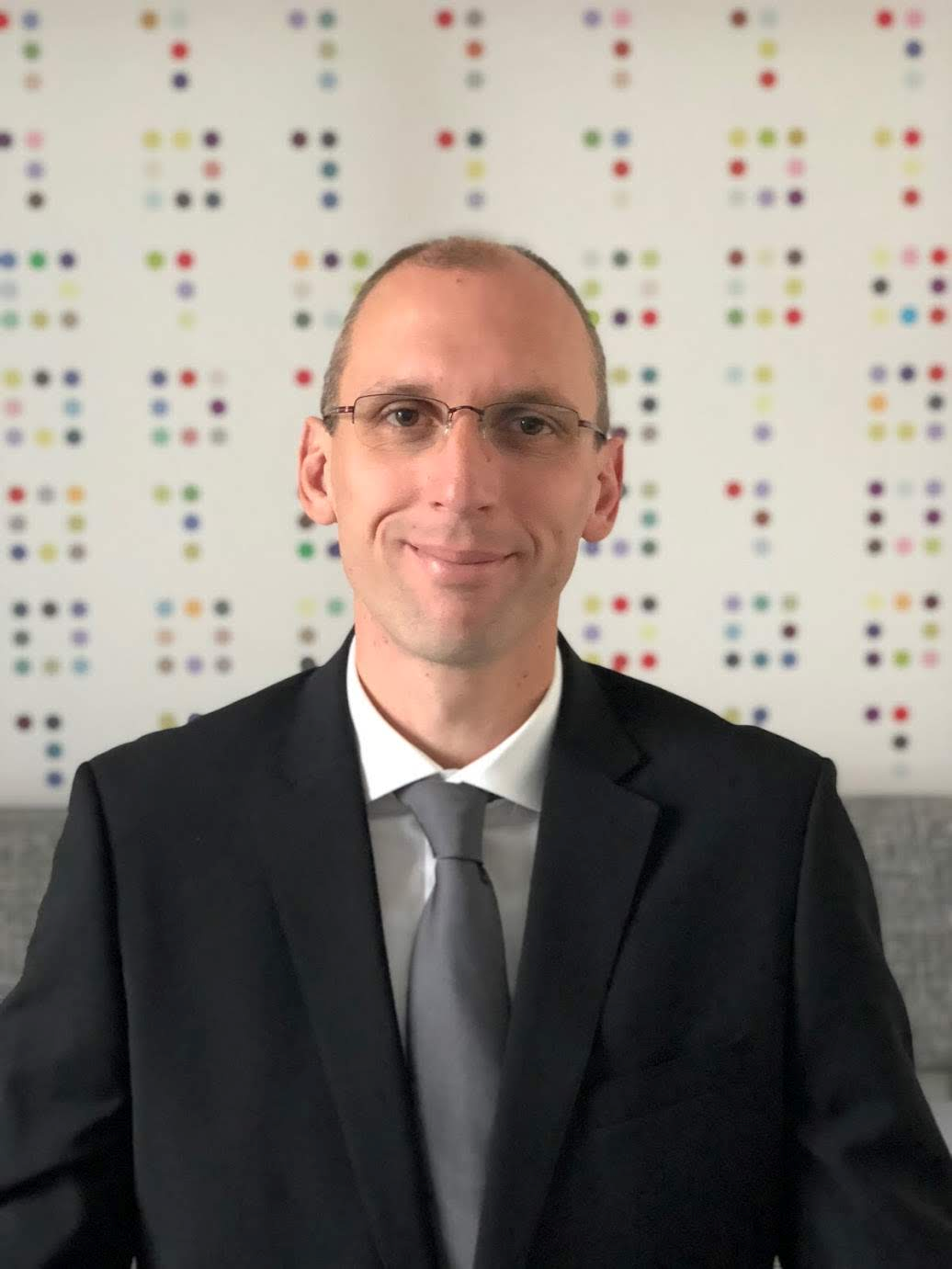}}]{Matja\v{z} Perc} (Member, IEEE) received his Ph.D. in physics from the University of Maribor in 2006. He is currently Professor of physics at the University of Maribor, staff researcher at the Community Healthcare Center Dr. Adolf Drolc Maribor, Adjunct Professor at Kyung Hee University and Korea University, and External faculty member at the Complexity Science Hub in Vienna. He is a member of Academia Europaea and the European Academy of Sciences and Arts, and among top 1\% most cited physicists according to 2020, 2021, 2022, 2023, and 2024 Clarivate Analytics data. He is also the 2015 recipient of the Young Scientist Award for Socio and Econophysics from the German Physical Society, and the 2017 USERN Laureate. In 2018 he received the Zois Award, which is the highest national research award in Slovenia. In 2019 he became Fellow of the American Physical Society. Since 2021 he is also Vice-Dean of Natural Sciences at the European Academy of Sciences and Arts.
\end{IEEEbiography}

\vspace{-3\baselineskip}
\begin{IEEEbiography}[{\includegraphics[width=1in,height=1.25in,clip,keepaspectratio]{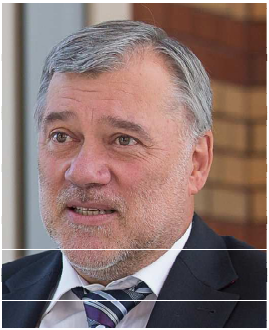}}]{J\"{u}rgen Kurths} received the B.S. degree in mathematics from the University of Rostock, Rostock, Germany, the Ph.D. degree from the Academy of Sciences, German Democratic Republic, Berlin, Germany, in 1983, the Honorary degree from N.I.Lobachevsky State University, Nizhny Novgorod, Russia in 2008, and the Honorary degree from Saratow State University, Saratov, Russia, in 2012. From 1994 to 2008, he was a Full Professor with the University of Potsdam, Potsdam, Germany. Since 2008, he has been a Professor of nonlinear dynamics with the Humboldt University of Berlin, Berlin, Germany, and the Chair of the Research Domain Complexity Science with the Potsdam Institute for Climate Impact Research, Potsdam, Germany. He has authored more than 700 papers, which are cited more than 60000 times (H-index: 111). His main research interests include synchronization, complex networks, time series analysis, and their applications. Dr. Kurths was the recipient of the Alexander von Humboldt Research Award from India, in 2005, and from Poland in 2021, the Richardson Medal of the European Geophysical Union in 2013, and the Eight Honorary Doctorates. He is a Highly Cited Researcher in Engineering. He is a member of the Academia 1024 Europaea. He is an Editor-in-Chief of CHAOS and on the Editorial Boards of more than ten journals. He is a Fellow of the American Physical Society, the Royal Society of 1023 Edinburgh, and the Network Science Society.
\end{IEEEbiography}

\end{document}


%
\title{Dynamic Evolution of Complex Networks: A Reinforcement Learning Approach Applying Evolutionary Games to Community Structure (Supplementary Material)}
%
%
%

\author{Bin Pi, ~\IEEEmembership{Student Member,~IEEE,}
        Liang-Jian Deng, ~\IEEEmembership{Senior Member,~IEEE,}
        Minyu Feng, ~\IEEEmembership{Senior Member,~IEEE,}
        Matja\v{z} Perc,~\IEEEmembership{Member,~IEEE,}
        and J\"{u}rgen Kurths

\thanks{B. Pi, L.-J. Deng, and M. Feng were supported in part by the National Nature Science Foundation of China (NSFC) under Grant Nos. 12271083, 62206230, and 62273077, in part by the Sichuan Province's Science and Technology Empowerment for Disaster Prevention, Mitigation, and Relief Project under Grant No. 2025YFNH0001, and in part by the Natural Science Foundation of Chongqing under Grant No. CSTB2023NSCQ-MSX0064. M.P. was supported by the Slovenian Research and Innovation Agency (Javna agencija za znanstvenoraziskovalno in inovacijsko dejavnost Republike Slovenije) under Grant Nos. P1-0403 and N1-0232.}

\thanks{Bin Pi and Liang-Jian Deng are with the School of Mathematical Sciences, University of Electronic Science and Technology of China, Chengdu 611731, China (e-mail: liangjian.deng@uestc.edu.cn).

Minyu Feng is with the College of Artificial Intelligence, Southwest University, Chongqing 400715, China (e-mail: myfeng@swu.edu.cn).

Matja\v{z} Perc is with Faculty of Natural Sciences and Mathematics, University of Maribor,
Koro{\v s}ka cesta 160, 2000 Maribor, Slovenia, with Community Healthcare Center Dr. Adolf Drolc Maribor, Ulica talcev 9, 2000 Maribor, Slovenia, with Department of Physics, Kyung Hee University, 26 Kyungheedae-ro, Dongdaemun-gu, Seoul 02447, Republic of Korea, with Complexity Science Hub, Metternichgasse 8, 1030 Vienna, Austria, and with University College, Korea University, 145 Anam-ro, Seongbuk-gu, Seoul 02841, Republic of Korea.

J\"{u}rgen Kurths is with the Department of Complexity Science, Potsdam Institute for Climate Impact Research, 14473 Potsdam, Germany, and also with the Institute of Physics, Humboldt University of Berlin, 12489 Berlin, Germany.}

\thanks{Corresponding authors: Liang-Jian Deng and Minyu Feng.}}


%
%

\markboth{IEEE TRANSACTIONS ON PATTERN ANALYSIS AND MACHINE INTELLIGENCE}%
{Shell \MakeLowercase{\textit{et al.}}: Bare Demo of IEEEtran.cls for IEEE Journals}
%



\IEEEtitleabstractindextext{

\begin{abstract}
\small
\justifying
This supplementary material provides the proofs for Theorems 1 and 2 from the main manuscript, as well as additional figures that support and complement the main findings. Additionally, further simulation results and analyses related to the SWBD model with uniform, exponential, and lognormal distributions are also provided for interested readers.

\end{abstract}

\begin{IEEEkeywords}
Complex networks, Evolutionary games, Reinforcement learning, Community structure, Stochastic process.
\end{IEEEkeywords}}

\maketitle

\IEEEdisplaynontitleabstractindextext

%
\IEEEpeerreviewmaketitle

\section{Proof of Theorem 1}
\label{Proof of Theorem 1}

\newtheorem{thm}{Theorem}
\begin{thm}
\color{black}
\label{Theorem 1}
For the continuous-time Markov chain $N(t)$ with the state space $E$, assume that the expectation of death process $\{G(t), t\geq 0\}$ exists, its limiting probability $\{\pi_i, i = 1, 2, \cdots\}$ exists and follows

\begin{equation}
\begin{aligned}
\pi _i&=\underset{t\rightarrow \infty}{\lim}p_i\left( t \right) =\underset{t\rightarrow \infty}{\lim}P\left\{ N\left( t \right) =i \right\} \\
&=\frac{\{\lambda E[G(t)]\}^i}{i!}e^{-\lambda E[G(t)]}.
\end{aligned}
\end{equation}

\end{thm}

\begin{proof}
\color{black}
As we stated, new individuals enter the system at time intervals following an exponential distribution with parameter $\lambda$ and have a lifetime that obeys a general distribution. We yield

\begin{equation}
\begin{aligned}
	p_i\left( t \right) &=P\left\{ N\left( t \right) =i \right\} =\sum_{n=i}^{\infty}{P_{n,i}(N\mid A,t)P\left[ A\left( t \right) =n \right]}\\
	&=\sum_{n=i}^{\infty}{\frac{n!\left[ X(t) \right] ^i[1-X(t)]^{n-i}}{i!(n-i)!}\frac{\left( \lambda t \right) ^n}{n!}e^{-\lambda t}}\\
	&=\frac{\left[ \lambda tX\left( t \right) \right] ^i}{i!}e^{-\lambda t}\sum_{n=i}^{\infty}{\frac{[\lambda t-\lambda tX(t)]^{n-i}}{(n-i)!}}\\
	&=\frac{\left[ \lambda tX\left( t \right) \right] ^i}{i!}e^{-\lambda tX\left( t \right)}.\\
\end{aligned}
\end{equation}

Obviously, the probability $p_i(t)$ of the number of individuals in the system at time $t$ is a non-homogeneous Poisson process. Therefore, under the condition that the expectation of the general distribution $\{G(t), t\geq 0\}$ exists, its limiting probability $\{\pi_i, i = 1, 2, \cdots\}$ exists and follows

\begin{equation}
\label{pi}
\begin{aligned}
	\pi _i&=\underset{t\rightarrow \infty}{\lim}p_i\left( t \right) =\underset{t\rightarrow \infty}{\lim}\frac{[\lambda tX\left( t \right)] ^i}{i!}e^{-\lambda tX\left( t \right)}\\
	&=\frac{\lambda ^i}{i!}\underset{t\rightarrow \infty}{\lim}\left[ tX\left( t \right) \right] ^ie^{-\lambda tX\left( t \right)}\\
	&=\frac{\lambda ^i}{i!}\left\{ \int_0^{\infty}{\left[ 1-G\left( t-x \right) \right]}dx \right\} ^ie^{-\lambda \int_0^{\infty}{\left[ 1-G\left( t-x \right) \right]}dx}\\
	&=\frac{\left\{ \lambda E\left[ G\left( t \right) \right] \right\} ^i}{i!}e^{-\lambda E\left[ G\left( t \right) \right]}.\\
\end{aligned}
\end{equation}

The results follow.
\end{proof}

\section{Proof of Theorem 2}
\label{Proof of Theorem 2}

\begin{thm}
\color{black}
\label{Theorem 2}
The average scale of the system is 
\begin{equation}
\label{EN}
E[N(t)]=\lambda E[G(t)],
\end{equation}
the variance of the scale is
\begin{equation}
\label{variance}
D[N(t)]=\lambda E[G(t)],
\end{equation}
and the average staying time of each individual is
\begin{equation}
E(T)=E[G(t)].
\end{equation}

\end{thm}

\begin{proof}
\color{black}
The average scale of the system is the expectation of the result in Thm. \ref{Theorem 1}, which can be denoted as

\begin{equation}
\begin{aligned}
E[N(t)]&=\sum_{i=0}^{\infty}{i\pi _i}=e^{-\lambda E[G(t)]}\sum_{i=1}^{\infty}{\frac{\{\lambda E[G(t)]\}^i}{(i-1)!}} \\
&=\lambda E[G(t)],
\end{aligned}
\end{equation}
the variance is expressed as

\begin{equation}
\begin{aligned}
    D[N(t)]&=E\left[ N^2(t) \right] -E^2[N(t)] \\
    &=\sum_{i=0}^{\infty}{i^2\pi _i}-\{\lambda E[G(t)]\}^2  \\
&=e^{-\lambda E[G(t)]}\sum_{i=1}^{\infty}{i\frac{\{\lambda E[G(t)]\}^i}{\left( i-1 \right)!}}-\{ \lambda E[G(t)]\}^2 \\
&=\lambda E[G(t)],
\end{aligned}
\end{equation}
and the average staying time of each individual is calculated by

\begin{equation}
E(T)=\frac{E[N(t)]}{\lambda}=E[G(t)].
\end{equation}

The results follow.
\end{proof}

\section{Further Simulation Results}

In this section, we present the evolution of cooperative behaviors and communities, the effect of payoff parameter and exploitation ratio on community structures, and the network structure and metrics in systems with the birth-death process (SWBD), where the death process obeys different distributions, including uniform (SWBD with uniform distribution), exponential (SWBD with exponential distribution), and lognormal (SWBD with lognormal distribution) distributions. Moreover, this section includes supplementary figures that provide additional support and clarification for the key results discussed in the main manuscript. We emphasize that in order to ensure comparability, all parameter settings in each simulation are the same as those in the corresponding simulation in the main manuscript, with the exception of the death process distribution. Furthermore, based on the theoretical results derived in subsection 2.1 of the main manuscript, we adjust the parameters for different death distributions to ensure that the system scales remain similar when the evolution reaches stability.

\subsection{Further Results for Evolution of Network Cooperation Behaviors}

\begin{center}
\begin{figure}[htbp]
\color{black}
\centering
\subfigure[Uniform distribution]{
\includegraphics[scale=0.28]{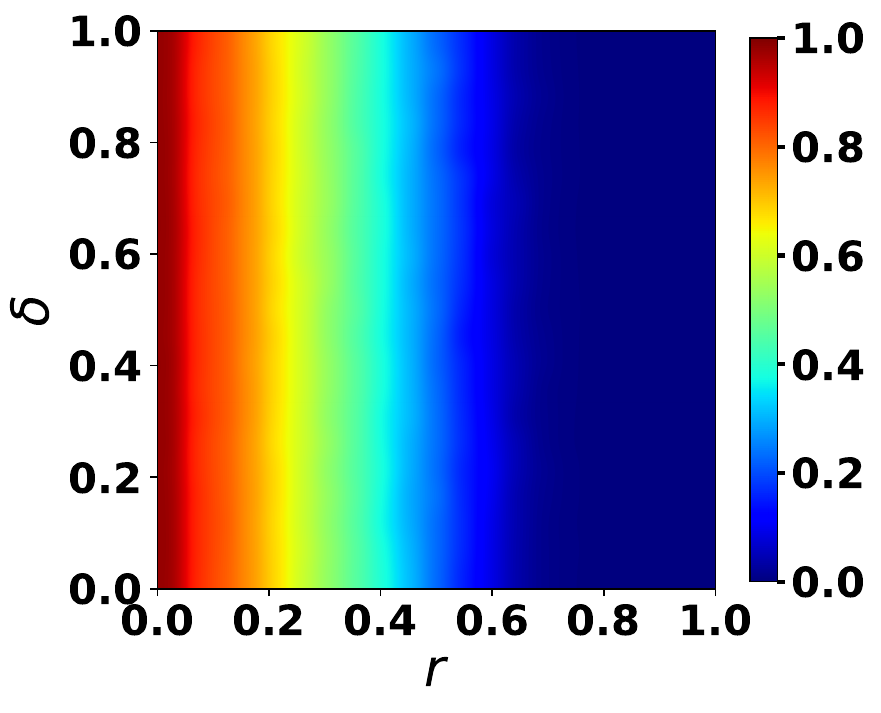}
\label{SWBD_delta_r_fc_uniform}}
\subfigure[Exponential distribution]{
\includegraphics[scale=0.28]{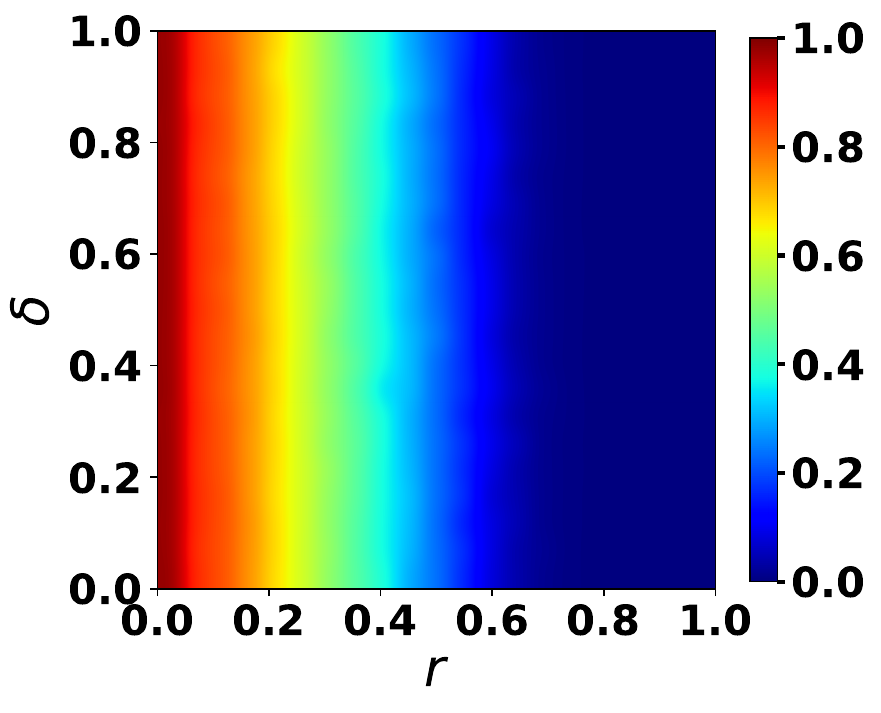}
\label{SWBD_delta_r_fc_exp}}
\subfigure[Lognormal distribution]{
\includegraphics[scale=0.28]{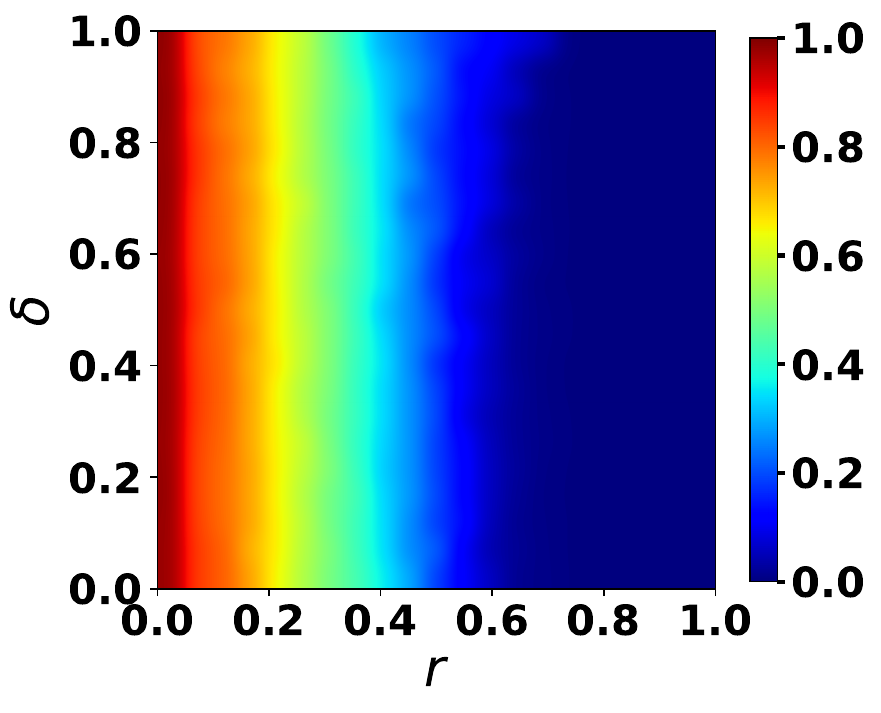}
\label{SWBD_delta_r_fc_lognormal}}
\caption{\textbf{Heat maps of cooperation fraction regarding payoff parameter $r$ and exploitation rate $\delta$ under different death processes.} The figure demonstrates the effect of $r$ and $\delta$ on cooperative behaviors when the death process follows uniform (in subplot (a)), exponential (in subplot (b)), and lognormal (in subplot (c)) distributions. The ranges of both the $x$-axis and $y$-axis are set to [0, 1], with the $x$-axis and $y$-axis respectively representing the payoff parameter $r$ and exploitation rate $\delta$.}
\label{heatmaps_fc}
\end{figure}
\end{center}

In this subsection, we illustrate the evolution of network cooperation behaviors under the death process following uniform, exponential, and lognormal distributions, and the results are shown in Fig. \ref{heatmaps_fc}. The learning rate, discount factor, and weight fading factor of the edge are set to 0.7, 0.3, and 2, respectively, consistent with the settings in the main manuscript. In addition, we set parameters for the different death distributions to ensure that the systems are similar in scale when evolutionary stabilization occurs. In particular, for the uniform distribution, we set the birth rate to $\lambda = 4$ and the starting and ending values to $a = 120$ and $b = 135$. The birth and death rates of the exponential distribution are set to $\lambda = 5.1$ and $\kappa = 0.01$. In the case of the lognormal distribution, the birth rate, mean, and standard deviation are set as $\lambda = 4, \upsilon = 2.85$, and $\phi = 2$, respectively. With these settings, the scale of the system at a steady state in all three cases is approximately 510, which aligns with the power-law distribution scenario discussed in the main manuscript.

\vspace{-1.5\baselineskip}

From Fig. \ref{heatmaps_fc}, we observe that the evolution of the cooperation ratio under the three death distributions is similar. Specifically, the cooperation ratio decreases with the increase of the payoff parameter $r$, while the exploitation rate $\delta$ has almost no effect on the survival of cooperators. The phenomenon is the same as the results in the main manuscript, where the death process obeys a power-law distribution. 

\vspace{-1\baselineskip}
\begin{center}
\begin{figure*}[htbp]
\centering
\subfigure[Network structure of SWBD with power-law distribution]{
\includegraphics[scale=0.19]{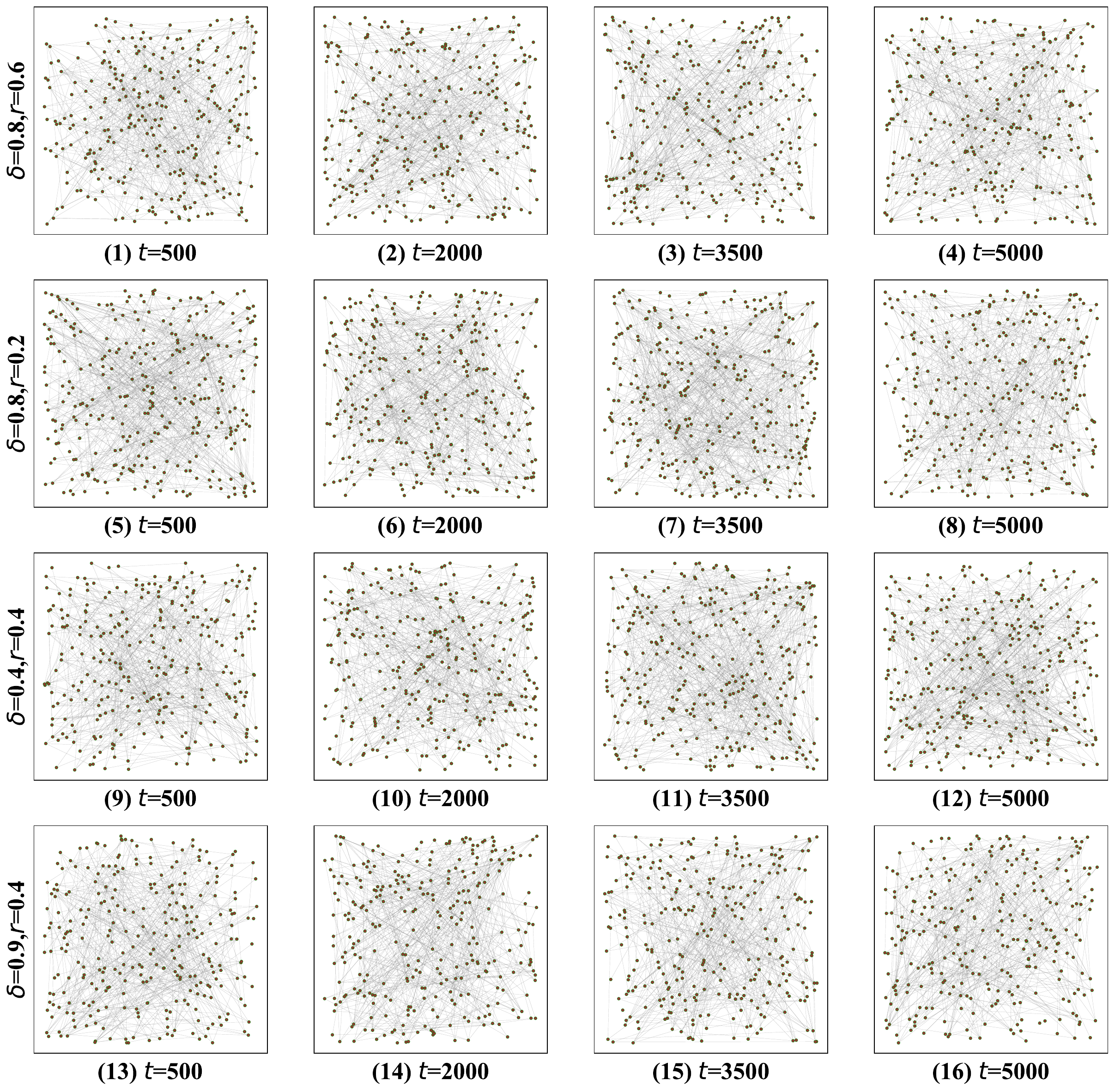}
\label{SWBD_snapshot_network}}
\subfigure[Network structure of SWOBD]{
\includegraphics[scale=0.19]{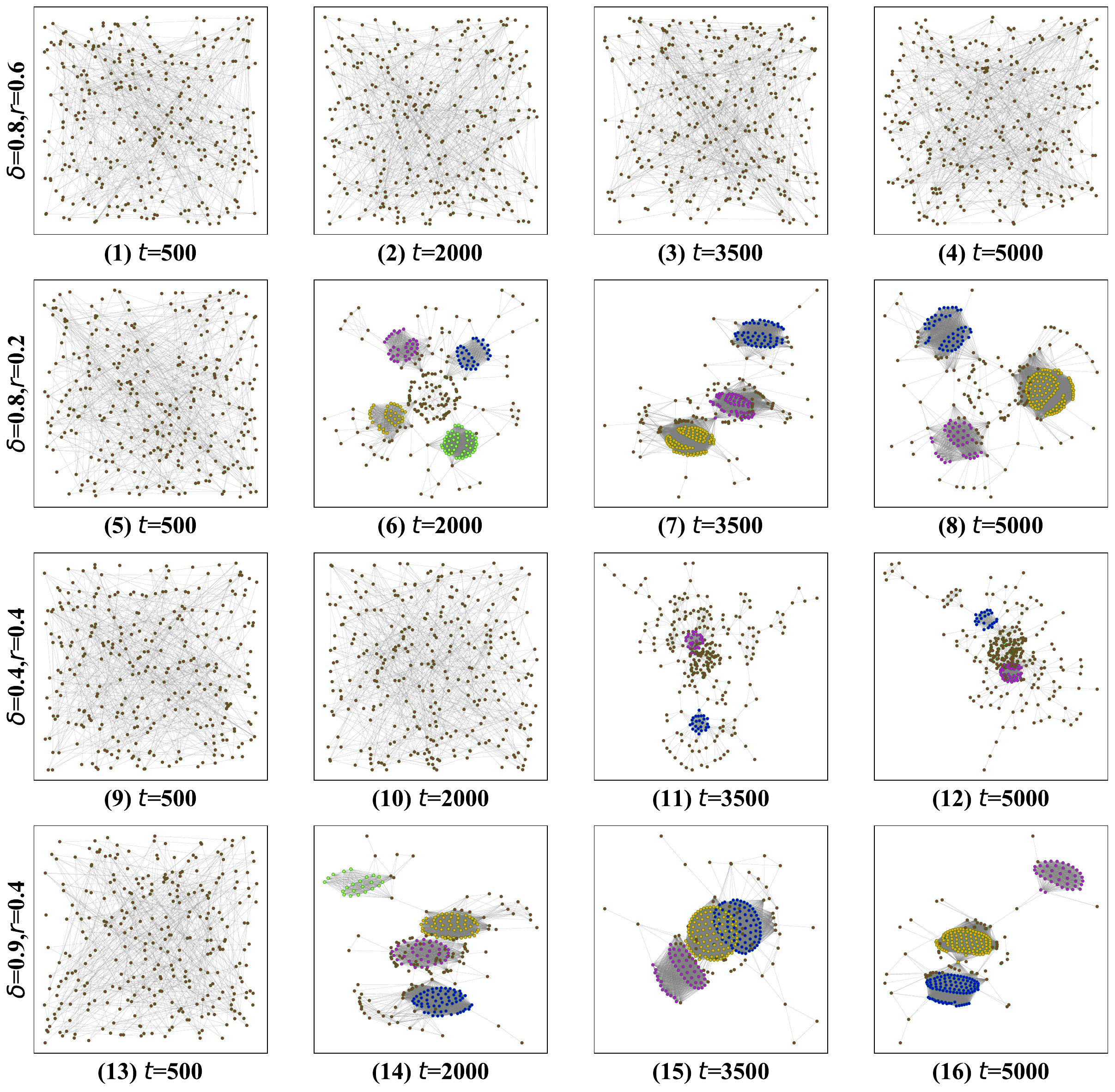}
\label{SWOBD_snapshot_network}}
\caption{\textbf{Network structures corresponding to the evolutionary snapshots in the two-dimensional space shown in Fig. 6 of the main manuscript.} The parameter pairs ($\delta$, $r$) are set to (0.8, 0.6), (0.8, 0.2), (0.4, 0.4), and (0.9, 0.4) from top to bottom. The time steps are fixed at 500, 2000, 3500, and 5000 from left to right.}
\label{snapshots_additional}
\end{figure*}
\end{center}
\vspace{-2.5\baselineskip}

\subsection{Further Results for Emergence and Evolution of Communities}
\label{Further Results for Emergence and Evolution of Communities}

Fig. \ref{snapshots_additional} illustrates the corresponding network structures for the evolutionary snapshots in the two-dimensional space depicted in Fig. 6 of the main manuscript, enabling a direct comparison between individual movements and the resulting network formations.

\vspace{-3.5\baselineskip}
\begin{center}
\begin{figure*}[htbp]
\color{black}
\centering
\subfigure[Individual distribution under uniform distribution]{
\includegraphics[width = 8.5cm, height = 7cm]{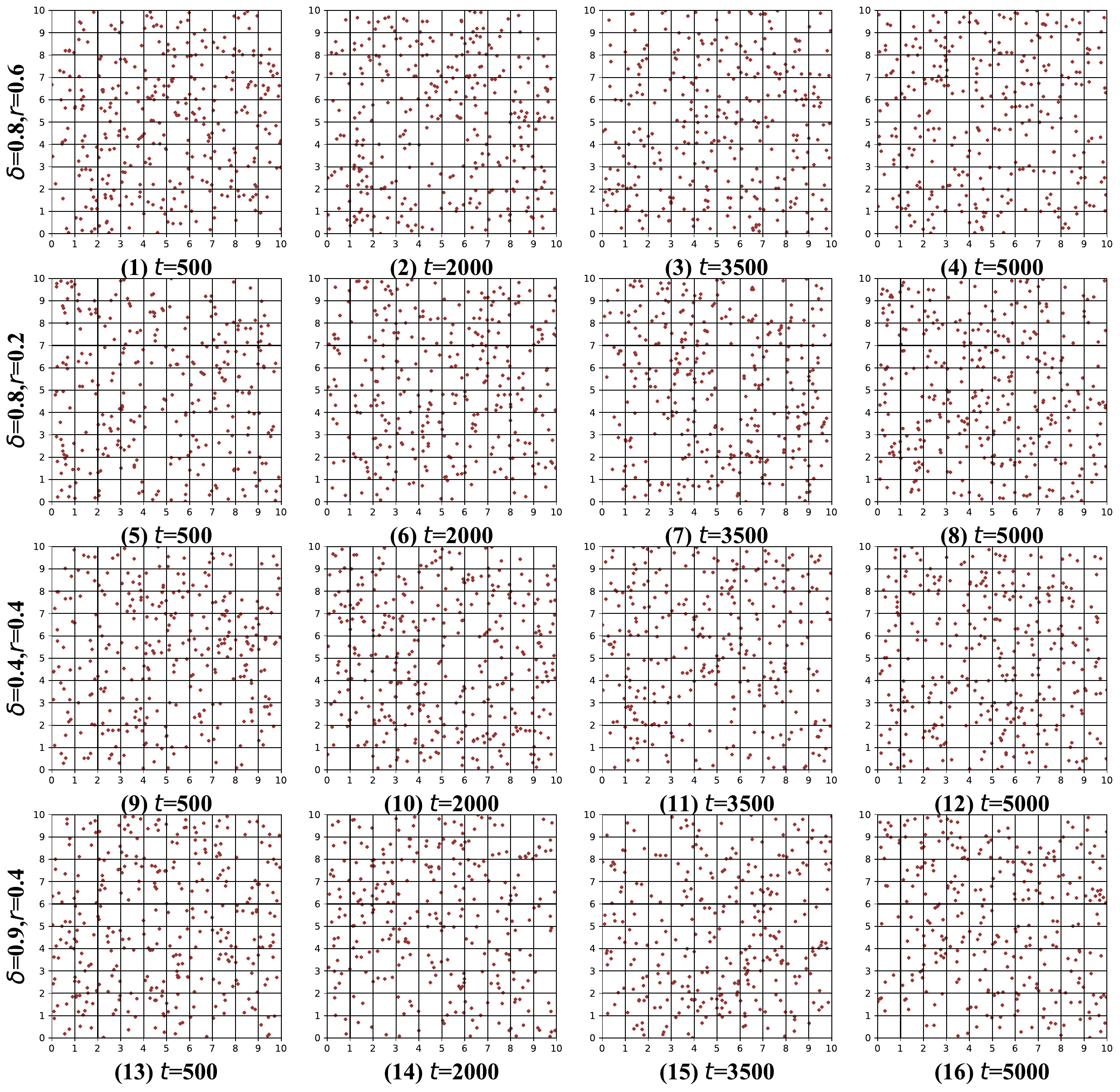}
\label{SWBD_snapshot_uniform}}
\subfigure[Network structure under uniform distribution]{
\includegraphics[width = 8.5cm, height = 7cm]{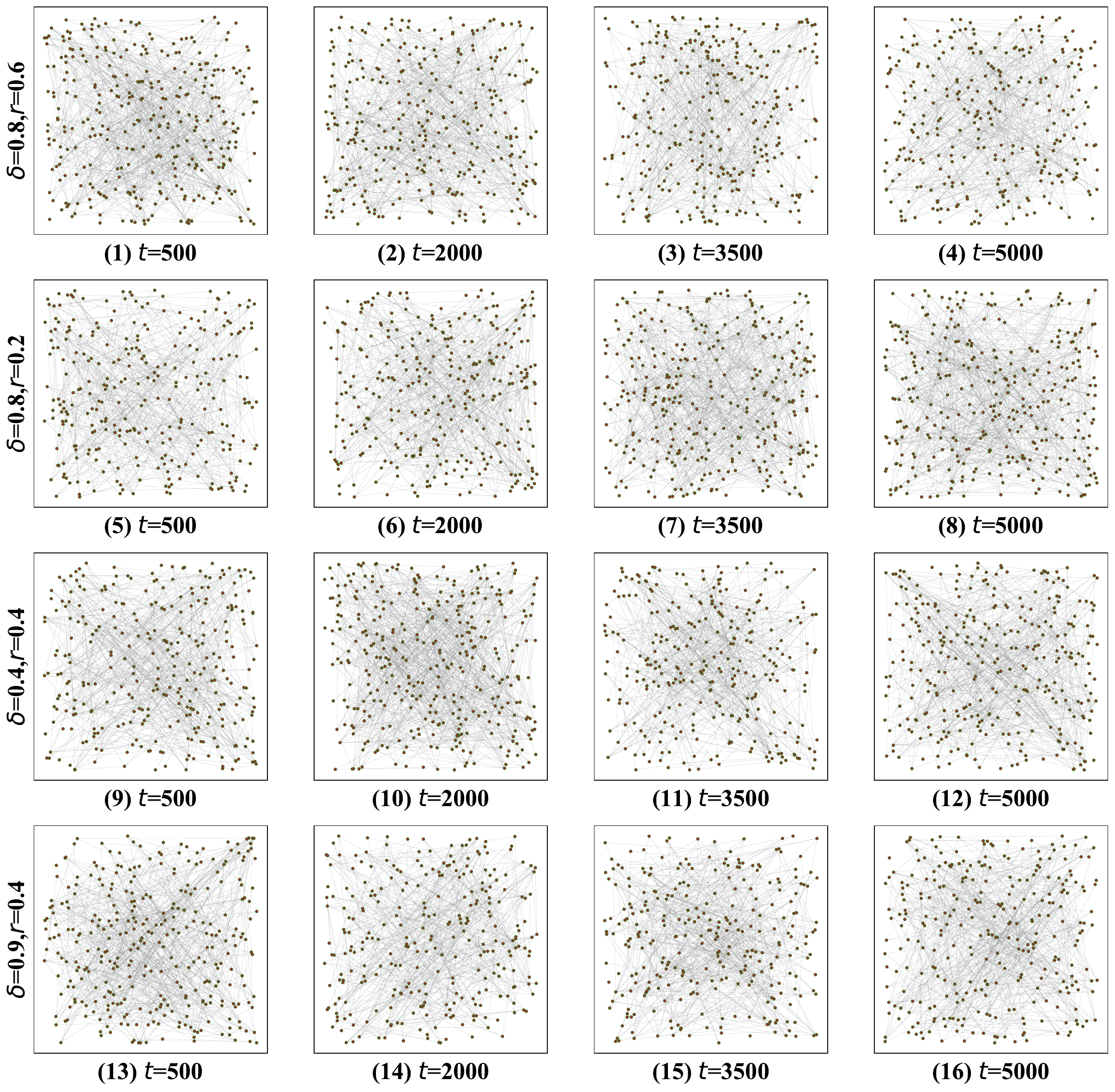}
\label{SWBD_snapshot_network_uniform}}
\subfigure[Individual distribution under exponential distribution]{
\includegraphics[width = 8.5cm, height = 7cm]{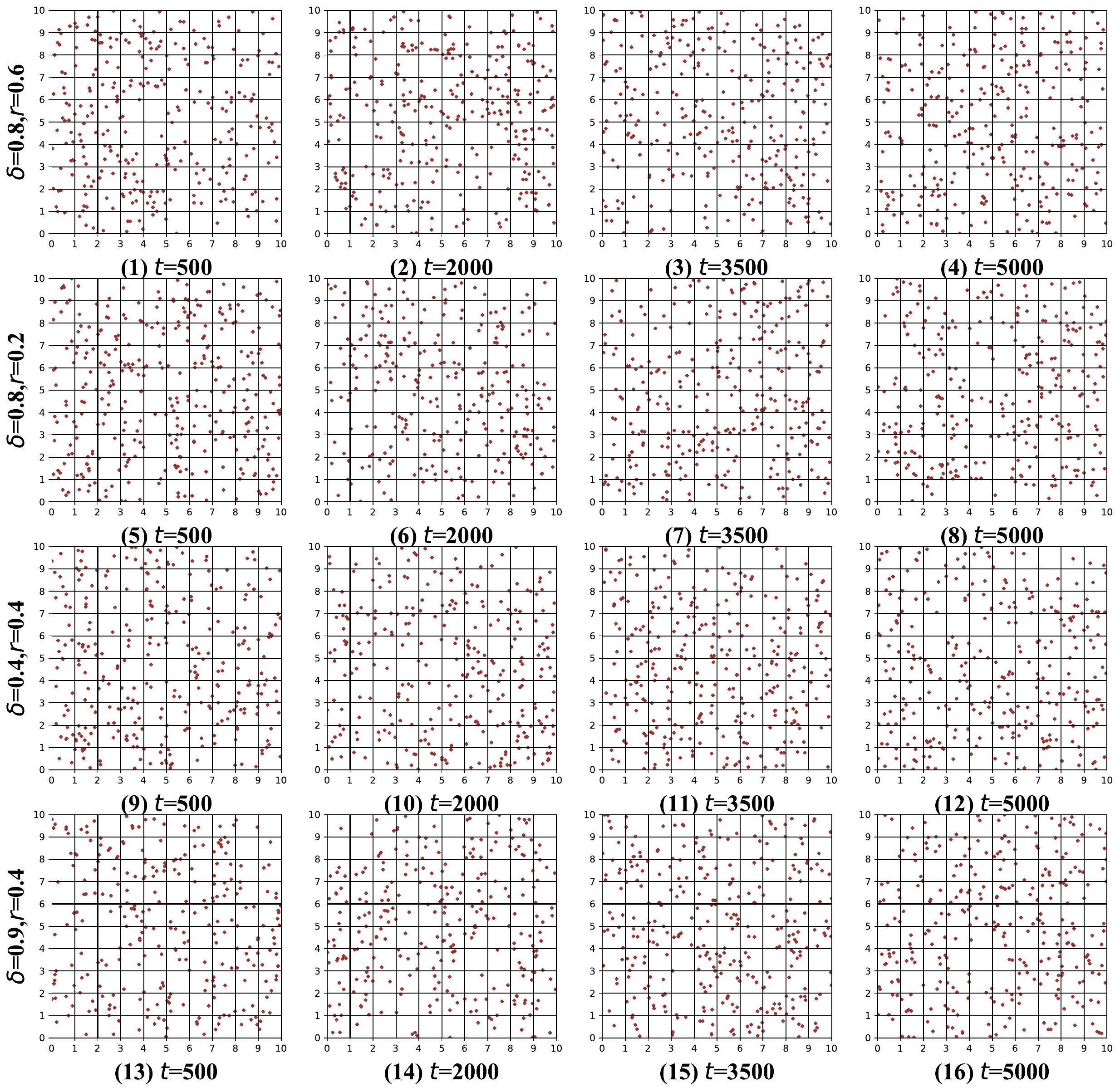}
\label{SWBD_snapshot_exp}}
\subfigure[Network structure under exponential distribution]{
\includegraphics[width = 8.5cm, height = 7cm]{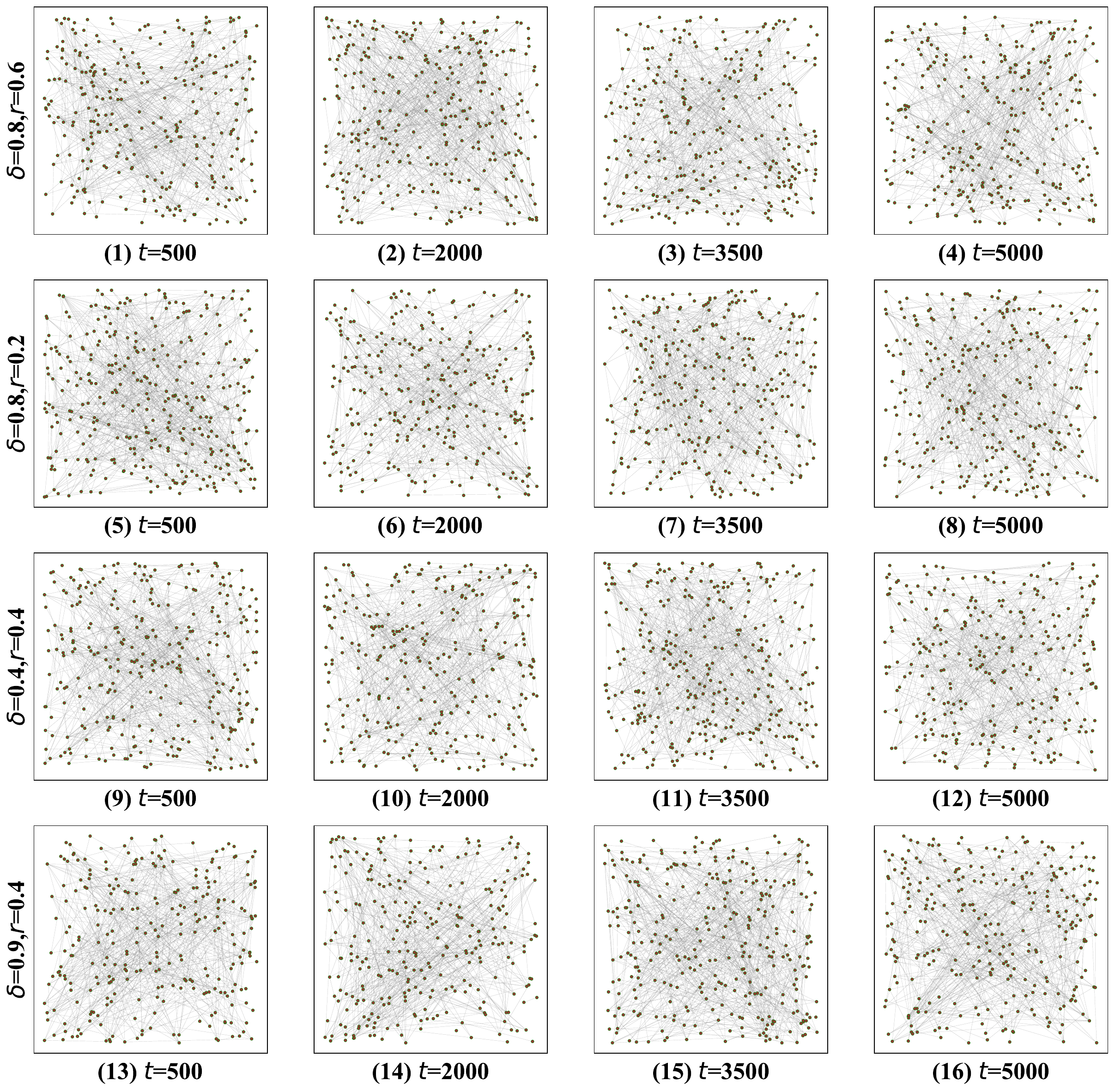}
\label{SWBD_snapshot_network_exp}}
\subfigure[Individual distribution under lognormal distribution]{
\includegraphics[width = 8.5cm, height = 7cm]{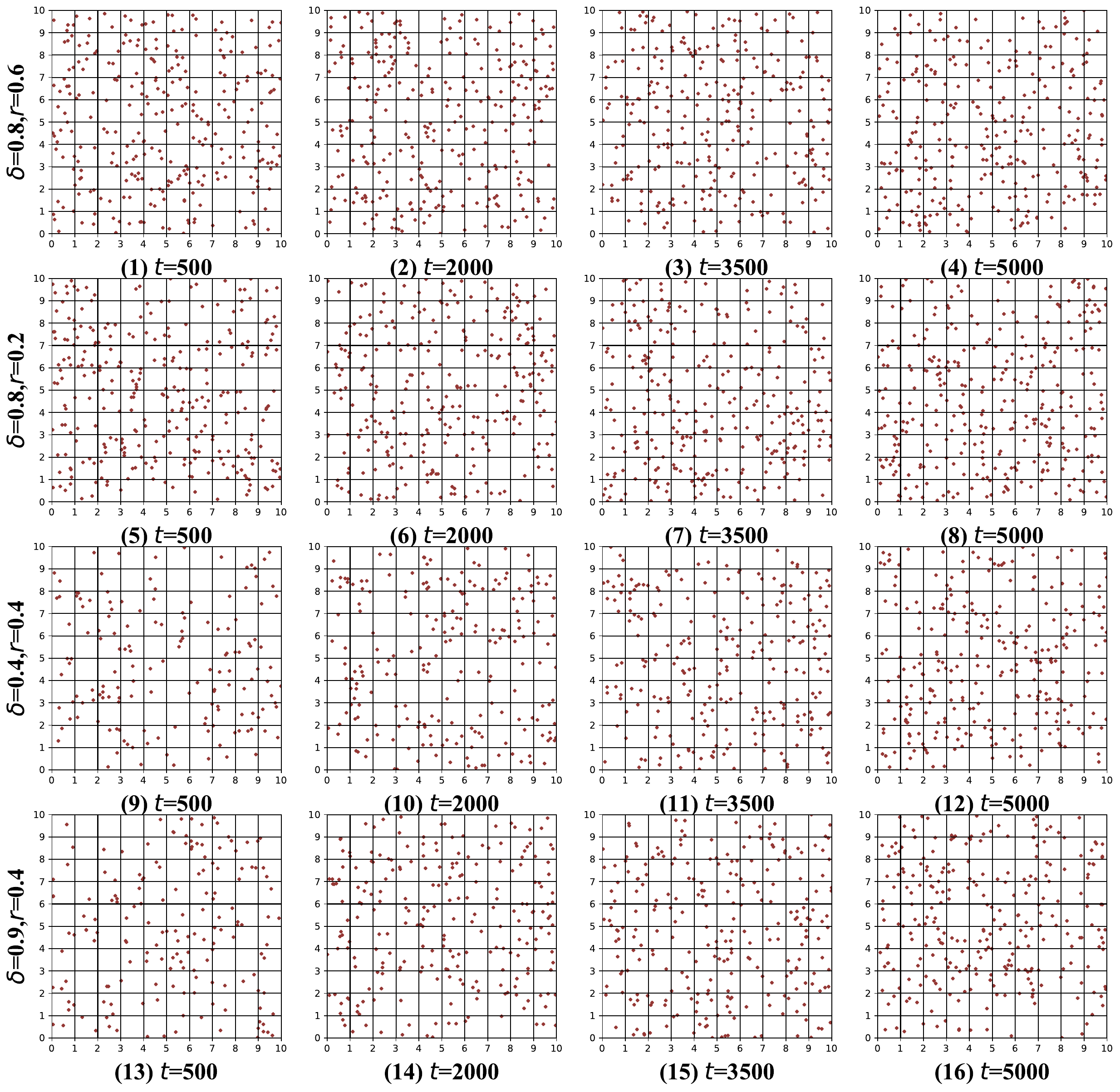}
\label{SWBD_snapshot_lognormal}}
\subfigure[Network structure under lognormal distribution]{
\includegraphics[width = 8.5cm, height = 7cm]{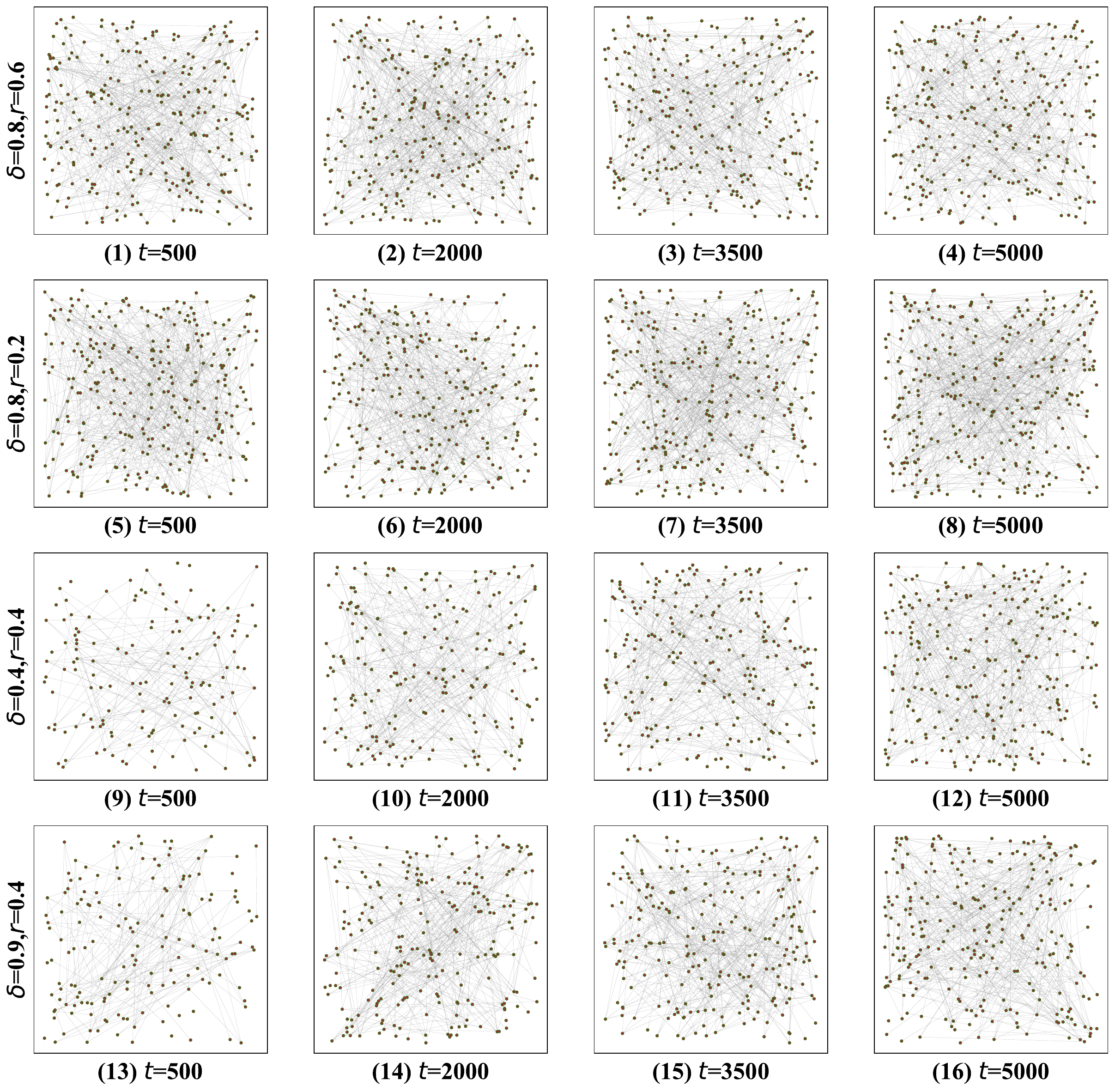}
\label{SWBD_snapshot_network_lognormal}}
\caption{\textbf{Evolutionary snapshots in the two-dimensional space and their corresponding network structures under different parameter pairs ($\delta$, $r$) and death processes.} By setting the parameter pairs ($\delta$, $r$) as (0.8, 0.6), (0.8, 0.2), (0.4, 0.4), and (0.9, 0.4) from top to bottom and fixing the time step at 500, 2000, 3500, and 5000 from left to right, we present the evolutionary snapshots of individual movement in the two-dimensional space on the system with the birth-death process, where the death process follows uniform (in panel (a)), exponential (in panel (c)), and lognormal (in panel (e)) distributions, respectively. The corresponding network structures are displayed in panels (b), (d), and (f). All other parameters in these subplots remain consistent with the main manuscript for comparison.}
\label{snapshots}
\end{figure*}
\end{center}

\begin{center}
\begin{figure*}[htbp]
\centering
\subfigure[Individual distribution of SWOBD within SHG]{
\includegraphics[width = 8.5cm, height = 7cm]{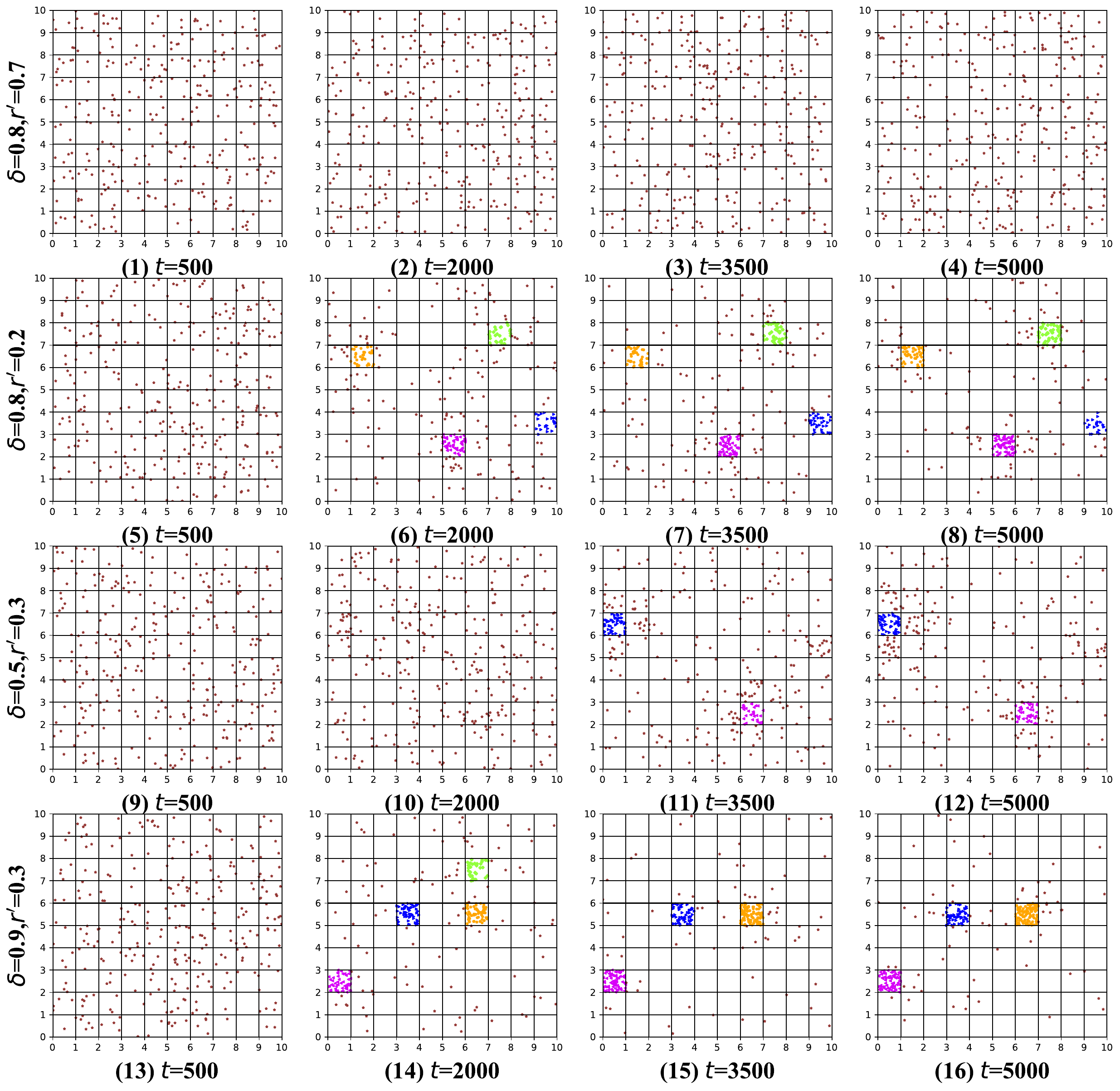}
\label{SWOBD_SHG}}
\subfigure[Network structure of SWOBD within SHG]{
\includegraphics[width = 8.5cm, height = 7cm]{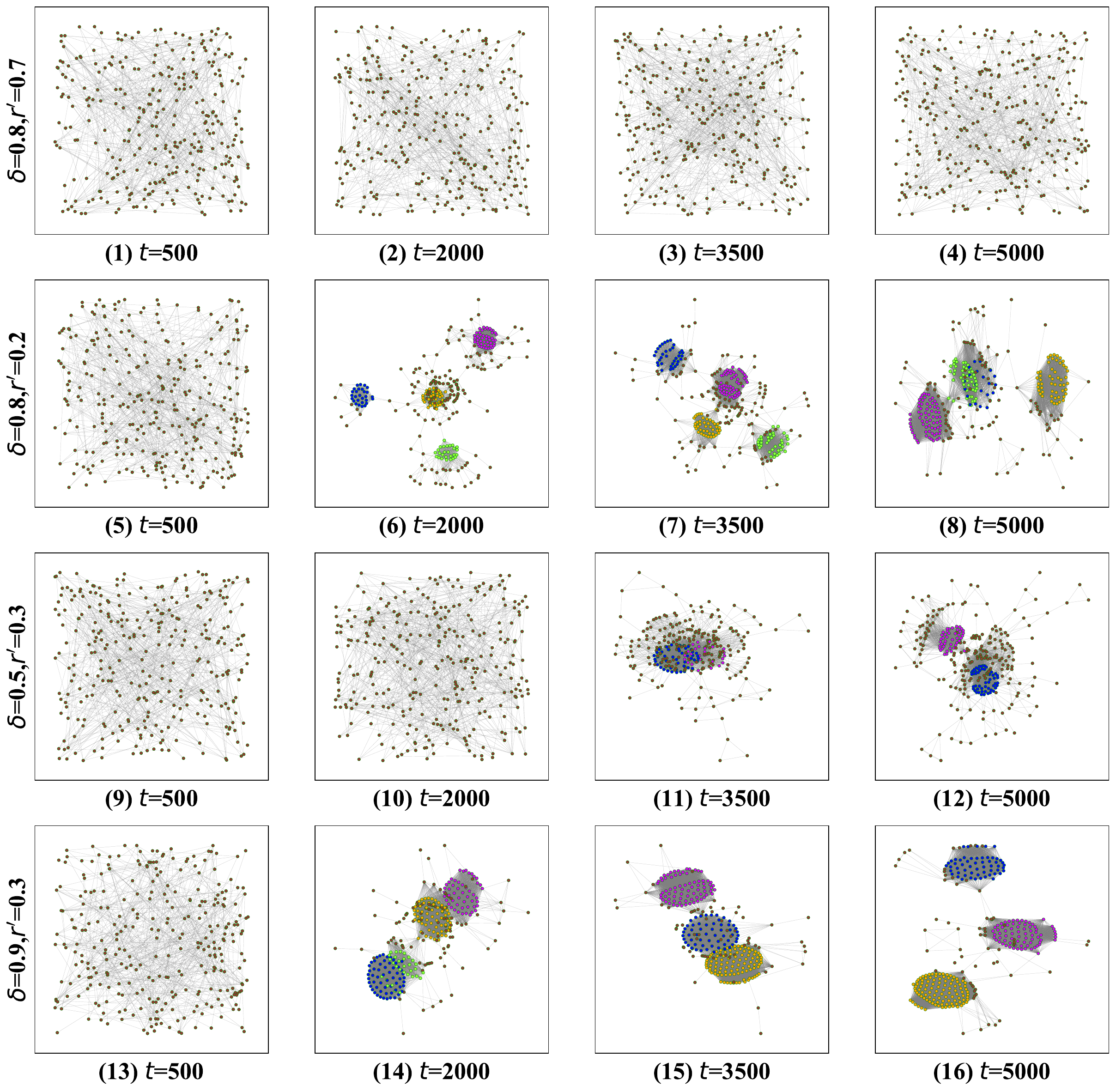}
\label{SWOBD_network_SHG}}
\caption{\textbf{Evolutionary snapshots in the two-dimensional space and corresponding network structures of SWOBD under different parameter pairs ($\delta$, $r'$) within the context of the stag hunt game.} Suplot (a) presents the evolutionary snapshots of individual movement in the two-dimensional space on the system without the birth-death process, under four different combinations of ($\delta$, $r'$): (0.8, 0.7), (0.8, 0.2), (0.5, 0.3), and (0.9, 0.3), arranged from top to bottom. For each combination, the system state is shown at time steps of 500, 2000, 3500, and 5000, from left to right. Subplot (b) displays the corresponding network structures. All other parameters are kept consistent with those used in the main manuscript to facilitate direct comparison.}
\label{snapshots_SHG}
\end{figure*}
\end{center}

To explore the distribution of individuals in two-dimensional space and their corresponding network structure under the death process obeying other distributions, we select the same four sets of parameters ($\delta$, $r$) as in the main manuscript, i.e., (0.8, 0.6), (0.8, 0.2), (0.4, 0.4), and (0.9, 0.4). The evolutionary snapshots in the two-dimensional space and their corresponding network structures under different parameter pairs ($\delta$, $r$) and death processes are demonstrated in Fig. \ref{snapshots}.

From these snapshots of Fig. \ref{snapshots}, it is evident that, regardless of the distribution followed by the death process, no distinct community structure emerges in the system. The distribution of individuals in the two-dimensional space remains random, which is consistent with the results observed when the death process follows a power-law distribution as presented in the main manuscript.

In addition to the snowdrift game, we further investigate the emergence and evolution of communities in SWOBD within the context of the stag hunt game (SHG), where the payoff matrix is defined as:
\begin{equation}
\label{payoff matrix}
M'=\left( \begin{matrix}
	1&		-r'\\
	r'&		0\\
\end{matrix} \right),
\end{equation}
where $r' \in [0,1]$ represents a tunable payoff parameter. Figs. \ref{SWOBD_SHG} and \ref{SWOBD_network_SHG} depict the evolutionary snapshots in the two-dimensional space and the corresponding network structures of SWOBD under different combinations of the exploitation rate $\delta$ and the payoff parameter $r'$ in the SHG setting, respectively. The results show that higher values of $\delta$ and lower values of $r'$ favor the rapid formation and sustained evolution of communities, consistent with the findings observed in the SWOBD within the snowdrift game scenario. These findings confirm that the snowdrift game employed in this study serves as a special case, and the proposed framework can be extended to other game scenarios by substituting the corresponding payoff matrix.

\subsection{Further Results for Effect of Payoff Parameter and Exploitation Ratio on Community Structures}

\begin{center}
\begin{figure}[htbp]
\centering
\subfigure[Uniform distribution]{
\includegraphics[scale=0.28]{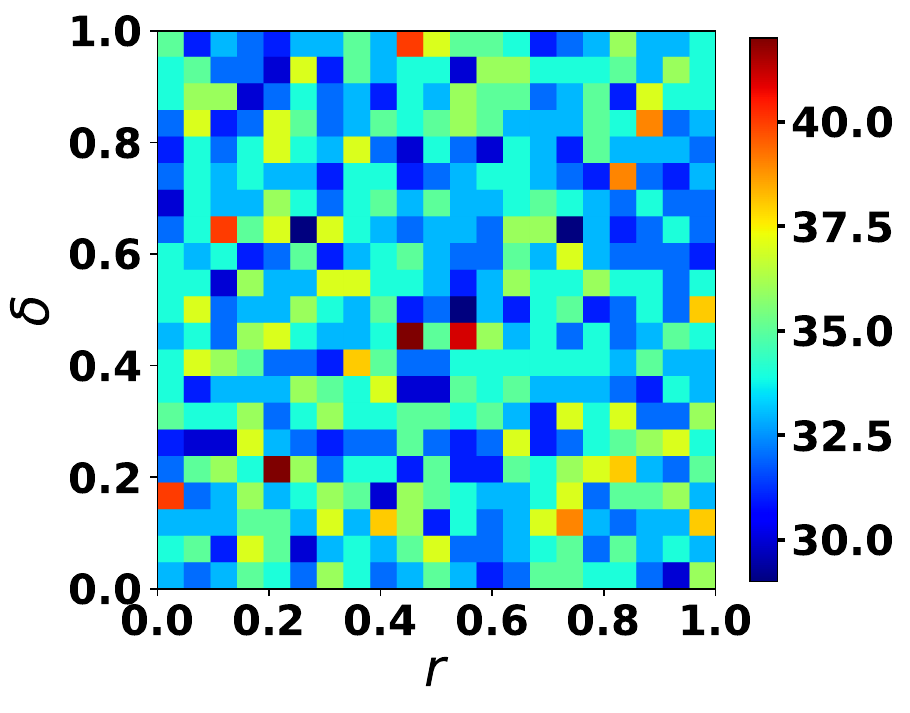}
\label{SWBD_delta_r_cluster_uniform}}
\subfigure[Exponential distribution]{
\includegraphics[scale=0.28]{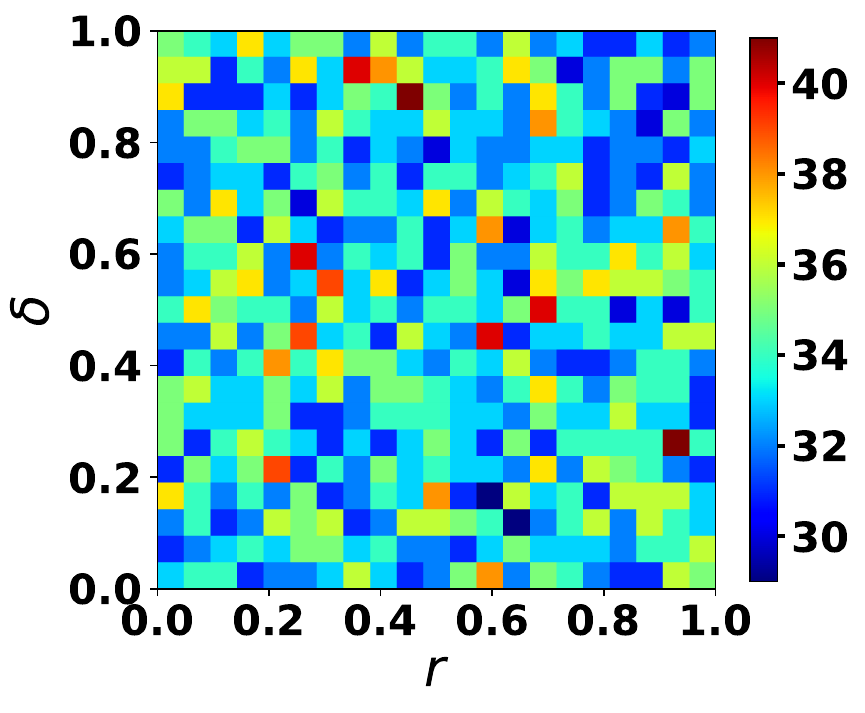}
\label{SWBD_delta_r_cluster_exp}}
\subfigure[Lognormal distribution]{
\includegraphics[scale=0.28]{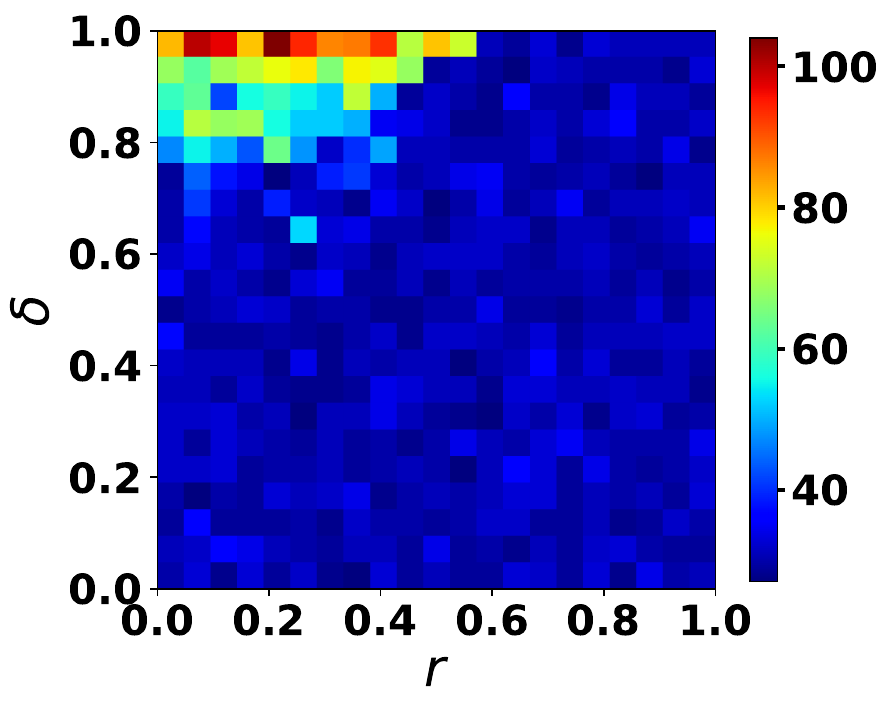}
\label{SWBD_delta_r_cluster_lognormal}}
\caption{\textbf{Heat maps of $N_c$ with respect to payoff parameter $r$ and exploitation rate $\delta$ under various death processes.} By setting the $y$-axis as the exploitation rate $\delta$ with a range [0, 1] and the $x$-axis as the payoff parameter $r$ with a range [0, 1], we demonstrate the heat maps of $N_c$ about payoff parameter $r$ and exploitation rate $\delta$ under the death process obeys uniform (in subplot (a)), exponential (in subplot (b)), and lognormal (in subplot (c)) distributions, respectively.}
\label{heatmaps_community}
\end{figure}
\end{center}

Although the results shown in subsection \ref{Further Results for Emergence and Evolution of Communities} indicate no community structure formation for four specific parameter combinations, this observation may not generalize across all parameter combinations. To further investigate the effect of the payoff parameter and the exploitation rate on community structures, we plot the heat maps of $N_c$ (the sum of the number of individuals in the four locations with the highest populations throughout the evolutionary process) with respect to $r$ and $\delta$ under three different death distributions, and the results are demonstrated in Fig. \ref{heatmaps_community}.

\vspace{-1.5\baselineskip}

From Figs. \ref{SWBD_delta_r_cluster_uniform} and \ref{SWBD_delta_r_cluster_exp}, we can get that the value of $N_c$ is relatively small when the death process obeys both uniform and exponential distributions, indicating a lack of community structure in these cases. Furthermore, the heat maps exhibit a highly uneven distribution of $N_c$, suggesting that community formation is not significantly influenced by the payoff parameter $r$ or the exploitation rate $\delta$. These findings align with the results obtained with the power-law distribution for the death process, as discussed in the main manuscript. In contrast to the results for other death distributions, the case where the death process follows a lognormal distribution shows a notable emergence of community structures. As illustrated in Fig. \ref{SWBD_delta_r_cluster_lognormal}, when the exploitation rate $\delta$ is large and the payoff parameter $r$ is small, $N_c$ can reach approximately 100, indicating the formation of community structures within the system.

\begin{center}
\begin{figure}[htbp]
\centering
\includegraphics[scale = 0.28]{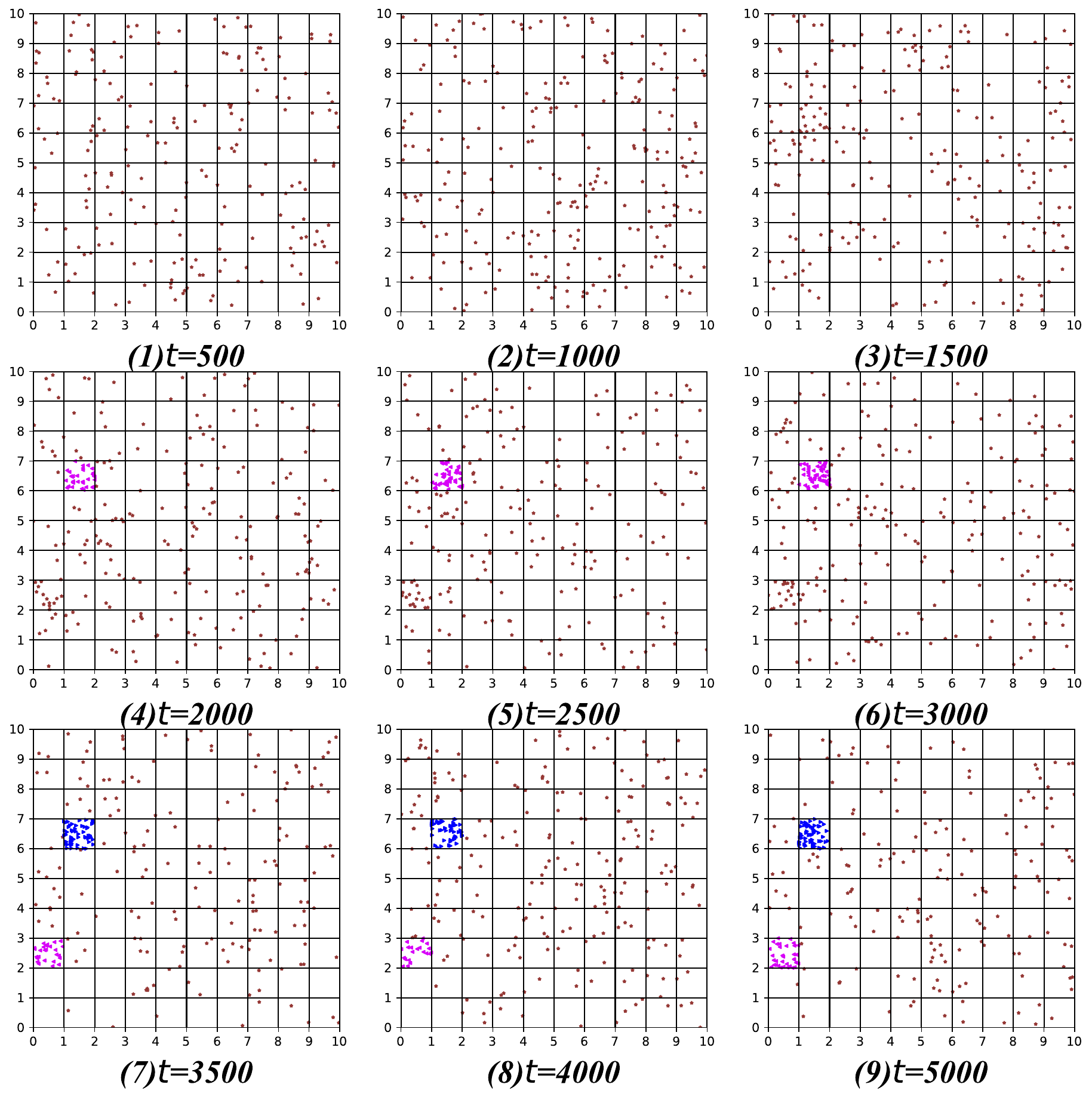}
\caption{\textbf{Evolutionary snapshots in the two-dimensional space for parameter pair ($\delta$, $r$) = (1.0, 0.1) with a lognormal death process.} The figure depicts the individual distribution in the two-dimensional space over time, where the death process follows a lognormal distribution with the parameters $\delta = 1.0$ and $r = 0.1$. The snapshots reveal the emergence of distinct community structures as time progresses.}
\label{snapshot_lognormal}
\end{figure}
\end{center}

To provide a more intuitive understanding, Fig. \ref{snapshot_lognormal} presents the distribution of individuals in the two-dimensional space over time, with the death process following a lognormal distribution and parameters set to $\delta = 1.0$ and $r = 0.1$, respectively. The snapshots reveal that a community structure begins to form around $t = 2,000$. Another community structure emerges around $t = 3,500$ and subsequently stabilizes, which is consistent with the results depicted in Fig. \ref{SWBD_delta_r_cluster_lognormal}. However, it is important to note that although community structures are present in this scenario, the value of $N_c$ remains smaller compared to systems without the birth-death process. Specifically, the number and size of community structures in systems without the birth-death process are larger, as demonstrated in Figs. 6(b) and 7(b) of the main manuscript.

\subsection{Further Results for Fit of the Degree Distribution using Reinforcement Learning on Real Networks}

\begin{center}
\begin{figure}[htbp]
\centering
\subfigure[HS-HT]{
\includegraphics[width = 4.2cm, height = 3cm]{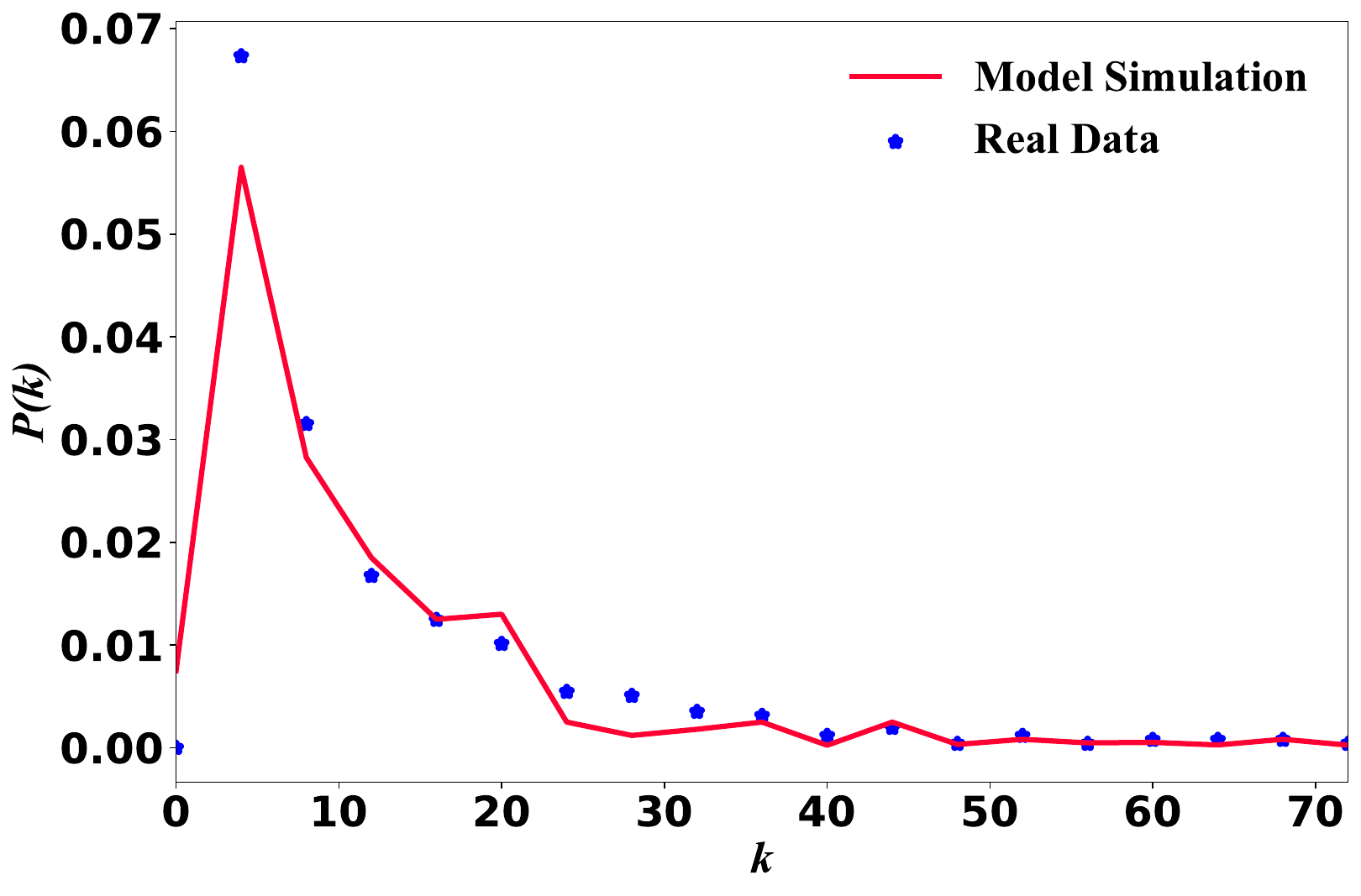}
\label{fit_SWOBD}}
\subfigure[TWITTER-COPEN]{
\includegraphics[width = 4.2cm, height = 3cm]{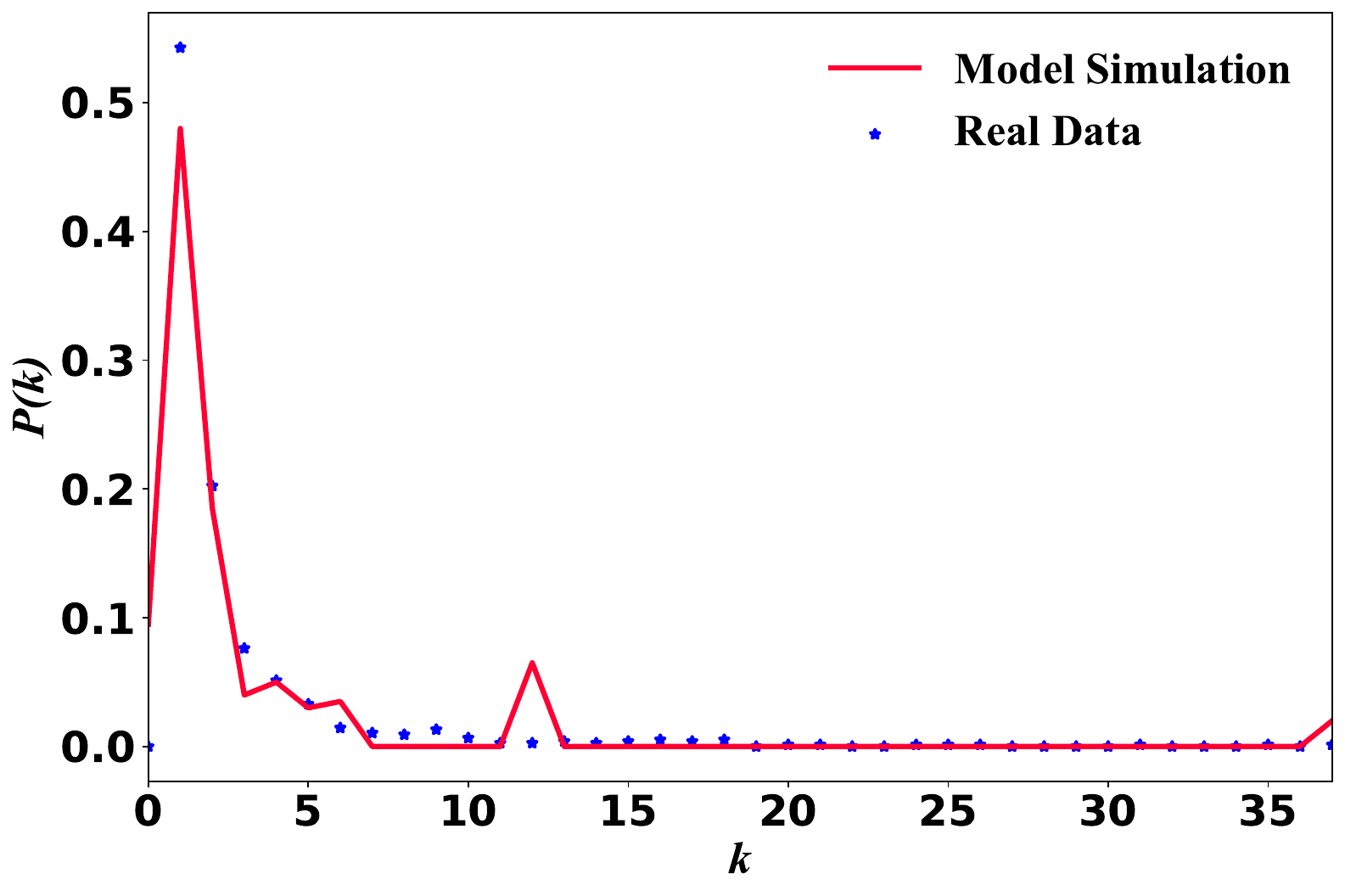}
\label{fit_SWOBD_2}}
\subfigure[AVES-WEAVER-SOCIAL]{
\includegraphics[width = 4.2cm, height = 3cm]{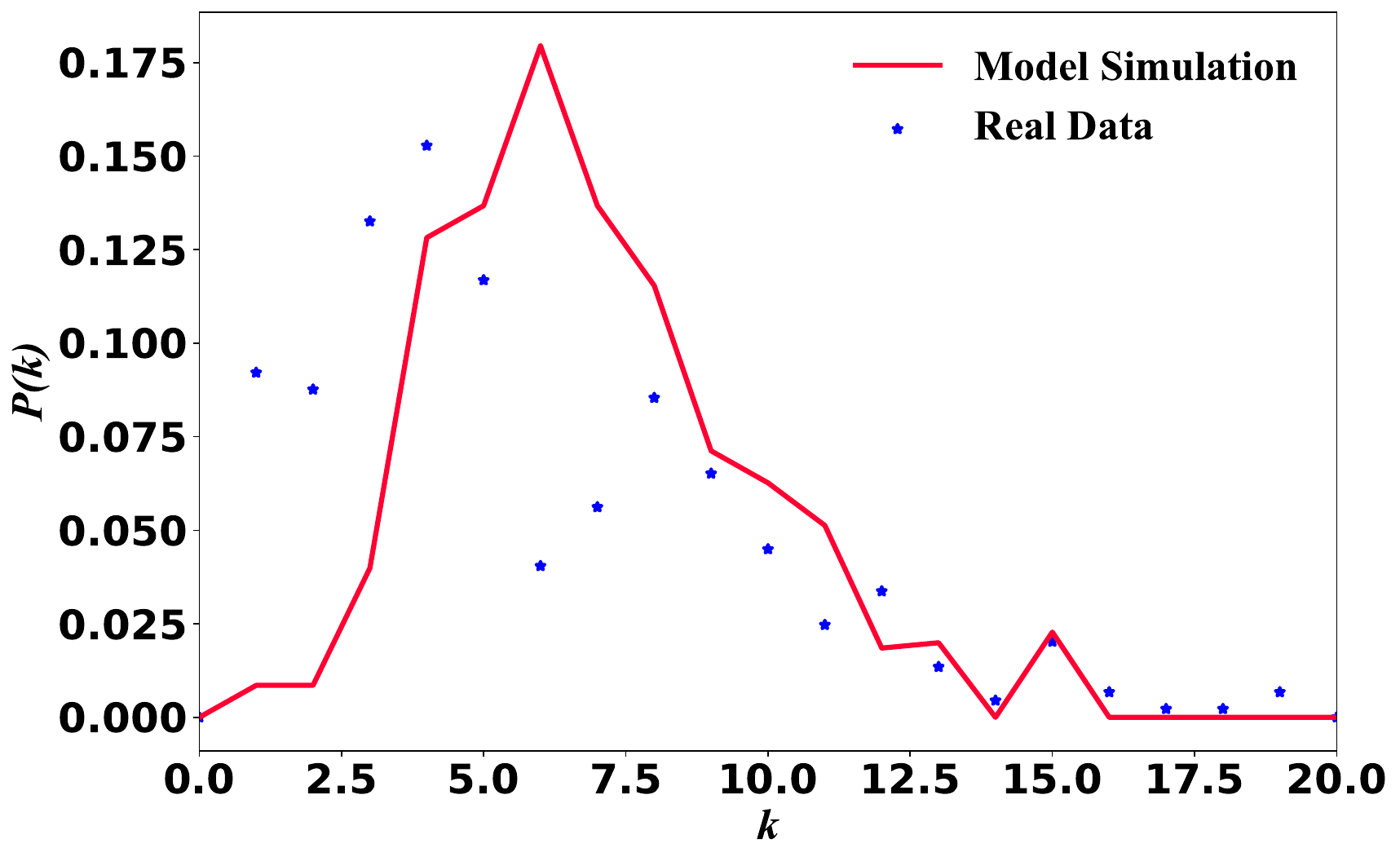}
\label{fit_SWBD}}
\subfigure[DWT-607]{
\includegraphics[width = 4.2cm, height = 3cm]{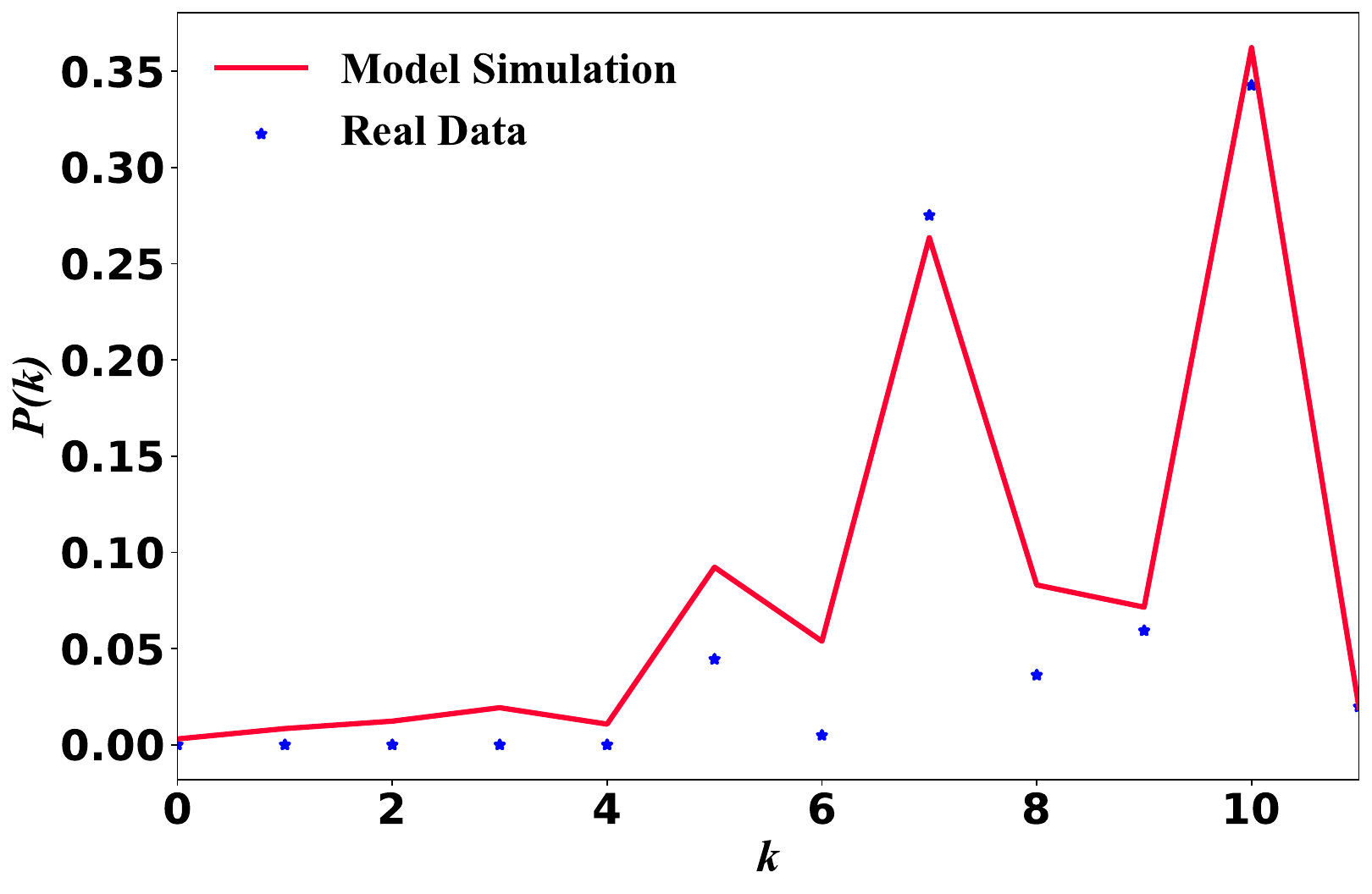}
\label{fit_SWOBD_uniform}}
\subfigure[MAMMALIA-DOLPHIN]{
\includegraphics[width = 4.2cm, height = 3cm]{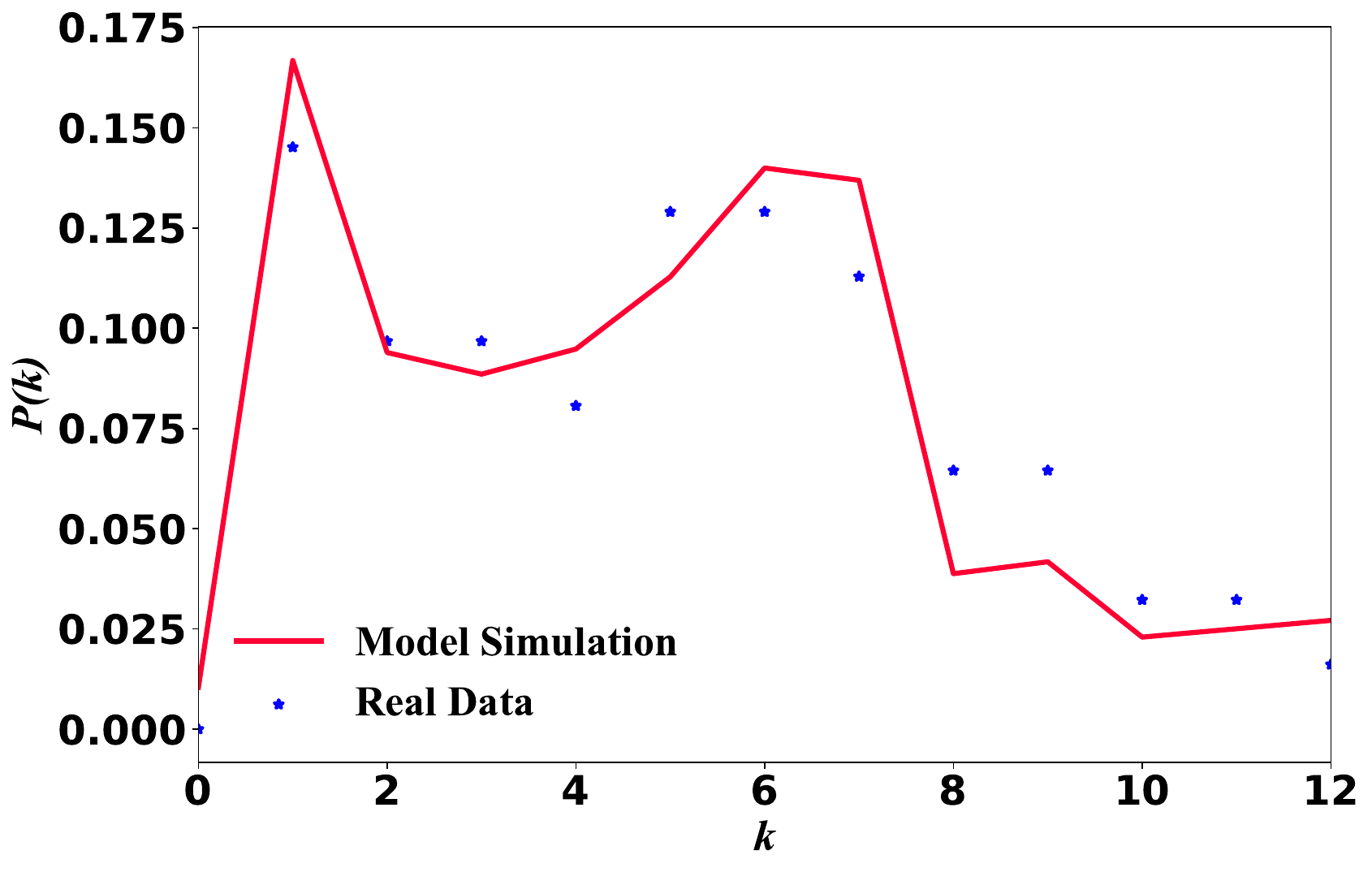}
\label{fit_SWOBD_exp}}
\subfigure[NETSCIENCE]{
\includegraphics[width = 4.2cm, height = 3cm]{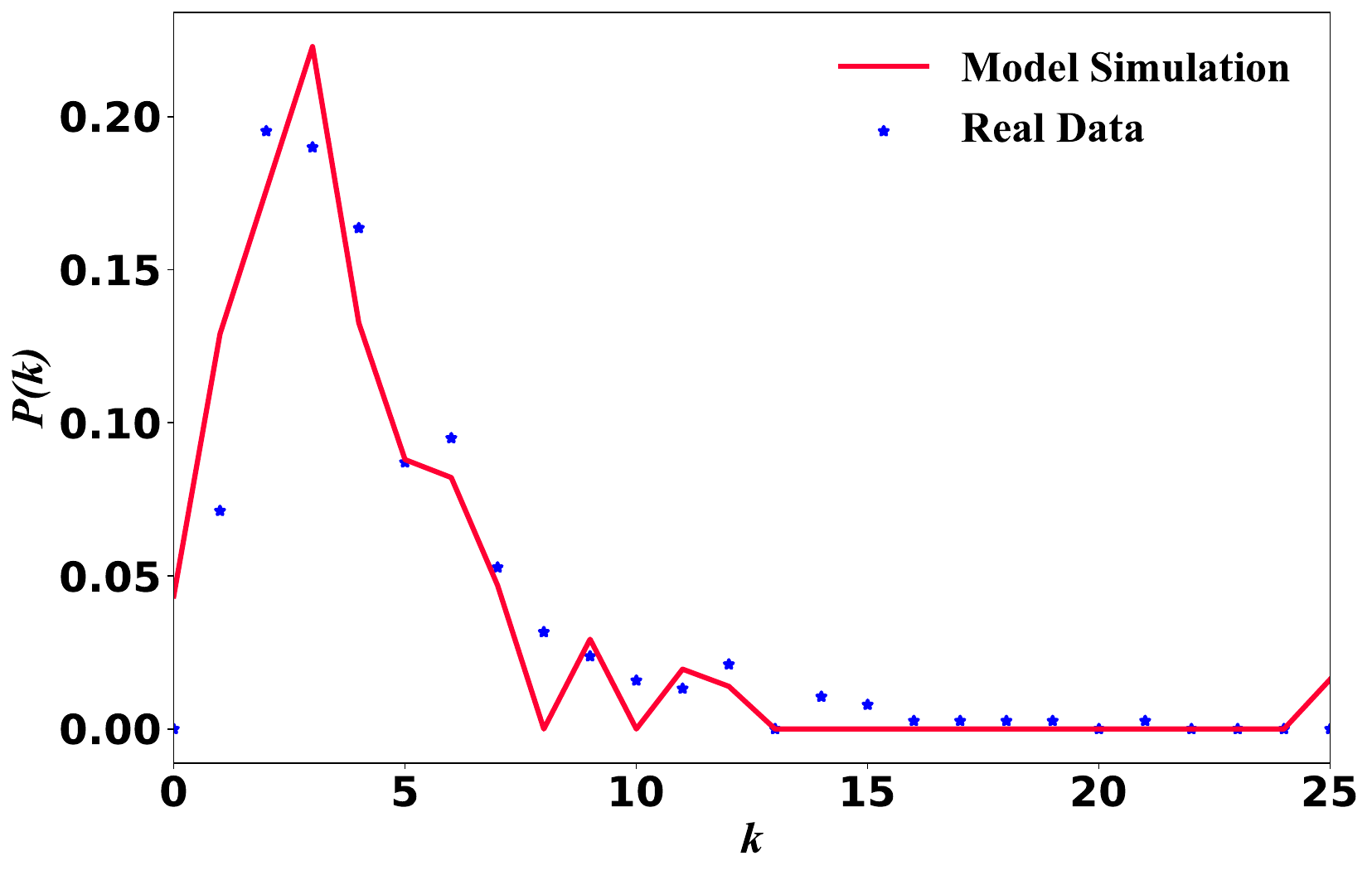}
\label{fit_SWOBD_lognormal}}
\caption{\textbf{Fit of the degree distribution on six different real networks.} The figure illustrates the degree distributions of networks generated by the SWOBD model, which are used to fit the degree distributions of the HS-HT network (panel (a)) and the TWITTER-COPEN network (panel (b)). Additionally, we fit the degree distributions of real networks using networks generated by the SWBD model under various death processes. Subplots (c) and (d) show the fitting results for the AVES-WEAVER-SOCIAL and DWT-607 networks, where the death processes follow power-law and uniform distributions. Subplots (e) and (f) present the fitting results for the MAMMALIA-DOLPHIN and NETSCIENCE networks, where the death processes follow exponential and lognormal distributions, respectively. The $x$-axis represents the degree of the network, while the $y$-axis indicates the probability of each degree.}
\label{fit_degree}
\end{figure}
\end{center}

In this subsection, we evaluate the fit of the degree distribution on six different real networks, and the results are shown in Fig. \ref{fit_degree}.

\vspace{-1.5\baselineskip}
\vspace{-3\baselineskip}

\subsection{Further Results for Comparison with Heuristic Model}

\begin{center}
\begin{figure}[htbp]
\centering
\subfigure[Degree distribution]{
\includegraphics[scale=0.3]{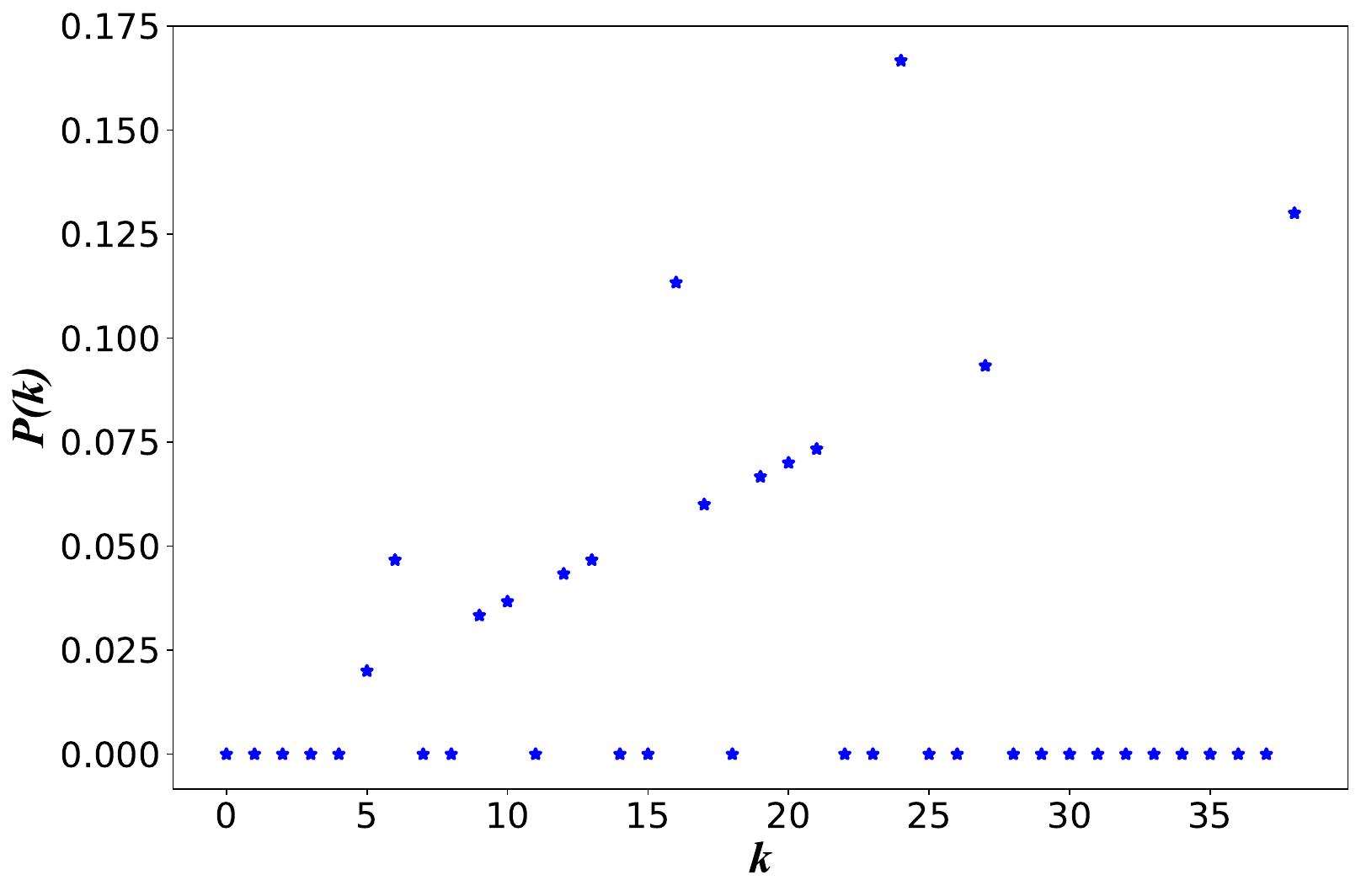}
\label{heuristic_degree}}
\subfigure[State transition ratio]{
\includegraphics[scale=0.3]{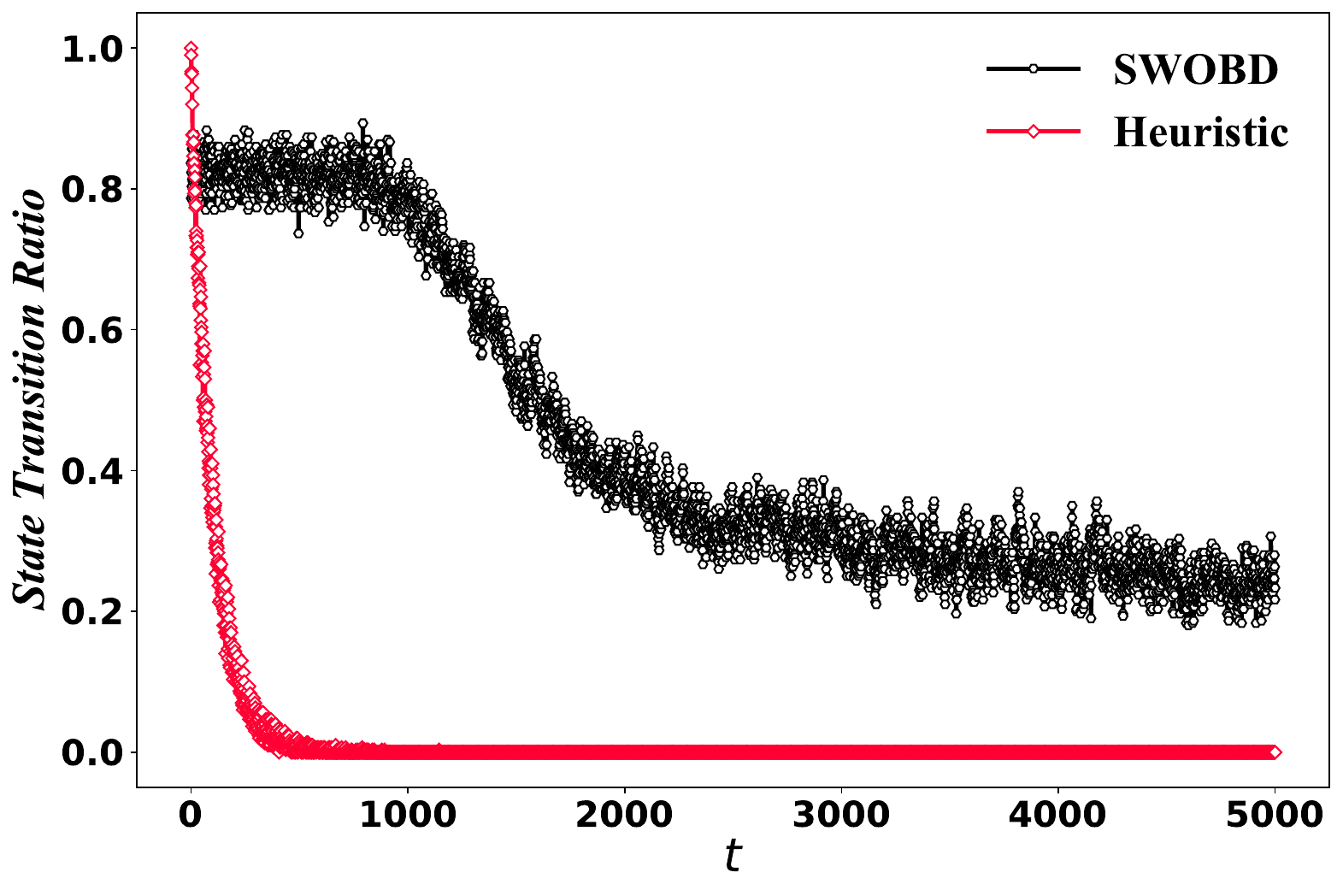}
\label{heuristic_state}}
\caption{\textbf{The performances of the heuristic model.} (a) is the degree distribution of the network generated by the heuristic model. (b) shows the evolutionary curves of the state transition ratio as time progresses for the SWOBD and heuristic models.}
\label{heuristic_performance}
\end{figure}
\end{center}

This subsection presents the performances of the heuristic model, and the results are illustrated in Fig. \ref{heuristic_performance}.

\vspace{-3\baselineskip}

\subsection{Further Results for Network Structure and Metrics}
\vspace{-1.5\baselineskip}

\begin{center}
\begin{figure*}[htbp]
\centering
\includegraphics[width=18cm,height=20cm]{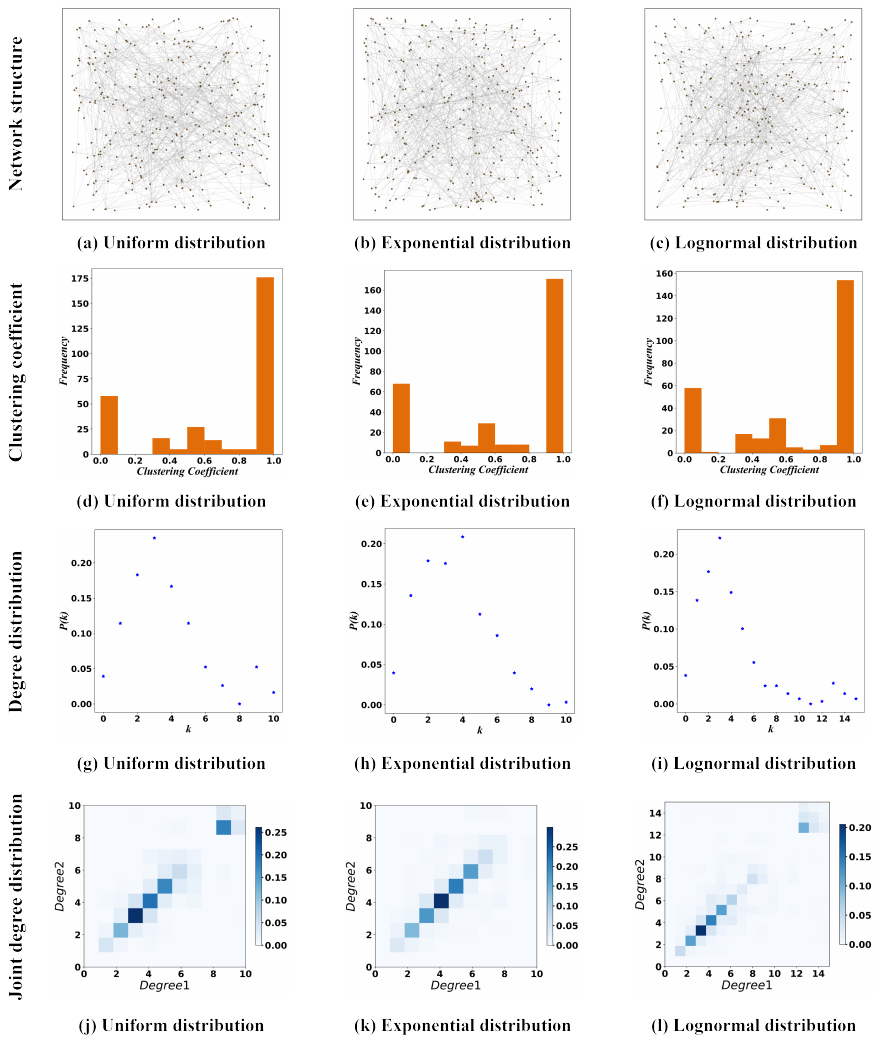}
\caption{\textbf{Some topological metrics of networks generated by SWBD with uniform, exponential, and lognormal distributions.} In this figure, we respectively demonstrate the networks and their corresponding topological metrics of SWBD with uniform (first column), exponential (second column), and lognormal (third column) distributions. The network structures are shown in subplots (a)-(c), clustering coefficients in subplots (d)-(f), degree distributions in subplots (g)-(i), and joint degree distributions in subplots (j)-(l).}
\label{classic_networks}
\end{figure*}
\end{center}

In this subsection, we study the network structure, clustering coefficient, degree distribution, and joint degree distribution of networks generated by the death process following uniform, exponential, and lognormal distributions, and the results are displayed in Fig. \ref{classic_networks}.

From Figs. \ref{classic_networks}(d)-(f), it is evident that most nodes in the networks generated by these three death processes exhibit high clustering coefficients, with values of 0.701, 0.676, and 0.671 for the uniform, exponential, and lognormal distributions, respectively. This suggests a significant level of aggregation among nodes in all three networks. The degree distributions illustrated in Figs. \ref{classic_networks}(g)-(i) reveal that while most nodes have relatively small degrees, there are also a few nodes with relatively larger degrees. Notably, the degree distributions for the death process obeying uniform and lognormal distributions are similar, both approximately following a power-law distribution. By plotting the joint degree distributions of the three networks, we observe from Figs. \ref{classic_networks}(j)-(l) that the numbers around and on the diagonal line in the heat maps are substantial. This indicates that all three networks exhibit assortative mixing and high degree correlation, consistent with findings obtained in the main manuscript where the death process adheres to a power-law distribution. Furthermore, the assortativity coefficients for networks generated by the uniform, exponential, and lognormal death processes are 0.874, 0.755, and 0.871, respectively. These relatively high values corroborate the observations from the joint degree distributions and align with our previous analysis.

\ifCLASSOPTIONcaptionsoff
  \newpage
\fi